\documentclass[aps,amsmath,amssymb,prd,nofootinbib,longbibliography,superscriptaddress]{revtex4-1}

\pdfoutput=1 

\usepackage{graphicx}
\usepackage{hyperref}
\usepackage{caption,subcaption} 
\usepackage[section]{placeins}
\usepackage[utf8]{inputenc}

\newcommand{\comment}[1]{}
\newcommand{\beq}[1]{\begin{equation}\label{#1}}
\newcommand{\eeq}{\end{equation}}

\begin{document}
\title{Phase Diagram of Stability for Massive Scalars in Anti-de Sitter
Spacetime}

\author{Brad Cownden}
\affiliation{Department of Physics \& Astronomy,\\ University of Manitoba\\
Winnipeg, Manitoba R3T 2N2, Canada}
\email{cowndenb@myumanitoba.ca}

\author{Nils Deppe}
\affiliation{Cornell Center for Astrophysics and Planetary Science and
Department of Physics,\\ Cornell University\\
122 Sciences Drive, Ithaca, New York 14853, USA}
\email{nd357@cornell.edu}

\author{Andrew R.~Frey}
\affiliation{Department of Physics \& Astronomy,\\ University of Manitoba\\
Winnipeg, Manitoba R3T 2N2, Canada}
\affiliation{Department of Physics and Winnipeg Institute for Theoretical
Physics,\\ University of Winnipeg\\
515 Portage Avenue, Winnipeg, Manitoba R3B 2E9, Canada}
\email{a.frey@uwinnipeg.ca}

\begin{abstract}
We diagram the behavior of 5-dimensional anti-de Sitter spacetime against 
horizon
formation in the gravitational collapse of a scalar field, treating the scalar
field mass and width of initial data as free parameters, which we call 
the stability phase diagram. We find that the class of stable initial data
becomes larger and shifts to smaller widths as the field
mass increases. In addition to
classifying initial data as stable or unstable, we identify two other classes
based on nonperturbative behavior.  The class of metastable initial data
forms a horizon over
longer time scales than suggested by the lowest order perturbation theory
at computationally accessible amplitudes,
and irregular initial data can exhibit non-monotonic and 
possibly chaotic behavior in
the horizon formation times. Our results include evidence for chaotic
behavior even in the collapse of a massless scalar field.
\end{abstract}


\maketitle

\section{Introduction}

Through the anti-de Sitter spacetime (AdS)/conformal field theory (CFT)
correspondence, string theory on AdS$_5\times X^5$ is
dual to a large $N$ conformal field theory in four spacetime dimensions
($\mathbb{R}\times S^3$ when considering global AdS$_5$).  The simplest
time-dependent system to study in this context is the gravitational dynamics
of a real scalar field with spherical symmetry, corresponding to the
time dependence of the expectation value of the zero mode of a single
trace operator in the gauge theory.  Starting with the pioneering work of
\cite{1104.3702,1106.2339,1108.4539,1110.5823}, numerical studies have
suggested that these dynamics may in fact be generically unstable
toward formation of (asymptotically) AdS$_{d+1}$ black holes
even for arbitrarily small amplitudes.  While perhaps surprising compared
to intuition from gravitational collapse in asymptotically flat spacetimes,
the dual picture of thermalization of small energies in a compact space
is more expected.  In terms of the scalar eigenmodes on a fixed AdS background,
the instability is a cascade of energy to higher frequency modes and shorter
length scales (weak turbulence), which eventually concentrates energy within
its Schwarzschild radius.  In a naive perturbation theory, this is evident
through secular growth terms.

However, some initial scalar field profiles lead to quasi-periodic evolution
(at least on the time scales accessible via numerical studies)
at small but finite amplitudes; even early work \cite{1104.3702,1109.1825}
noted that it is possible to remove the secular growth terms in the
evolution of a single perturbative eigenmode.  A more sophisticated
perturbation theory \cite{1403.6471,1407.6273,Basu:2014sia,1410.1880,1412.4761,1412.3249,Evnin:2015gma,1507.02684,1507.08261,1508.04943,1508.05474,1510.07836}
supports a broader class of quasi-periodic solutions that can contain
non-negligible contributions from many modes, and other stable solutions
orbit the basic quasi-periodic solutions \cite{1507.08261}.  Stable
solutions exhibit inverse cascades of energy from higher frequency to lower
frequency modes due to conservation laws following from the high symmetry
of AdS (integrability of the dual CFT).  Stable behavior also appears
in the full non-perturbative dynamics for initial profiles with widths near
the AdS length scale \cite{1304.4166,1307.2875,1308.1235}; however,
analyses of the perturbative and full dynamics in the literature have not
always been in agreement at fixed small amplitudes.  For example, some
perturbatively stable evolutions at finite amplitude actually form black holes
in numerical evaluation of the full dynamics
\cite{1403.6471,1410.2631,1506.07907}.  Understanding the breakdown of the
approximations used in the perturbative theory, as well as its region of
validity, is an active and important area of research
\cite{1506.03519,1606.02712,1607.08094,Dimitrakopoulos:2016euh,Liebling:2017gfn}.

Ultimately, the main goal of this line of inquiry is to determine whether
stability or instability to black hole formation (or both) is generic on
the space of initial data, so the extent of the ``islands of stability''
around single-mode or other quasi-periodic solutions and how it varies
with parameters of the physics on AdS are key questions of interest.  The
biggest changes occur in theories with a mass gap in the black hole spectrum,
such as AdS$_3$ and Einstein-Gauss-Bonnet gravity in AdS$_5$, which cannot
form horizons at small amplitudes.  While small-amplitude evolution in
AdS$_3$ appears to be quasi-periodic \cite{1306.0317,1412.6002}, there is
some evidence to point toward late-time formation of a naked singularity
in AdS$_5$ Einstein-Gauss-Bonnet gravity \cite{1608.05402,1410.1869}
(along with a power law energy spectrum similar to that at horizon
formation).  Charged scalar and gauge field matter \cite{1606.00830}
also introduces a qualitative change in that initial data may lead to
stable evolution or instability toward either Reissner-Nordstr\"om black holes
or black holes with scalar hair.

In this paper, we extend the study of massive scalar matter initiated in
\cite{1504.05203,1508.02709}.  Specifically, using numerical evolution of
the full gravitational dynamics, we diagram classes of gravitational
collapse behavior as a function of scalar field mass and initial scalar profile
width, which we call a stability phase diagram 
in analogy to a phase diagram for phases of matter.  
This is the first systematic study of behavior for classes of initial data
in AdS gravitational collapse using two tuning parameters.
By considering the time to horizon formation as a function of
the initial profile's amplitude at finite amplitude, 
we identify several different classes
of behavior and indicate them on the phase diagram.  Finally, we
analyze and characterize these different behaviors, presenting evidence for
chaotic behavior, including the first evidence for chaotic behavior in
the horizon formation time of
massless scalar collapse, which has no length scale other than the AdS
radius.  Throughout,
we work in AdS$_5$, due to its relevance to strongly coupled gauge theories
in four dimensions and because previous literature has indicated massless
scalars lead to greater instability than in AdS$_4$ (the main other case
considered), which makes the effects of the scalar field mass more visible.

We note briefly two caveats for the reader.  First, horizon formation
always takes an infinite amount of time on the AdS conformal boundary
due to the usual time dilation effects associated with horizons; this agrees
with the understanding of thermalization in the CFT as an asymptotic process.
Horizon formation times discussed in this paper correspond to an approximate
notion of horizon formation that we will describe below, but alternate
measures of thermalization may be of interest.  Second, the black holes
we discuss are smeared on the compact $X^5$ dimensions of the gravitational
side of the duality, as in most of the literature concerning stability of
AdS, and we are particularly interested in small initial amplitudes that
lead to black holes small compared to the AdS scale. As described in
\cite{hep-th/0202189,1502.01574,1509.07780}, small black holes in this
situation suffer a Gregory-Laflamme-like instability toward localization
on $X^5$ (which may in fact lead to formation of a naked singularity).  At
the same time, certain light stable solutions for charged scalars (boson
stars) are stable against localization on $X^5$ \cite{1509.00774}.  We
therefore provisionally assume that the onset of the Gregory-Laflamme-like
instability occurs only at horizon formation, not at any point of the
earlier horizon-free evolution.

The plan of this paper is as follows: in section \ref{s:review}, we review
the time scales associated with horizon formation with an emphasis on
the behavior of massive scalars and briefly discuss our methods.  Then,
in section \ref{s:phases}, we present the phase diagram of different
stability behaviors, and an attempt at heuristic analytic understanding 
appears in \ref{s:analysis}.
We close with a discussion of our results.

\section{Review}\label{s:review}
In this section, we review results on the stability of scalar field initial
data as well as our methods (following the discussion of \cite{1508.02709}).

\subsection{Massive scalars, stability, and time scales}

As in most of the literature, we work in Schwarzschild-like coordinates,
which have the line element (in asymptotic AdS$_{d+1}$)
\beq{metric}
  ds^2=\frac{1}{\cos^2(x)}\left(-Ae^{-2\delta}dt^2+
  A^{-1}dx^2+\sin^2(x)d\Omega^{d-1}\right)
\eeq
in units of the AdS scale.  In these coordinates, a horizon appears at
$A(x,t)=0$, but reaching zero takes an infinite amount of time (measured either
in proper time at the origin or in conformal boundary time); following
the standard approach, we define a horizon as having formed at the earliest
spacetime point (as measured by $t$) where $A$ drops below a specified
threshold defined in \S\ref{s:methods} below.  Of course, horizon formation
represents a coarse-grained description since the pure initial state of
the dual CFT cannot actually thermalize; a more precise indicator of
approximate thermalization may be the appearance of a power law energy spectrum
(exponentially cut off) in the perturbative scalar eigenmodes.  This
indicator is tightly associated with horizon formation (though see
\cite{1608.05402,1410.1869} for some counterexamples).  

A key feature of any perturbative formulation of the gravitational
collapse is that deviations from $A=1,\delta=0$ appear at order $\epsilon^2$,
where $\epsilon$ is the amplitude of initial data.  As a result,
horizons can form only after a time $t\sim\epsilon^{-2}$; in the
multiscale perturbation theory of \cite{1403.6471,1407.6273,1410.1880,1412.4761,1412.3249,1507.02684,1507.08261,1508.04943,1508.05474,1510.07836},
there is in fact a scaling symmetry
$\epsilon\to\epsilon',t\to t(\epsilon/\epsilon')^2$ that enforces the
proportionality $t_H\propto \epsilon^{-2}$, where $t_H$ is the (approximate)
horizon formation time for unstable initial data at small amplitude.

At this point, it is worth making a small clarification.  If the collapsing
matter takes the form of a well-defined pulse, horizon formation occurs when
the pulse nears the origin.  For massless matter, that means that
$t_H$ is piecewise continuous as a function of $\epsilon$;
each continuous ``step'' has approximately constant $t_H$
and is separated from the next step by a time of approximately
$\pi$, the light crossing time for a round trip from the origin to the boundary
of AdS.  Massive matter does not reach the boundary, so the steps are not
always separated by $\pi$, and may in fact not be separated at all if the pulse
spreads out in radius.  In any case, though, the width of the steps decreases
drastically as amplitude decreases, so it becomes very difficult to find
the transition amplitudes numerically.  In fact, adjacent amplitudes in a
numerical sample are typically multiple steps apart once the evolution is
already long, which justifies using the perturbative scaling
$t_H\propto \epsilon^{-2}$.

\begin{figure}[!t]
\centering
\begin{subfigure}[t]{0.47\textwidth}
\includegraphics[width=\textwidth]{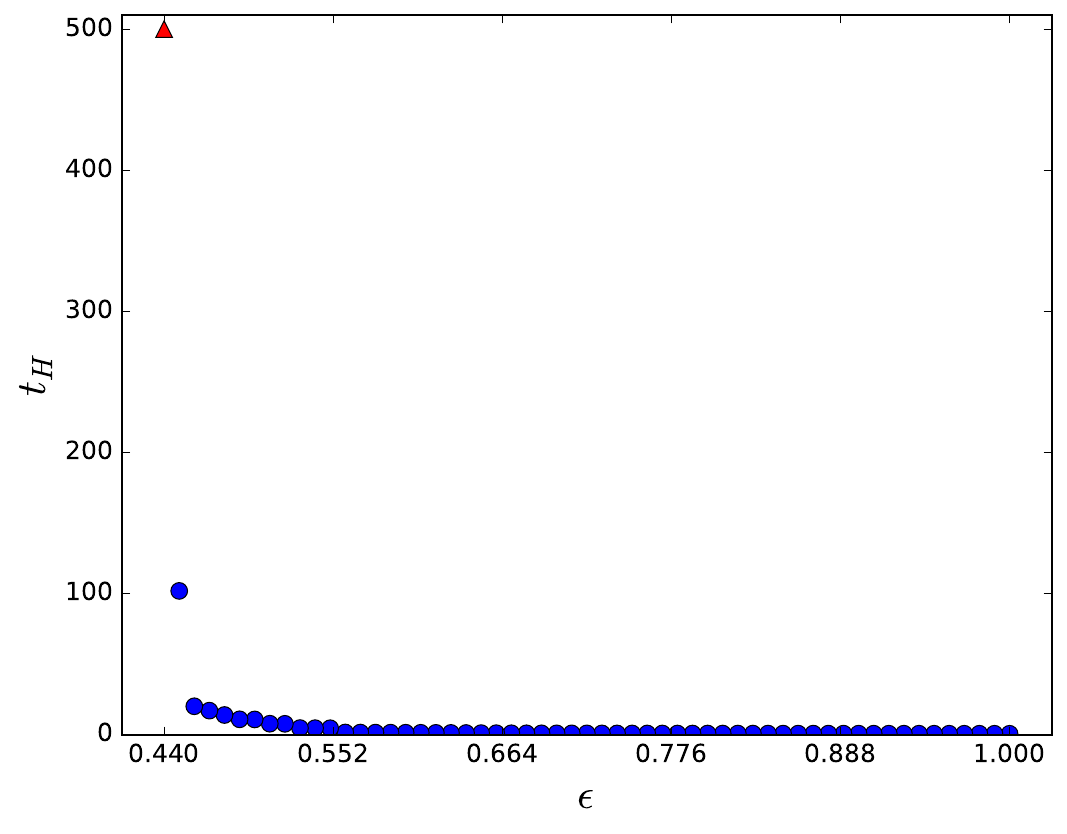}
\caption{\textit{Stable} initial data for $\sigma=1.5$}
\label{f:m0w15}
\end{subfigure}\hfill
\begin{subfigure}[t]{0.47\textwidth}
\includegraphics[width=\textwidth]{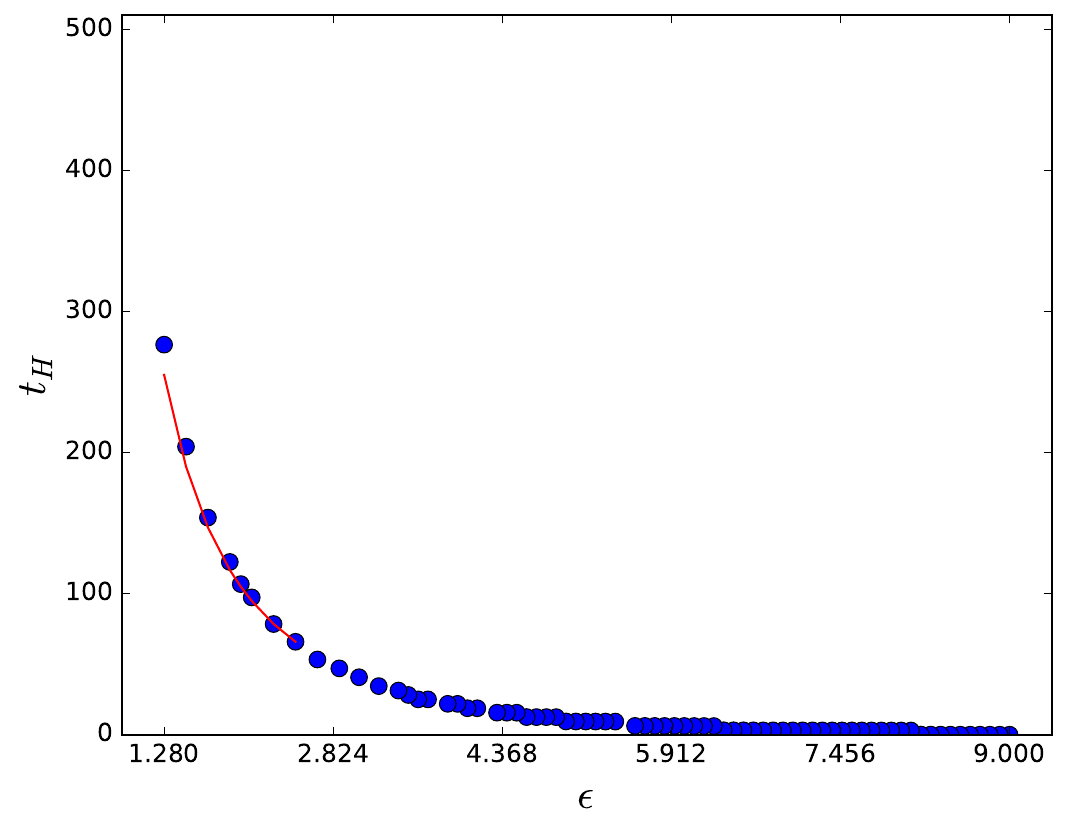}
\caption{\textit{Unstable} initial data for $\sigma=0.25$}
\label{f:m0w025}
\end{subfigure}
\begin{subfigure}[t]{0.47\textwidth}
\includegraphics[width=\textwidth]{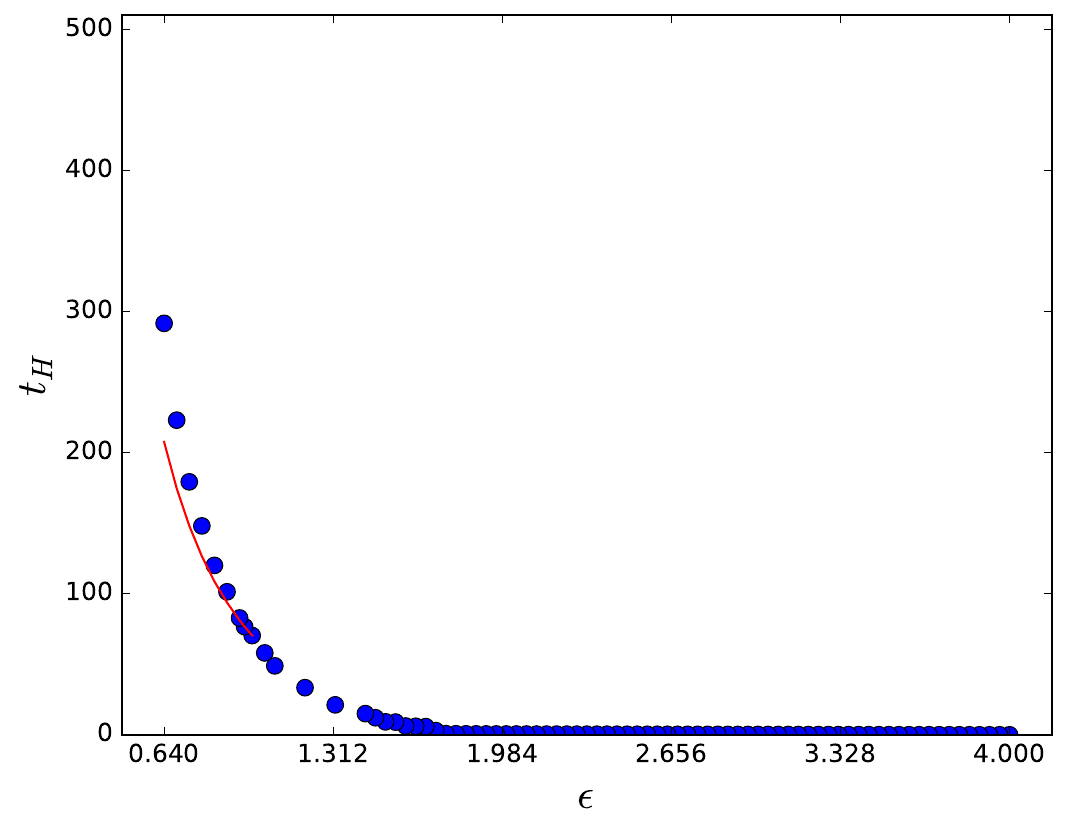}
\caption{\textit{Metastable} initial data for $\sigma=0.85$}
\label{f:m0w085}
\end{subfigure}\hfill
\begin{subfigure}[t]{0.47\textwidth}
\includegraphics[width=\textwidth]{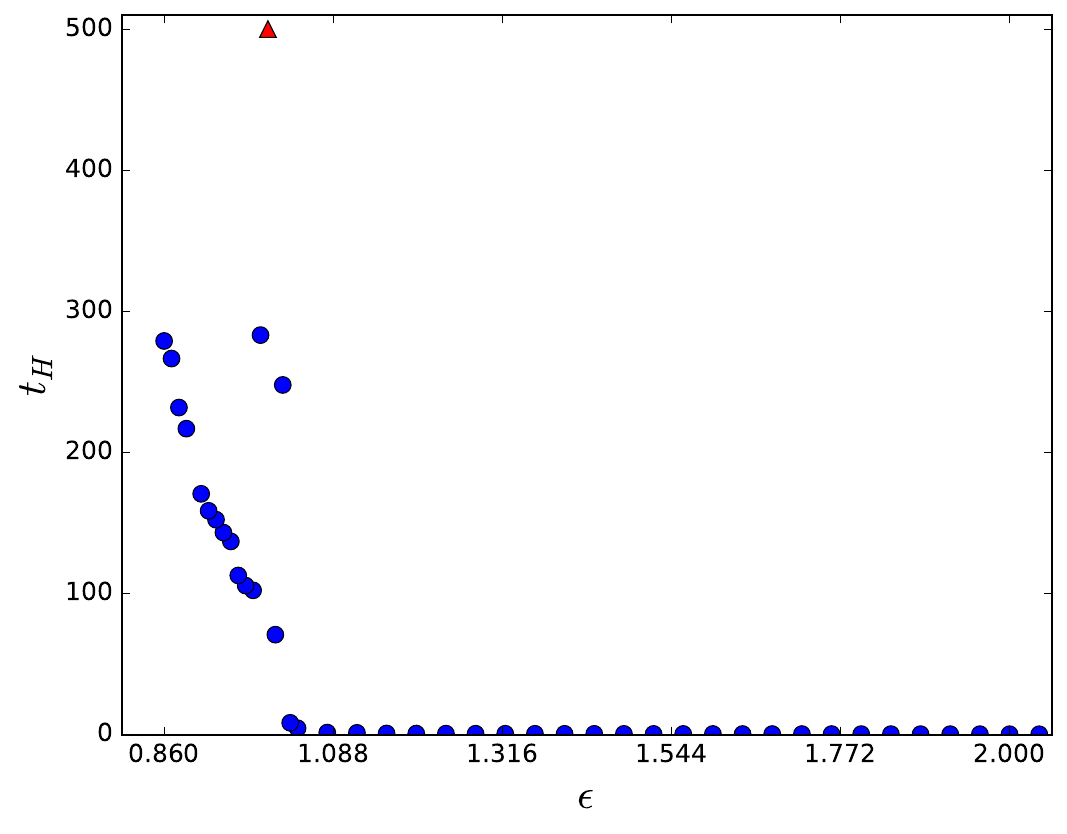}
\caption{\textit{Irregular} initial data for $\sigma=1.1$}
\label{f:m0w11}
\end{subfigure}
\caption{Classes of initial data for massless scalars and initial width
$\sigma$.  Blue dots represent horizon formation;
red triangles indicate a lower limit for $t_H$.  Red curves in subfigures
\ref{f:m0w025} and \ref{f:m0w085} are $t_H=a\epsilon^{-2}+b$ matched to the 
largest
two amplitudes in the curve.}
\label{f:classes}
\end{figure}

Based on the perturbative scaling relation,
initial data can be divided into several classes with respect
to behavior at low amplitudes, as illustrated in figure \ref{f:classes}
for massless scalars.  \textit{Stable} initial data evolve
indefinitely without forming a horizon.  In practice, we identify this
type of behavior in numerical evolutions by noting rapid horizon formation
at high amplitude with a vertical asymptote in $t_H$ just above some
critical amplitude.  In our numerical results, we see
a sudden jump at the critical amplitude to evolutions with no horizon formation
to a large time $t_{lim}$, possibly with a small
window of amplitudes with large $t_H$ just above the critical amplitude.
In a few cases, we have captured a greater portion of the asymptotic region.
See figure \ref{f:m0w15}.
\textit{Unstable} initial data, in contrast, forms a horizon at all
amplitudes following the perturbative scaling relation $t_H\propto \epsilon^{-2}$
as $\epsilon\to 0$. In our analysis, we will verify this scaling by
fitting $t_H$ to a power law as
described in section \ref{s:methods} below; if we limit the fit to smaller
values of $\epsilon$, the scaling becomes more accurate.  Figure \ref{f:m0w025}
shows unstable data. The red curve is of the form $t_H=a\epsilon^{-2}+b$
with $a,b$ determined by matching the curve to the data for the largest
two amplitudes with $t_H\geq 60$ (not a best fit); note that the data
roughly follow this curve.
The categorization of different initial data profiles with similar
characteristic widths into stable and unstable is robust for massless and
massive scalars \cite{1508.02709}; small and large width initial data are
unstable, while intermediate widths are stable.  One of the major results
of this paper is determining how the widths of initial data in these
``islands of stability'' vary with scalar mass.

A priori, there are other possible types of behavior, at least beyond
the first subleading order in perturbation theory, that is, at finite
$\epsilon$.
\textit{Metastable} initial data collapse with $t_H\propto\epsilon^{-p}$ with
$p>2$ at small but not arbitrarily small amplitudes 
(or another more rapid growth of $t_H$ with decreasing amplitude).  
We will find this type of behavior common on the
``shoreline'' of islands of stability where stable behavior transitions to
unstable.  As we will discuss further below, metastable
behavior may or may not continue as $\epsilon\to 0$;
in principle, as higher order terms in perturbation theory become less
important, the behavior may shift to either stable or unstable as described
above.  In principle, initial data that are stable at third order in
perturbation theory but unstable at higher order could have metastable
scaling even in the $\epsilon\to 0$ limit, though our numerical study cannot
address this case.
We in fact find circumstantial evidence in favor of the different
possibilities. In the case that the $\epsilon\to 0$ behavior is perturbatively
unstable, the perturbative scaling $t_H\propto \epsilon^{-2}$ only appears
for larger $t_H$ than the typical unstable case; it may therefore be
reasonable for the reader to consider metastable initial data as part of a
second order transition between unstable and stable classes of initial data.
Figure \ref{f:m0w085} shows metastable initial data that continue to collapse
to times $t_H\sim 0.6t_{lim}$ but more slowly than $\epsilon^{-2}$;
note that $t_H$ for collapsed evolutions at small amplitudes lies significantly
above the curve $t_H=a\epsilon^{-2}+b$ (which is determined as in
figure \ref{f:m0w025}).
There was one additional type of behavior identified by \cite{1508.02709},
which was called ``quasi-stable'' initial data at the time
since the low-amplitude
behavior was not yet clear.  We find here that these initial data are typically
stable at small amplitude but exhibit irregular
behavior in $t_H$ as a function of $\epsilon$, so we will denote them as
\textit{irregular} initial data; irregular behavior may be strongly 
non-monotonic or even exhibit some evidence of chaos.  
Figure \ref{f:m0w11} shows an example of
irregular initial data.  Later, we will see more striking examples of
this behavior for massive scalars.

We emphasize that we are not claiming that metastable or irregular behavior
persist to arbitrarily small amplitudes (though a priori metastable behavior
could). In that sense, the multiscale perturbation theory suggests  
that the only two classes of stability
behavior are stable and unstable with $t_H\propto\epsilon^{-2}$ scaling
as $\epsilon\to 0$. However, it is also important to understand physics
outside the perturbative regime, and classifying the behavior of AdS when
higher-order or nonperturbative effects contribute is still of interest.
For example, it is clear that metastable initial data (as defined precisely
below) do not exhibit perturbatively unstable behavior for $t_H$ values
as small as other unstable initial data, even in the cases where it may at all.
This may help understanding the breakdown of the multiscale perturbation theory.
Similarly, irregular initial data lead to qualitatively different behavior
even visually and suggests that nonperturbative dynamics are important.
It is in the spirit of looking beyond the multiscale perturbation theory that
we call metastable and irregular initial data independent classes of behavior, 
even if 
they are not quite on the same standing as perturbatively stable or unstable
classes. This paper presents the first systematic mapping of where these
distinct behaviors appear.

\subsection{Methods}\label{s:methods}

For spherically symmetric motion,
the Klein-Gordon equation for scalar mass $\mu$ can be written in first order
form as
\begin{align}
\phi_{,t}=&Ae^{-\delta}\Pi,\quad\Phi_{,t}=\left(Ae^{-\delta}\Pi\right)_{,x},
\label{evoleqns}\\
\Pi_{,t}=&\frac{(Ae^{-\delta}\tan^{d-1}(x)\Phi)_{,x}}{\tan^{d-1}(x)}-
\frac{e^{-\delta}\mu^2\phi}{\cos^2(x)}\ ,\label{KGequation}
\end{align}
where $\Pi$ is the canonical momentum and $\Phi=\phi_{,x}$ is an auxiliary
variable.  The Einstein equation reduces to constraints, which can be
written as
\begin{align}
  \label{deltaDeriv}
  \delta_{,x}=&-\sin(x)\cos(x)(\Pi^2+\Phi^2)\\
  \label{massDeriv}
  M_{,x}=&\left(\tan(x)\right)^{d-1}\left[
    A\frac{\left(\Pi^2+\Phi^2\right)}{2}+
    \frac{\mu^2\phi^2}{2\cos^2(x)}\right],\\
  A=&1-2\frac{\sin^2(x)}{(d-1)}\frac{M}{\tan^{d}(x)},
\end{align}
where the mass function $M$ asymptotes to the conserved ADM mass at the
boundary $x=\pi/2$.  We will restrict to $d=4$ spatial dimensions.
Since results are robust against changes in the type of
initial data \cite{1508.02709},
we can take the initial data to be a Gaussian of the areal radius in the
canonical momentum and trivial in the field.  Specifically,
\beq{PiGaussianID}
\Pi(t=0,x)=\epsilon\exp\left(-\frac{\tan^2(x)}{\sigma^2}\right),
\quad\phi(t=0,x)=0.
\eeq
The width $\sigma$ and field mass $\mu$ constitute the parameter space
for our stability phase diagram.

We solve the Klein-Gordon evolution equations (\ref{evoleqns},\ref{KGequation})
and Einstein constraint equations (\ref{deltaDeriv},\ref{massDeriv})
numerically using methods similar to those of \cite{1308.1235} on a spatial
grid of $2^n+1$ grid points; we discuss the
convergence properties of our code in the appendix.
We denote the approximate horizon position $x_H$ and formation time $t_H$
by the first point such that $A(x_H,t_H)\leq 2^{7-n}$.
In detail, we evolve the system in time using a 4th-order Runge-Kutta
stepper and initially use a 4th-order Runge-Kutta spatial integrator at
resolution $n=14$.  If necessary, we switch to a 5th-order Dormand-Prince
spatial integrator and increase resolution near horizon formation.  Due to
time constraints, we do not increase the resolution beyond $n=21$ for any
particular calculation; if a higher resolution would be required to track
horizon formation for a given amplitude, we exclude that amplitude.

To determine the stability class of initial data with a given width $\sigma$,
we allow evolutions to run to a maximum time of $t_{lim}=500$ in AdS units, so
$t_{lim}$ is a lower limit for $t_H$ for amplitudes that do not form a horizon
within that time.  Normally, however, if the initial data appear unstable,
we only evolve amplitudes with $t_H\lesssim 0.6t_{lim}$; this is partly to
save computational resources and partly to distinguish stable evolutions from
collapsing ones.  For unstable or metastable initial data, we find the
best fit of the form $t_H=a \epsilon^{-p}+b$ to evolutions with $t_H>t_{fit}$,
where $t_{fit}$ is a constant time chosen such that amplitudes with
evolutions that last longer are usually roughly perturbative;\footnote{The
power law plus constant fits the leading and first subleading contribution
to $t_H$ in a perturbative expansion in $\epsilon$, and we have found that
the subleading term is typically not negligible in the computationally 
accessible regime.}
in practice, $t_{fit}=60$ gives results close to the perturbative result
$p=2$ for evolutions expected to be unstable from the literature, but we will
also consider $t_{fit}=80,100$ as described below. In other words, since
a given amplitude $\epsilon$ may be in the perturbative scaling regime for
one set of initial data but nonperturbative for another, we compare 
initial data at similar horizon formation times (addressing the onset of 
perturbative behavior). Choosing $t_{fit}$ as above
gives consistent values of the fit parameters for the three values of $t_{fit}$
for the largest and smallest initial data widths, which are unstable.

\section{Phase Diagram of Stability}\label{s:phases}
Here we give our main result, the phase diagram of stability classes
as a function of initial profile width and scalar mass, along with a
more detailed discussion of the scaling of horizon formation time with
amplitude for varying initial data.

\begin{figure}[!t]
\centering
\includegraphics[width=\textwidth]{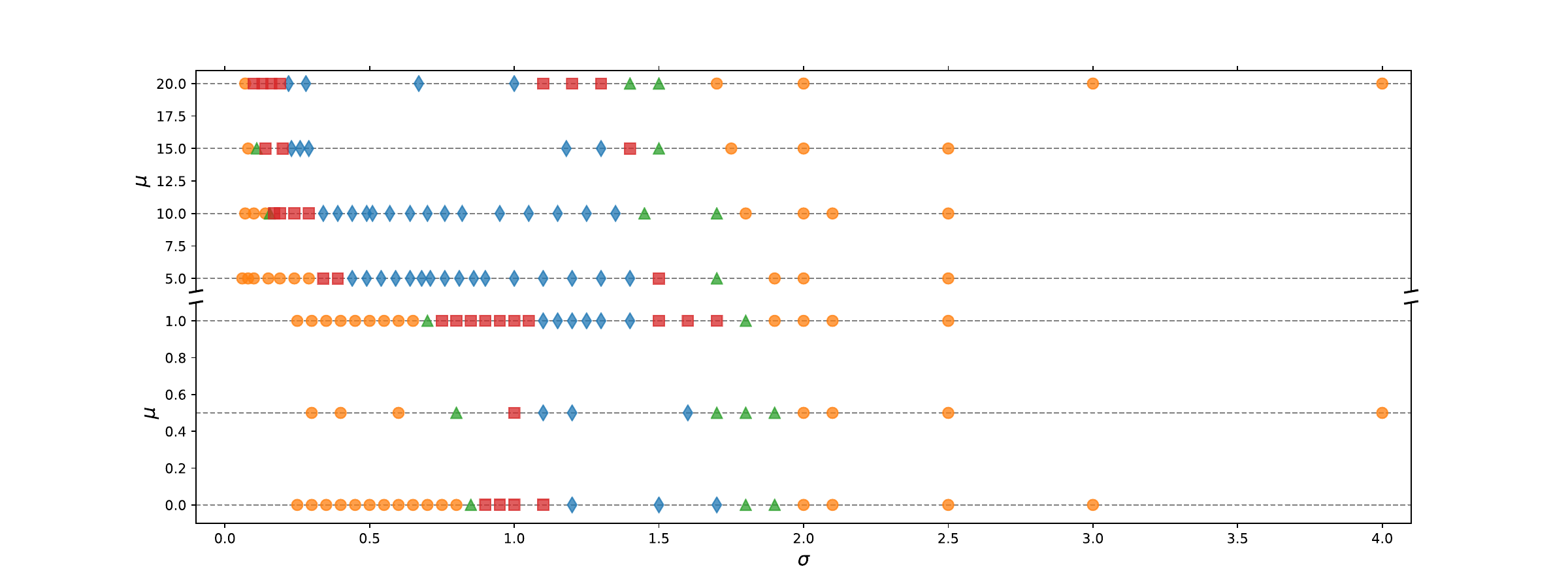}
\caption{Stability phase diagram as a function of initial data width 
$\sigma$ and scalar mass $\mu$.  Blue diamonds represent stable initial 
data, orange circles unstable initial data, green triangles
metastability, and red squares irregular behavior.}
\label{f:phase}
\end{figure}

The stability phase diagram for spherically
symmetric scalar field collapse in AdS$_5$, treating the width $\sigma$
of initial data and scalar field mass $\mu$ as tunable parameters,
appears in figure \ref{f:phase}.  Each $(\mu,\sigma)$ combination that
we evolved numerically is indicated by a point, with blue diamonds and 
orange circles representing stable and unstable initial data respectively.  The
metastable class is represented by green triangles, while
the irregular class is represented by red squares. Note that the graph has been
divided into two regions with different scales, separated by a break: 
$0 \leq \mu \leq 1$ is pictured on the bottom, while $5 \leq \mu \leq 20$ 
is pictured on the top.  At a glance, two features
of the stability phase diagram are apparent: as $\mu$ increases, the island
of stability moves toward smaller values of $\sigma$ and takes up a
gradually larger range of $\sigma$.  To be specific, the stable class of initial
data is centered at $\sigma=\bar\sigma\sim 1.4$ and has a width of
$\Delta\sigma\sim 0.7$ for $\mu=0,0.5$, with $\bar\sigma\sim 1.2$ for $\mu=1$.
$\Delta\sigma$ increases to $\sim 1.1$, and the island of
stability is centered at
$\bar\sigma\sim 0.9$ for $\mu=5,10$, while $\Delta\sigma\sim 1.2$
for $\mu=15,20$ with the stable class centered at $\bar\sigma\sim 0.8$.
Note that the transition between ``light field'' and ``heavy field'' 
behavior occurs for $\mu>1$ in AdS units.

The metastable and irregular classes appear at the shorelines of the island
of stability, the boundary between unstable and stable classes.  In particular,
the slope of the power law $t_H\sim \epsilon^{-p}$ as $\epsilon\to 0$
increases as the width moves toward the island of stability, leading to 
metastable behavior.  We find metastability at the large $\sigma$ shoreline
for all $\mu$ values considered and also at the small $\sigma$ shoreline
for several scalar masses.  It seems likely that metastable behavior appears
in only a narrow range of $\sigma$ for larger $\mu$, which makes it harder
to detect in a numerical search, leading to its absence in some parts of the
stability phase diagram.  We find irregular behavior 
at the small $\sigma$ shoreline
for every mass and at the large $\sigma$ boundary for large $\mu$,
closer to stable values of $\sigma$ than metastable initial data.  This class
of initial data
includes a variety of irregular and non-monotonic behavior, as detailed below.
Evidence for chaotic behavior especially becomes more prominent at larger 
values of $\mu$, as we will discuss below.

\subsection{Metastable versus unstable initial data}\label{s:metastable}

While stable and irregular initial data are typically apparent by eye in a plot
of $t_H$ vs $\epsilon$, distinguishing the unstable from metastable classes
is a quantitative task.  As we described in section \ref{s:methods}, we
find the least squares fit of $t_H=a\epsilon^{-p}+b$ to all evolutions
with $t_H>t_{fit}$ for the given $(\mu,\sigma)$, running over values
$t_{fit}=60,80,100$.  Using the covariance matrix of the fit, we also find
the standard error for each fit parameter.  We classify a width as having
unstable evolution if the best fit value of $p$ is within two standard
errors of $p=2$ for $t_{fit}=60,80$ or one standard error for $t_{fit}=100$
(due to a smaller number of data points, the standard errors for $t_{fit}=100$
tend to be considerably larger).  In contrast, we classify a width
as having metastable evolution if the best fit $p$ is statistically
significantly different from $2$ (in that the best fit value is more than 2
standard errors from $p=2$ for $t_{fit}=60,80$ and more than 1 standard error
from $p=2$ for $t_{fit}=100$). This indicates that either further subleading
contributions in a perturbative expansion of $t_H$ are non-negligible in
this regime for metastable initial data or that possibly metastable initial
data are stable at the first nontrivial order in perturbation theory.
Considering larger values of $t_{fit}$
helps to ensure that the leading perturbative terms do not come to 
dominate for particular initial profile at the smallest computationally
accessible amplitude values.  In the case that the fit
to $t_H=a\epsilon^{-p}+b$ has large reduced $\chi^2$ or is sensitive to
fitting algorithm, the data are not well-described by our fitting function,
so we classify it as irregular (see the next subsection).

\begin{figure}[t]
\centering
\includegraphics[width=0.7\textwidth]{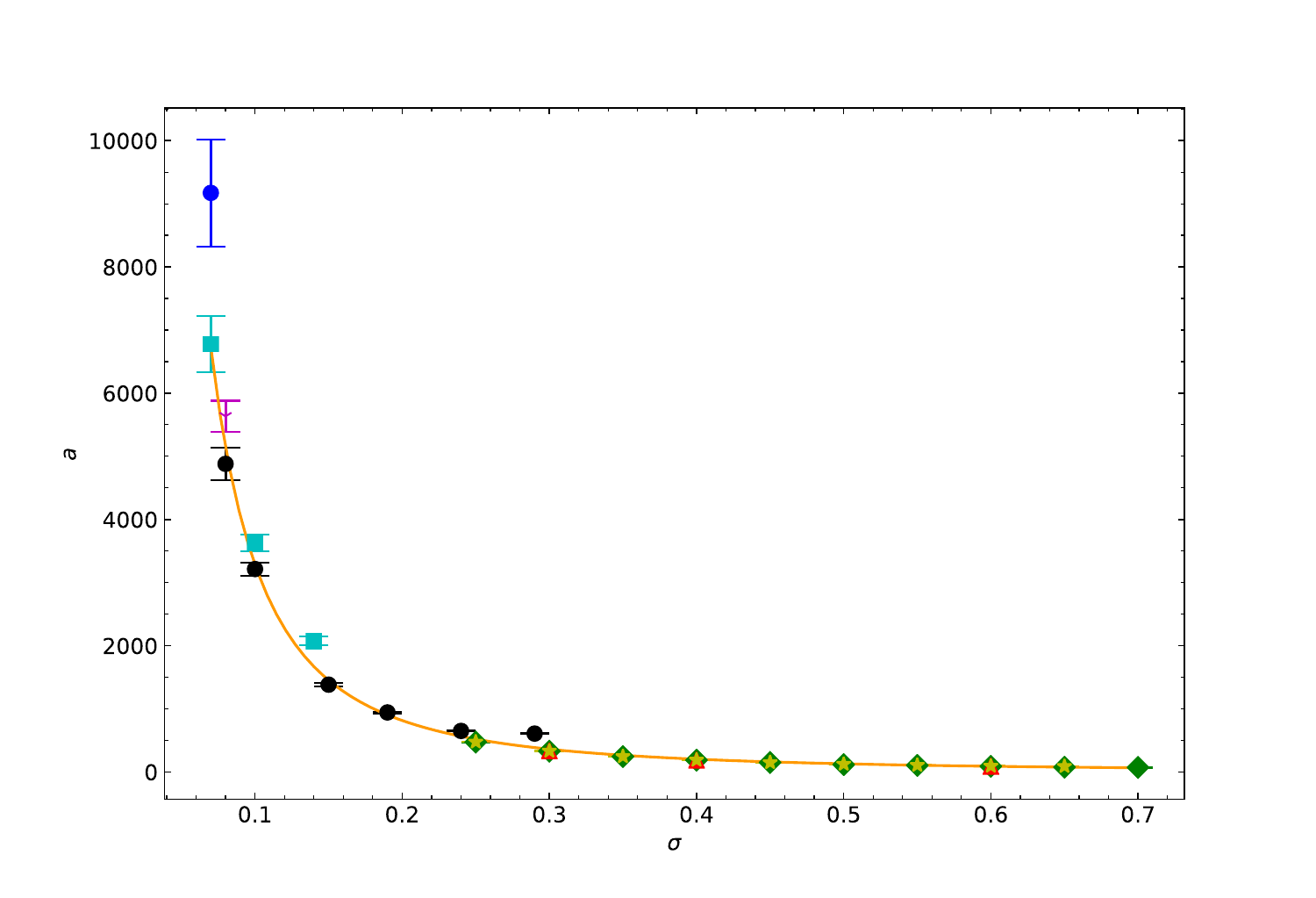}
\caption{Coefficient $a$ from the fit $t_H=a\epsilon^{-p}+b$ as a function
of width $\sigma$ using $t_{fit}=60$. 
Shows data for $\mu=0$ (green diamonds), $0.5$ 
(red triangles), $1$ (yellow stars), 5 (black circles), 10 (cyan squares),
15 (magenta Y), and 20 (blue circles).  The orange line is the best 
power law fit.}
\label{f:unstable}
\end{figure}

The fits $t_H=a\epsilon^{-p}+b$ allow us to explore the time scale of 
horizon formation across the stability phase diagram, 
for example through a contour
plot of one of the coefficients vs $\sigma$ and $\mu$.  In most cases,
this has not been informative, but an intriguing feature emerges if we
plot the normalization coefficient $a$ vs $\sigma$ for unstable initial 
data at small $\sigma$, as shown in figure \ref{f:unstable} for $t_{fit}=60$.  
By eye, the coefficient is reasonably well described by the fit
$a=(32.0\pm 0.3) \sigma^{-(2.01\pm 0.02)}$ (values following $\pm$ 
are standard errors of
the best fit values) \textit{independent of scalar field mass}.
This is not born out very well quantitatively; the reduced $\chi^2$ for the
fit is $\chi^2$/degrees of freedom (d.o.f.)$=180$, 
indicating a poor fit.  However, the large
$\chi^2$ seems largely driven by a few outlier points with large
scalar mass, so it is tempting
to speculate that the gravitational collapse in this region of parameter
space is driven by gradient energy, making all fields effectively massless
at narrow enough initial $\sigma$.  The picture is qualitatively similar
if we consider the parameter $a$ for $t_{fit}=80,100$ instead.

\begin{figure}[!t]
\centering
\begin{subfigure}[t]{0.47\textwidth}
\includegraphics[width=\textwidth]{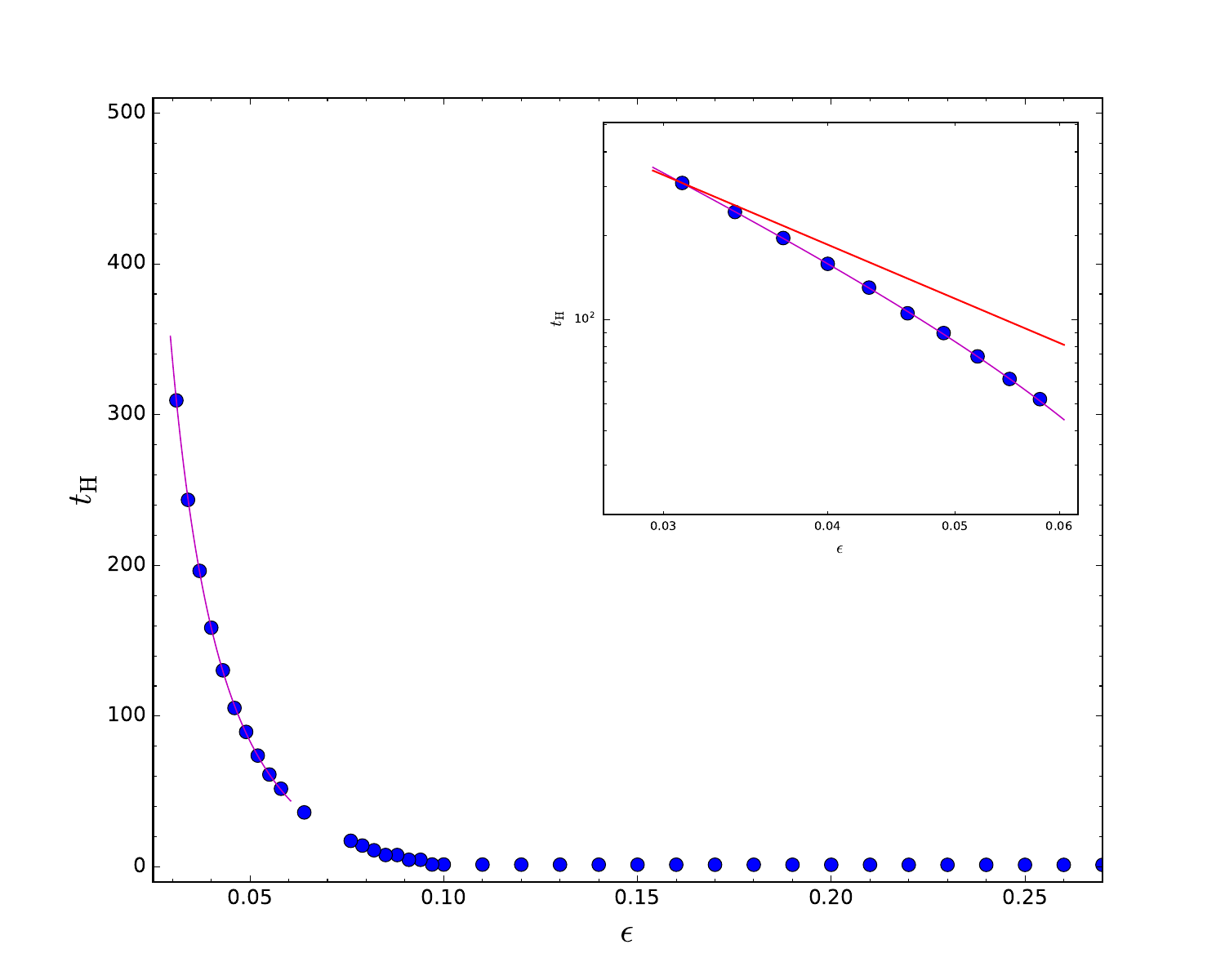}
\caption{$\mu=15,\sigma=1.5$}
\label{f:m15w150}
\end{subfigure}\hfill
\begin{subfigure}[t]{0.47\textwidth}
\includegraphics[width=\textwidth]{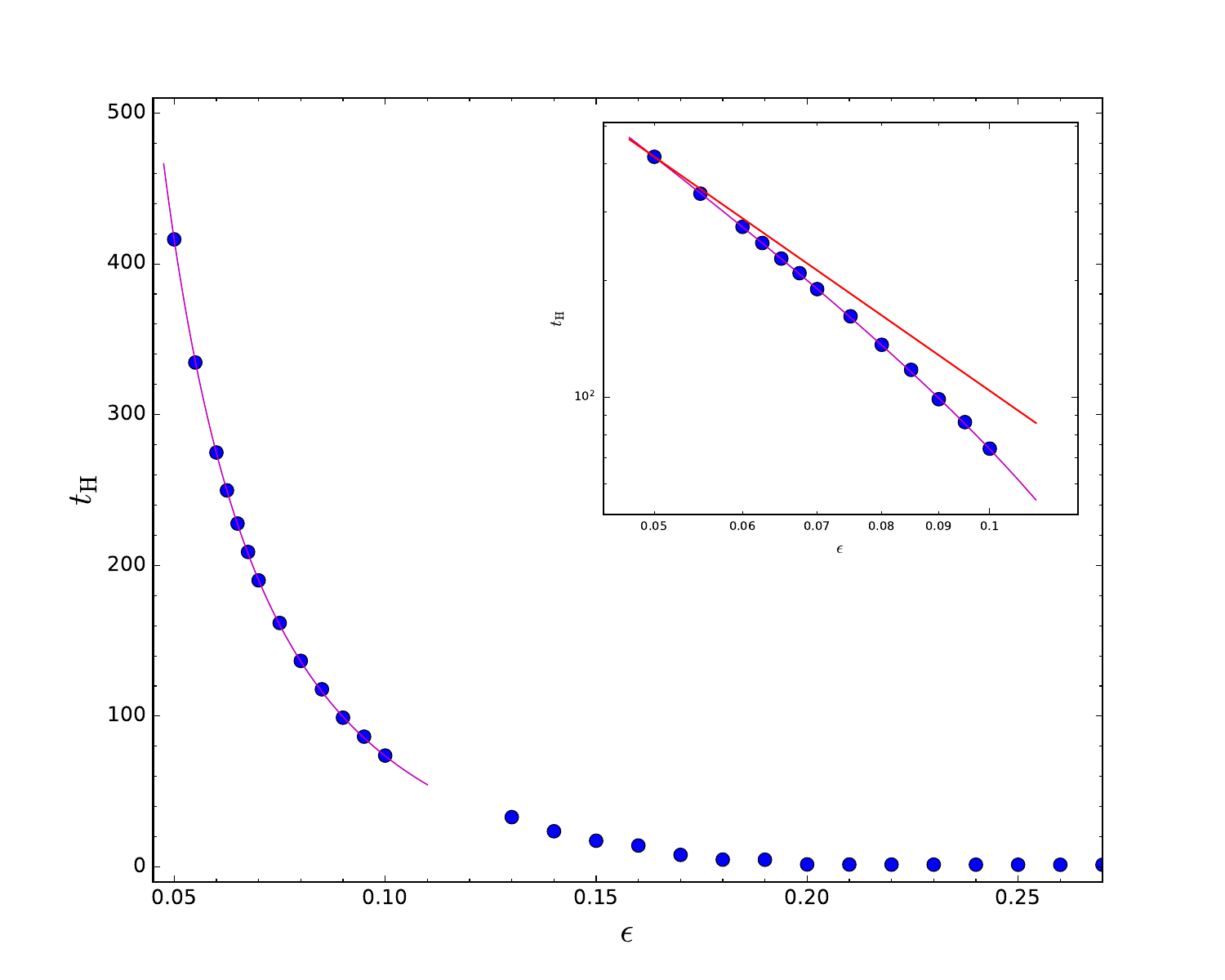}
\caption{$\mu=5,\sigma=1.7$}
\label{f:m5w170}
\end{subfigure}
\begin{subfigure}[t]{0.47\textwidth}
\includegraphics[width=\textwidth]{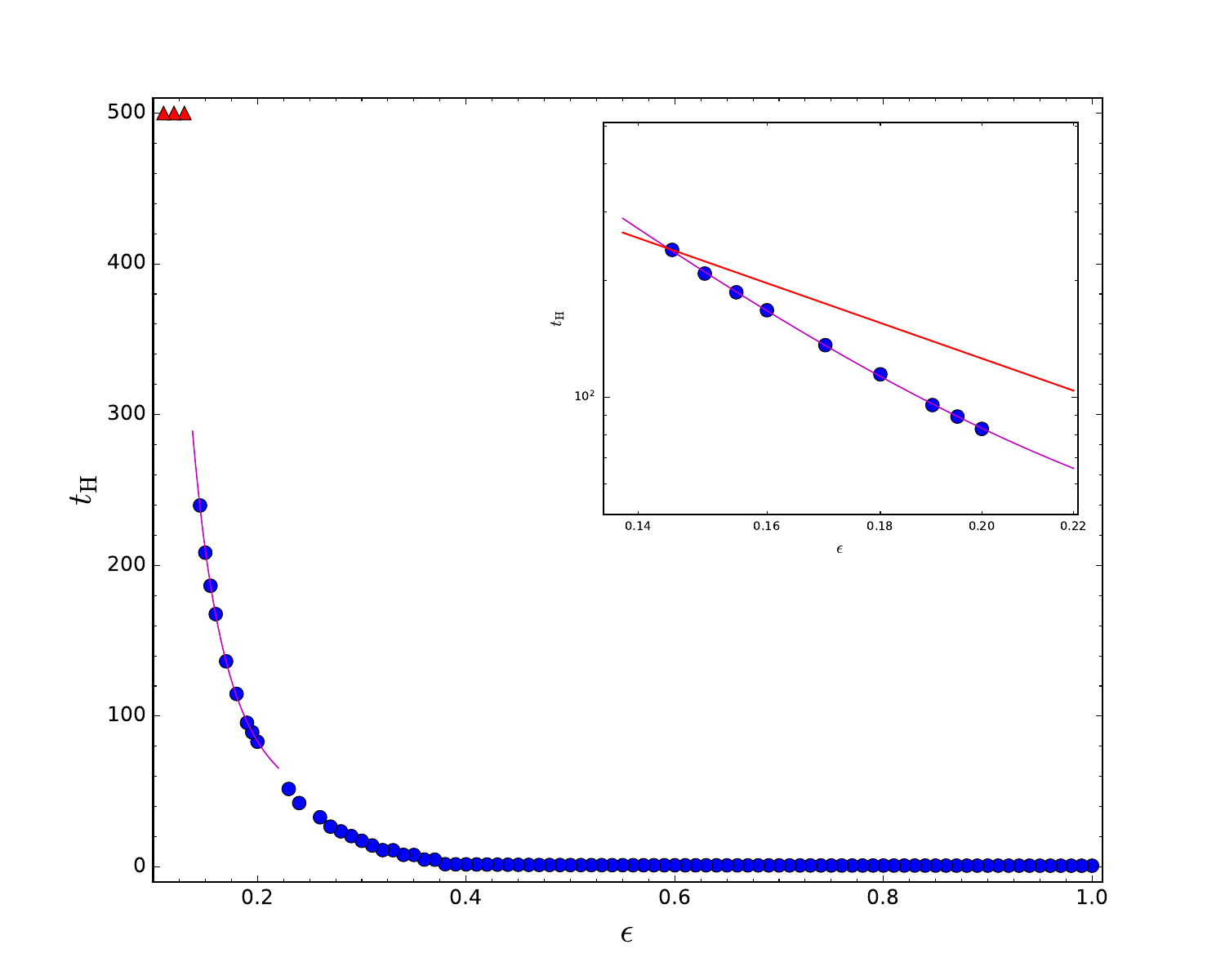}
\caption{$\mu=0,\sigma=1.8$}
\label{f:m0w18}
\end{subfigure}\hfill
\begin{subfigure}[t]{0.47\textwidth}
\includegraphics[width=\textwidth]{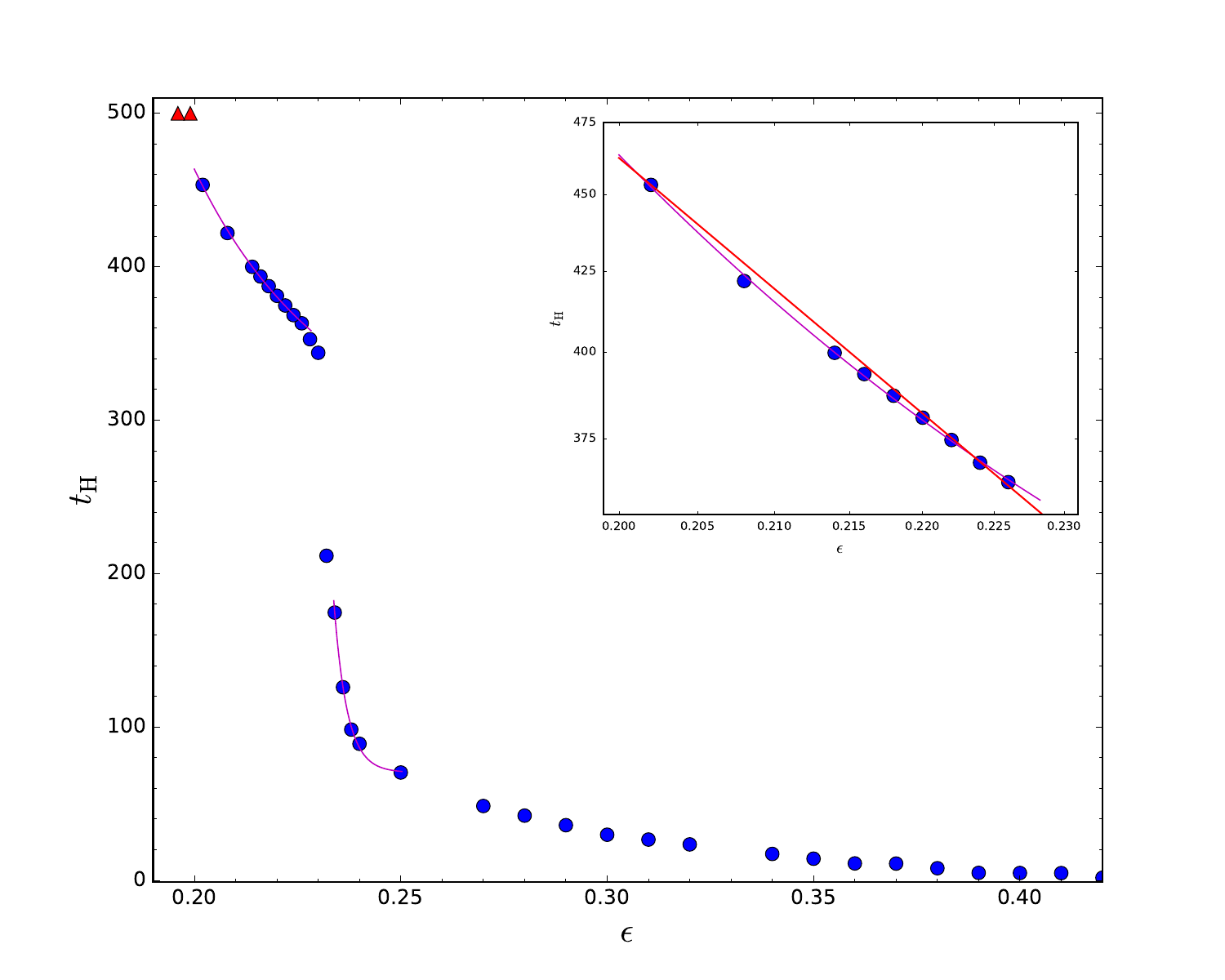}
\caption{$\mu=0.5,\sigma=1.7$}
\label{f:m05w17}
\end{subfigure}
\caption{Metastable behavior: blue dots represent horizon formation and
red triangles a lower limit on $t_H$.  Magenta curves are fits
$t_H=a\epsilon^{-p}+b$ over the shown range of amplitudes. Insets show the
fit region with log-log scale; note that the fit is not strictly a power
law, so the fits are not straight lines. See table \ref{t:figfits} for best 
fit parameters. Red lines in insets are $\epsilon^{-2}$ power laws normalized
to the $t_H$ of the smallest amplitude shown.}
\label{f:metastable}
\end{figure}

Several examples of metastable behavior appear in figure \ref{f:metastable}.
These figures show both data from the numerical evolutions (blue dots and
red triangles) and fits of the form $t_H=a\epsilon^{-p}+b$ for points with
$t_H>t_{fit}=60$ (magenta curves).  The best fit parameters are given in
table \ref{t:figfits} along with the standard errors (listed following
$\pm$ for the fit values) and $\chi^2$ values. The insets show the fit 
region with a log-log scale and an additional line (red) showing an 
$\epsilon^{-2}$ power law normalized to fit the smallest amplitude shown in
the inset. It is visually clear that $t_H$ grows faster than $\epsilon^{-2}$
for all these examples as $\epsilon$ decreases in the fit region (there is
a significant constant offset in figure \ref{f:m05w17}).

Figures \ref{f:m15w150},\ref{f:m5w170} demonstrate behavior typical of most
of the instances of metastable initial data we have found; specifically,
the initial data continue to collapse through horizon formation times of
$t_H\sim 0.6 t_{lim}$ but with $p$ significantly greater than the perturbative
value of $p=2$. Note that the evolutions of figure \ref{f:m5w170} have
been extended to larger
values of $t_H$ to demonstrate that the evolutions continue to collapse to
somewhat smaller amplitude values.  Figure \ref{f:m5w170} is also of
interest because its best fit value $p\approx 2.07\pm 0.02$ 
is approximately as close
to the perturbative value as several stable sets of initial data but has a
smaller standard error for the fit, so the difference from the perturbative
value is more significant (again, the value following the $\pm$ is the 
standard error).

\begin{table}[!t]
\begin{center}
\begin{tabular}{|c|c|c|c|c|}
\hline
&$a$&$p$ &$b$ & $\chi^2/$d.o.f.\\
\hline
$\mu=15,\sigma=1.5$& $0.10\pm 0.01$&$2.33\pm 0.05$& $-27\pm 4$&0.7736\\
\hline
$\mu=5,\sigma=1.7$&$0.91\pm 0.06$ & $2.07\pm 0.02$& $-33\pm 2$&0.5070\\
\hline
$\mu=0,\sigma=1.8$&$0.06\pm 0.02$&$4.3\pm 0.2$& $30\pm 5$& 1.502\\
\hline
$\mu=0.5,\sigma=1.7$ ($t_H<0.4t_{lim}$)&$(4\pm 32)\times 10^{-45}$ &$73\pm 5$ 
&$70\pm 2$ &5.409 \\
($t_H>0.72t_{lim}$)& $0.02\pm 0.03$ & $5.6\pm 0.8$ & $260\pm 20$ &1.078 \\
\hline
\end{tabular}
\end{center}
\caption{Best fit parameters for the cases shown in figure \ref{f:metastable}
restricting to $t_H>t_{fit}=60$ and as noted.  Listed errors ($\pm$ values) are
standard errors. $\chi^2/$d.o.f. is the
reduced $\chi^2$ value used as a measure of goodness-of-fit. }
\label{t:figfits}
\end{table}

Figure \ref{f:m0w18} shows metastable evolution to $t_H\lesssim 0.6t_{lim}$
but then a sudden jump to stability until $t=t_{lim}$.  In the figure, the
fit has been extended to the largest non-collapsing amplitude,
which demonstrates that
there is no collapse over a time period significantly longer than the fit
predicts. This example argues that metastable data
may in fact become stable at the smallest amplitudes.  On the other hand,
figure \ref{f:m05w17} shows a similar jump in $t_H$ to values $t_H<t_{lim}$;
evolution at lower amplitudes shows metastable scaling with
$p\approx 5.6\pm 0.8$ 
for $360<t_H<t_{lim}$.  The figure also shows a metastable fit 
with larger reduced $\chi^2$ at
larger amplitudes corresponding to $t_{fit}<t_H<0.4 t_{lim}$.
So this is another option: metastable behavior
may transition abruptly to metastable behavior with different scaling
(or possibly even perturbatively unstable behavior) at sufficiently
small amplitudes. It is also reasonable to classify this case as irregular
due to the sudden jump in $t_H$; we choose metastable due to the clean 
metastable behavior at low amplitudes.

Our point of view is that initial data in the metastable class are distinct
from the unstable class at finite amplitudes corresponding to $t_{fit}<t_H<300$; 
they take longer to collapse
at a fixed small value of $\epsilon$ than would be expected by the perturbative
scaling. An alternate point of view is to ask whether we can determine if
a given set of initial data is perturbatively unstable in the $\epsilon\to 0$
limit. We have already seen that metastable initial data do not follow
the perturbative scaling when fit to $t_H=a\epsilon^{-p}+b$, the first
two terms of the perturbative expansion. However, it is possible that a
perturbative description applies but requires a further subleading term.
To test this hypothesis, we fit unstable and metastable initial data to
$t_H=a\epsilon^{-p}+b+c\epsilon^2$; as described earlier in this section,
we determine if $p$ is within two standard errors of the perturbative value
$p=2$ (or one standard error for $t_{fit}=100$). 

The unstable class of initial data is instructive. For the new fits of
unstable initial data, $p$ is statistically equal to 2, and the new values
of $a,p,b$ are consistent with the values from the old fits to within two
standard errors (or sometimes slightly more). The fit value of $c$ is 
uniformly within a standard error of zero, and, for the amplitude values
in the fit region, the $\epsilon^2$ term is small compared to the constant
and $\epsilon^{-2}$ terms. What is more, for some unstable initial data near
the island of stability, the original $t_H=a\epsilon^{-p}+b$ fits for 
$t_{fit}=60$ have $p>2$ statistically; on the other hand, the new fits have 
$p=2$ within statistical error. In other words, the perturbative expansion
is still valid but requires more terms. Part of the metastable class of
initial data also behaves in this manner and could therefore be reasonably
considered to be perturbatively unstable. Of the metastable initial
data we found, these are $\sigma=1.9$ for $\mu=0$, $\sigma=0.8$ and 1.9 
for $\mu=0.5$, $\sigma=0.7$ for $\mu=1$, $\sigma=1.7$ for $\mu=5$, 
$\sigma=0.155$ for $\mu=10$, $\sigma=0.11$ and 1.5 for $\mu=15$, and 
$\sigma=1.5$ for $\mu=20$. In addition, $\mu=1,\sigma=1.8$ and 
$\mu=10,\sigma=1.7$ initial data have similar behavior, but $p$ is not
statistically consistent with 2 for any of the fit regions, though it is
closer than in the original fits. On the other hand, the other metastable
initial data ($\sigma=0.85$ and 1.8 for $\mu=0$, $\sigma=1.7$ and 1.8 for 
$\mu=0.5$, $\sigma=1.45$ for $\mu=10$, and $\sigma=1.4$ for $\mu=20$) show 
no evidence for perturbative behavior. Specifically, $p$ remains statistically
larger than 2 for all fits, the $\epsilon^2$ term in the new fit is roughly
the same magnitude as the other terms, and the $a,p,b$ values in the new
fits are not statistically consistent with the original fits.

To check if perturbative scaling might be masked by numerical errors, we
have also fit these remaining metastable data ($\sigma=0.85,1.8$ for $\mu=0$,
$\sigma=1.8$ for $\mu=0.5$, $\sigma=1.8$ for $\mu=1$, $\sigma=1.45,1.7$ for 
$\mu=10$, and $\sigma=1.4$ for $\mu=20$) with 
$t_H=a\epsilon^{-p}+b\epsilon^{-1}+c$. Of these, only the $\mu=1,\sigma=1.8$ and 
$\mu=10,\sigma=1.7$ initial data have best fit $p$ values statistically 
consistent with $p=2$. However, except for $\mu=1,\sigma=1.8$ initial data,
the best fit $p$ values are all further from $p=2$ in absolute terms (usually
substantially); the
main effect of including the $\epsilon^{-1}$ term is to increase the standard
error on the best fit for $p$. Therefore, it is not clear that potential
numerical errors alone can be responsible for the observed deviation from
perturbative scaling. We would 
also point out that, even if the extra $\epsilon^{-1}$ term turns out to be
important for these initial data, the fact that it is only important at 
the boundary of the island of stability indicates a change in behavior for
these mass/width combinations as compared to those farther from the stable
region. This justifies a separate classification related to the slower entry
of these mass/width combinations into the perturbative regime (as measured
by horizon formation time).

\subsection{Irregular behaviors}

\begin{figure}[!t]
\centering
\begin{subfigure}[t]{0.47\textwidth}
\includegraphics[width=\textwidth]{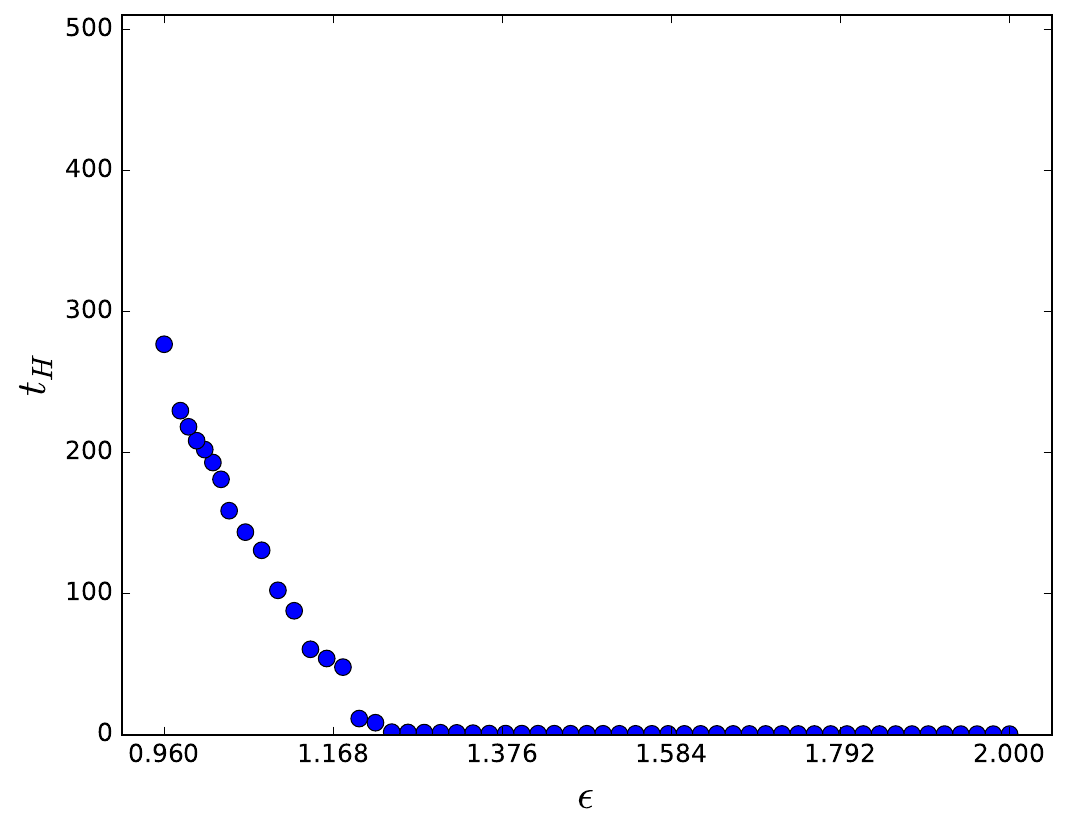}
\caption{$\mu=0.5,\sigma=1$}
\label{f:m05w1}
\end{subfigure}\hfill
\begin{subfigure}[t]{0.47\textwidth}
\includegraphics[width=\textwidth]{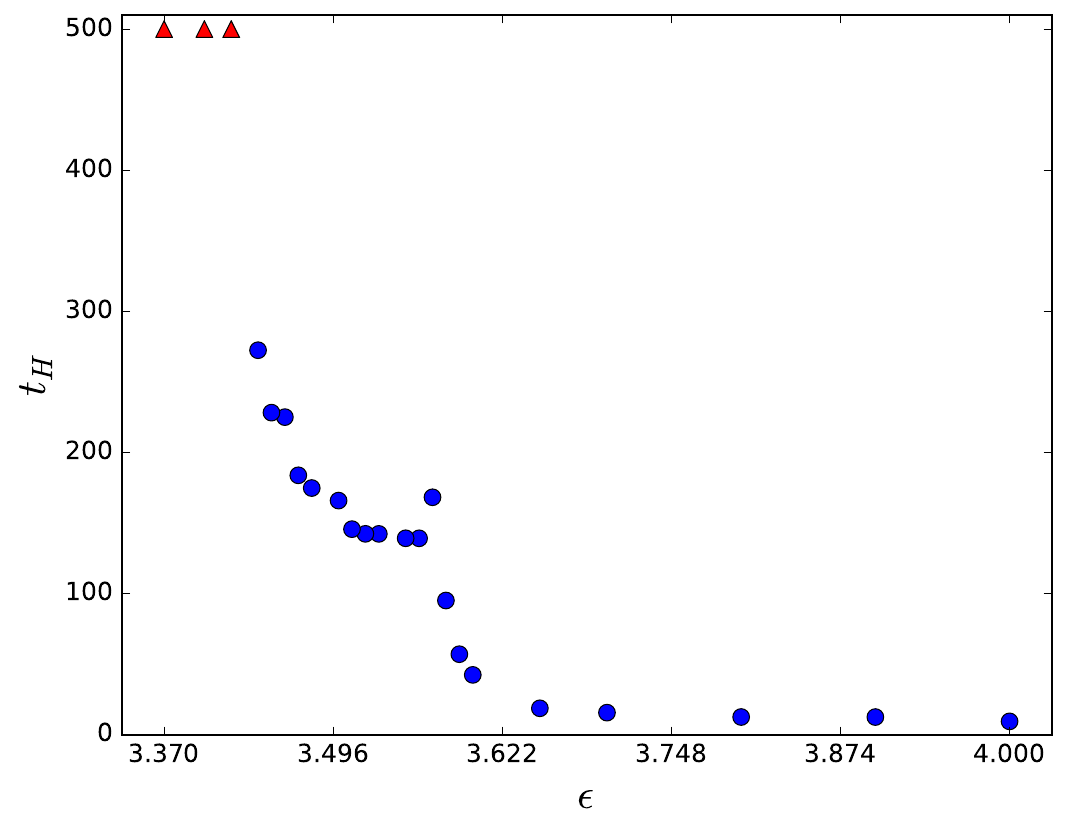}
\caption{$\mu=5,\sigma=0.34$}
\label{f:m5w034}
\end{subfigure}
\begin{subfigure}[t]{0.47\textwidth}
\includegraphics[width=\textwidth]{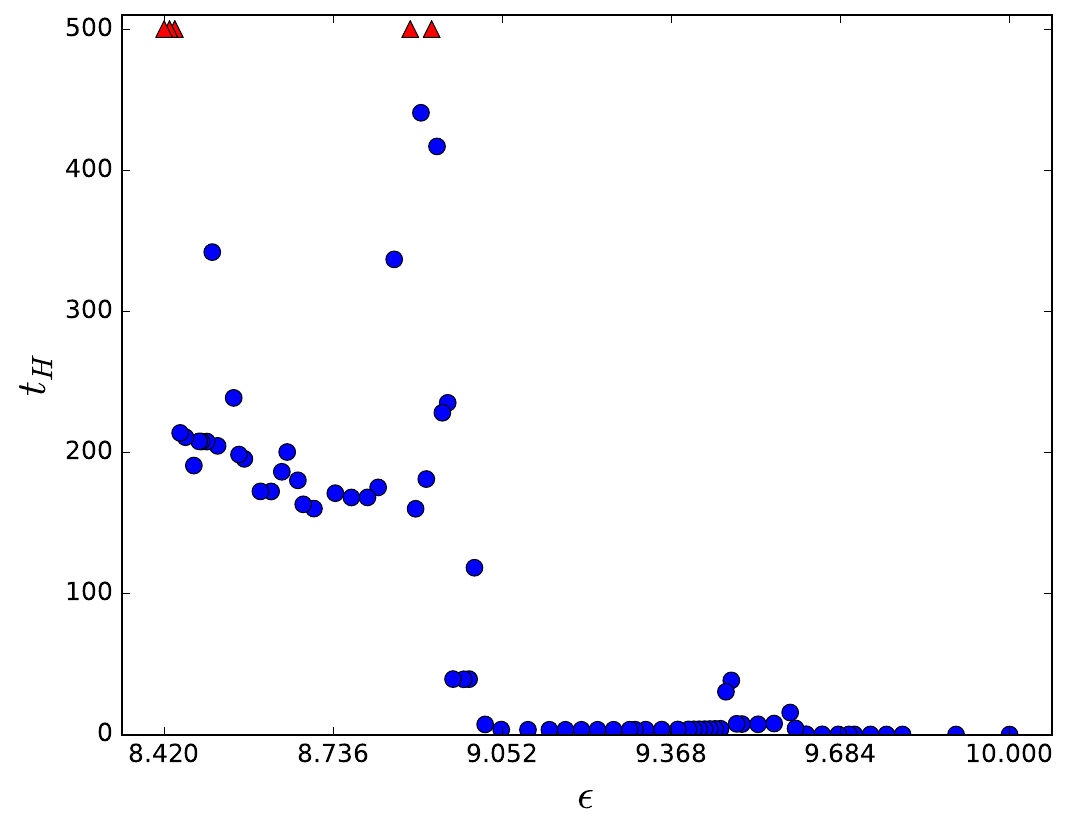}
\caption{$\mu=20,\sigma=0.16$}
\label{f:m20w016}
\end{subfigure}\hfill
\begin{subfigure}[t]{0.47\textwidth}
\includegraphics[width=\textwidth]{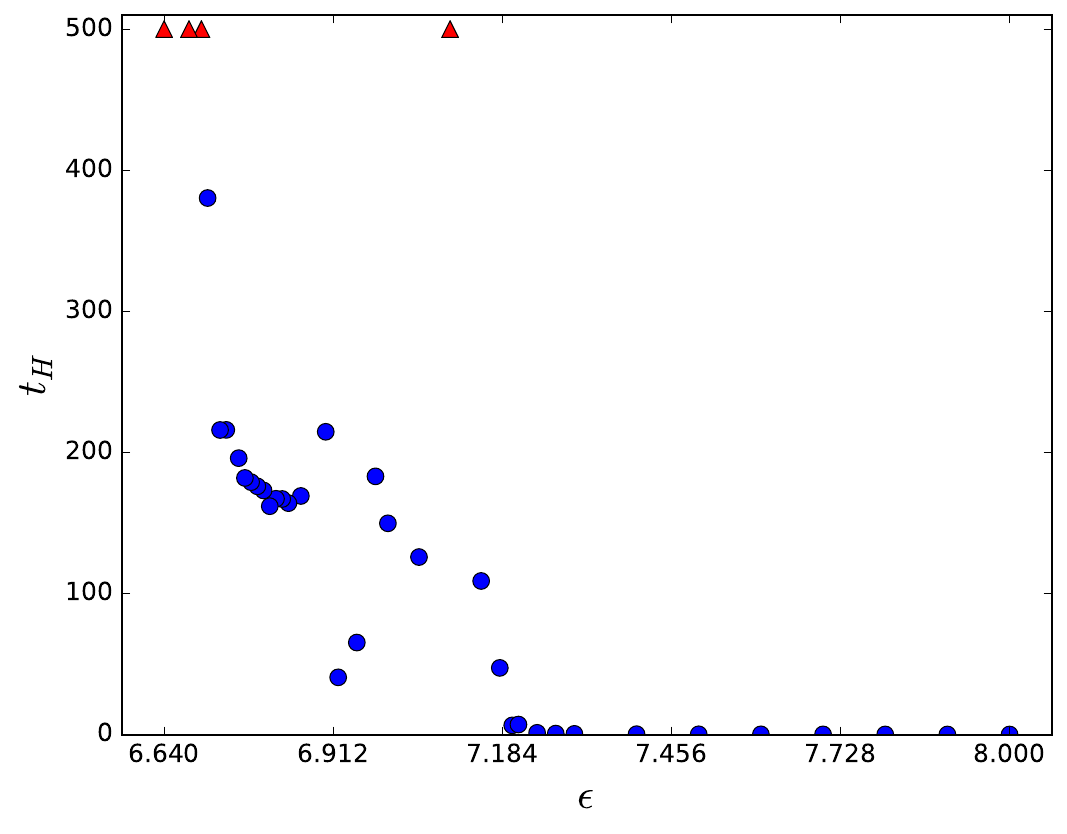}
\caption{$\mu=20,\sigma=0.19$}
\label{f:m20w019}
\end{subfigure}
\caption{Irregular behavior: blue dots represent horizon formation and
red triangles a lower limit on $t_H$.
}
\label{f:irregular}
\end{figure}

We have found a variety of irregular behaviors at the transition
between the metastable and stable classes which we have classified together
as irregular initial data; however, it may be better to describe them as
separate classes.  The stability phase diagram \ref{f:phase} indicates that the
irregular class extends along the ``inland'' side of the small $\sigma$
shoreline and at least part of the large $\sigma$ shoreline
of the island of stability.  What is not clear from our evolutions up to
now is whether each type of behavior appears along the entire shoreline
or if they appear in pockets at different scalar field masses. Examples
of each type of behavior that we have found appear in figure \ref{f:irregular}.

The first type of irregular behavior, shown in figure \ref{f:m05w1}, is
monotonic ($t_H$ increases with decreasing $\epsilon$ as usual), but it is
not well fit by a power law.  In fact, this behavior would classify as
metastable by the criterion of section \ref{s:metastable} in that the
power law of the best fit $t_H=a\epsilon^{-p}+b$ is significantly different
from $p=2$, except for the fact that the reduced $\chi^2$ value for the
fit is very large (greater than 10)
and also that different fitting algorithms can return
significantly different fits, even though the data may appear to the eye
like a smooth power law.  In any case, this type of behavior apparently
indicates a breakdown of metastable behavior and hints at the appearance of
non-monotonicity.  So far, our evolutions have not demonstrated sudden jumps
in $t_H$ typical of stability at low amplitudes, however.

Figure \ref{f:m5w034} exemplifies non-monotonic behavior in the irregular
class.  This type of behavior, which was noted already by \cite{1304.4166},
involves one or more sudden jumps in $t_H$ as $\epsilon$ decreases,
which may be followed by a sudden decrease in $t_H$ and then resumed smooth
monotonic increase in $t_H$.  There are suggestions that this type of
initial data is stable at low amplitudes due to the usual appearance of
non-collapsing evolutions, but it is worth noting that these amplitudes could
instead experience another jump and decrease in $t_H$, just at $t_H>t_{lim}$.
Finally, \cite{1508.02709} studied this type of behavior in some detail,
denoting it as ``quasi-stable.''

Some irregular initial data demonstrate evidence of chaotic behavior,
in that $t_H$ appears to be sensitive to initial conditions 
(ie, value of amplitude) over some range of amplitudes. 
This type of behavior appears over the range
of masses (see figure \ref{f:m0w11} for a mild case for massless scalars),
but it
is more common and more dramatic at larger $\mu$. Figures
\ref{f:m20w016},\ref{f:m20w019} represent the most extreme behavior of
this type
among the initial data that we studied with collapse at $t_H<50$ not very
far separated from amplitudes that do not collapse for $t<t_{lim}$ along with
an unpredictable pattern of variation in $t_H$.  This type of 
evidence for chaotic behavior
has been seen previously in the collapse of transparent but gravitationally
interacting thin shells in AdS \cite{Brito:2016xvw} as well as in the
collapse of massless scalars in AdS$_5$ Einstein-Gauss-Bonnet gravity
\cite{1410.1869,1608.05402}; these references speculated that the 
$t_H$ vs $\epsilon$ curve is fractal. In both cases, this type of behavior is
hypothesized to be due to the transfer of energy between two infalling shells,
with horizon formation only proceeding when one shell is sufficiently
energetic.  In the latter case, the extra scale of the theory
(given by the coefficient of the Gauss-Bonnet term in the action) leads the
single initial pulse of scalar matter to break into two pulses.

We should therefore ask two questions: does this irregular behavior show
evidence of true chaos, and is a similar mechanism at work here? We note first
that \cite{1608.05402} found evidence (using a modified box test) that the
$t_H$ vs $\epsilon$ curve has a non-integer fractal dimension 
for plots visually similar to our figures \ref{f:m20w016},\ref{f:m20w019}.
Here, to quantify the presence of chaos, 
we examine the difference in time evolution between
similar initial conditions (nearby amplitudes), which diverge exponentially
in chaotic systems. Specifically, any quantity $\Delta$ should satisfy
$|\Delta| \propto \exp(\lambda t)$ for Lyapunov coefficient $\lambda$.
Our characteristic will be the
upper envelope of the Ricci scalar at the origin per light crossing time,
$\bar{\mathcal R}(t)$. We consider three sets of irregular initial data: 
a massless scalar of width $\sigma = 1.1$ with amplitudes
$\epsilon=1.02,1.01,1.00$ (see figure \ref{f:m0w11}),
a $\mu = 5$ massive scalar of width $\sigma = 0.34$ and
$\epsilon=3.52,3.51,3.50$, and a $\mu = 20$ scalar of width $\sigma = 0.19$
and $\epsilon=6.98,6.95,6.92$ (figure \ref{f:m20w019}). 
We also determined the Lyapunov coefficient for unstable initial
data with $\mu=0.5$, $\sigma=0.3$, and $\epsilon=1.22,1.20,1.18$ for 
comparison.

\begin{figure}[!t]
\centering
\begin{subfigure}[t]{0.47\textwidth}
\includegraphics[width=\textwidth]{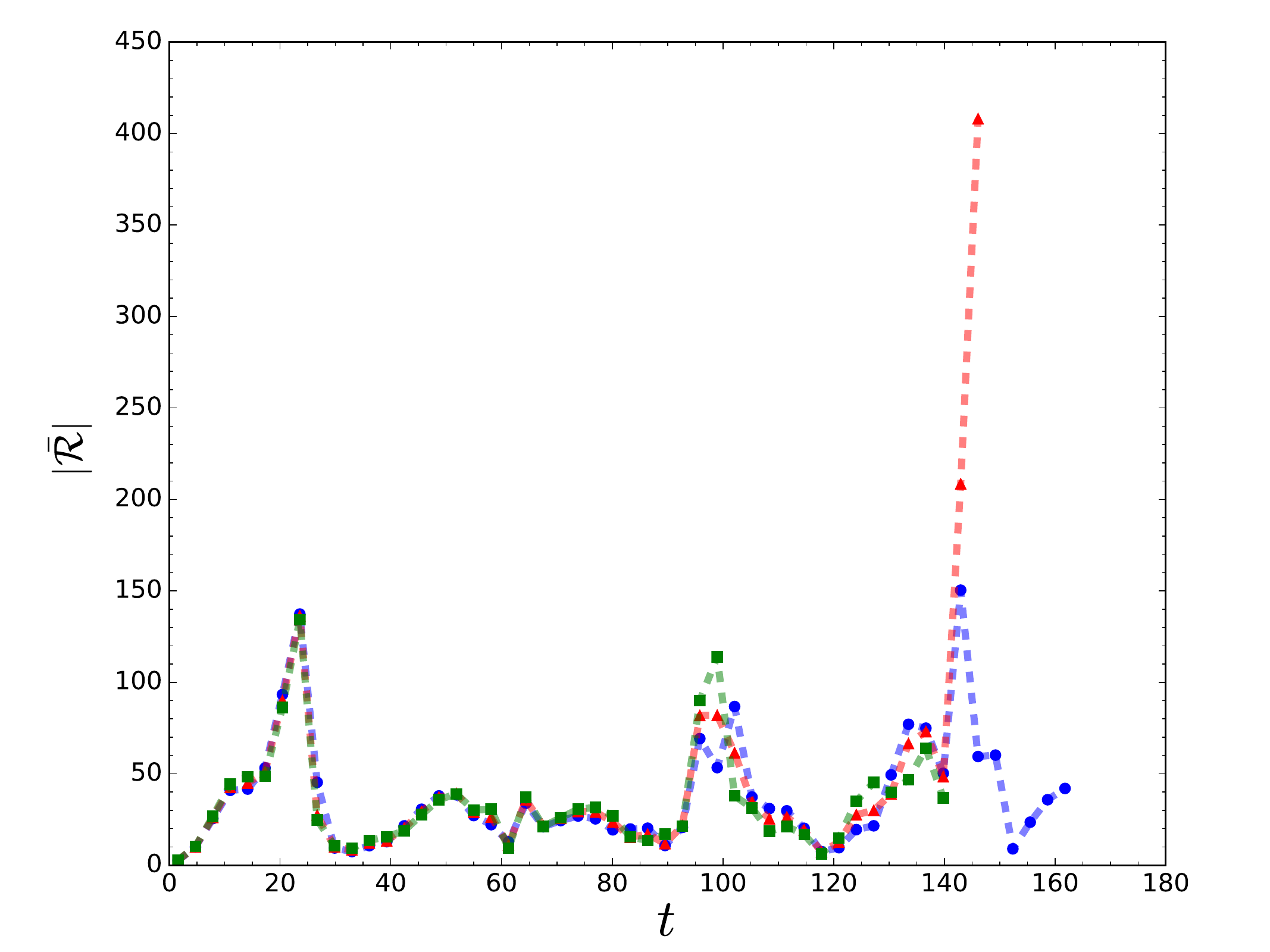}
\caption{Upper envelope of Ricci scalar at origin}
\label{f:m5Ricci}
\end{subfigure}\hfill
\begin{subfigure}[t]{0.46\textwidth}
\includegraphics[width=\textwidth]{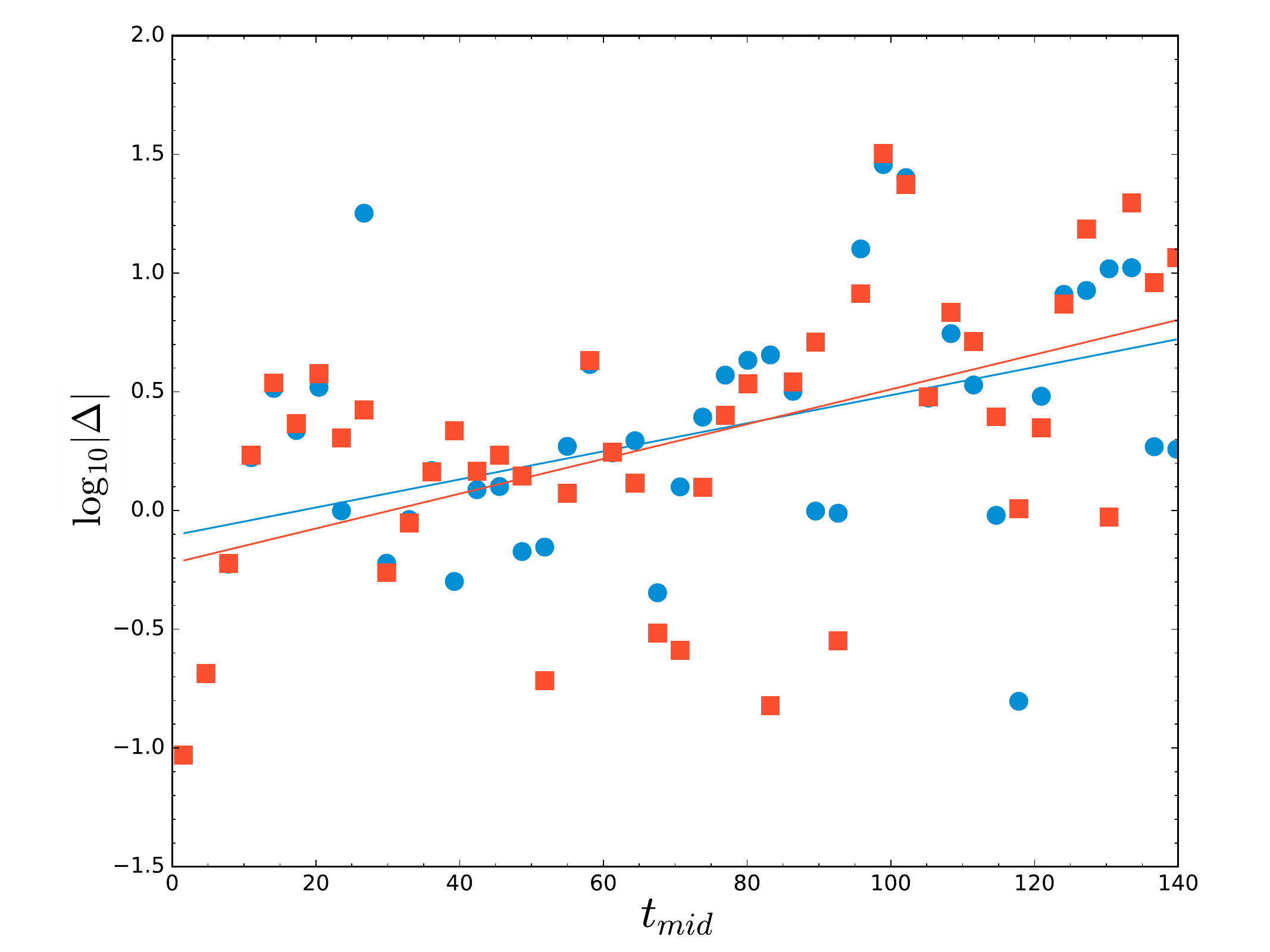}
\caption{$\log | \Delta |$ vs. $t_{mid}$}
\label{f:m5Lyapunov}
\end{subfigure}
\caption{
Left: The upper envelope of the Ricci scalar for amplitudes
$\epsilon_1=3.50$ (blue circles), $\epsilon_2=3.51$ (red triangles), and
$\epsilon_3=3.52$ (green squares) for $\mu=5,\sigma=0.34$.
Right: $\log(|\Delta_{12}|)$ and best fit (blue
circles and line) and $\log(|\Delta_{23}|)$ and best fit (red squares and line),
calculated as a function of the midpoint $t_{mid}$ of the time interval.}
\label{f:m5chaotic}
\end{figure}

Figure \ref{f:m5chaotic} details evidence for chaotic evolution in the
$\mu=5,\sigma=0.34$ case; figure \ref{f:m5Ricci} shows our characteristic
function $\bar{\mathcal{R}}(t)$ for the amplitudes $\epsilon_1 = 3.50$,
$\epsilon_2 = 3.51$, and $\epsilon_3 = 3.52$. By eye, $\bar{\mathcal{R}}$ shows
noticeable differences after a long period of evolution. These are more
apparent in figure \ref{f:m5Lyapunov}, which shows the log of the differences
$\Delta_{ab}\equiv\bar{\mathcal{R}}_{\epsilon_a}-\bar{\mathcal R}_{\epsilon_b}$,
along with the best fits. Although there is considerable noise --- or
oscillation around exponential growth --- in the
differences (leading to $R^2$ values $\sim 0.2,0.26$ for the fits), the
average slope gives Lyapunov coefficient $\lambda=0.007$ (within the error
bar of each slope), and each slope differs from zero by more than 3 standard
errors. One interesting point is that the $t_H$ vs $\epsilon$ curve
in figure \ref{f:m5w034} does not appear chaotic to the eye, even though it
shows some of the mathematical signatures of chaos at least for
$\epsilon_1<\epsilon<\epsilon_3$ (the visible spike in $t_H$ is at 
$\epsilon\sim 3.57$).

The story is similar for the massless and $\mu=20$ cases we
studied, which exhibit $\lambda$ values that
differ from zero by at least 1.9 standard errors; see table \ref{t:lyap}.
This is a milder version of the behavior noted by
\cite{Brito:2016xvw,1410.1869,1608.05402}, especially for the $\mu=5$ case
studied.  
One thing to note is that the strength of oscillation in
$\log(|\Delta|)$ around the linear fit increases with increasing
mass, so that the two best fit Lyapunov exponents for $\mu = 20$ are no
longer consistent with each other at the 1-standard deviation level.
We should note, however that the unstable initial data with 
$\mu=0.5,\sigma=0.3$ also exhibits a statistically positive Lyapunov
exponent, though we should note that the value of $\lambda$ quoted in table
\ref{t:lyap} includes the time shortly before horizon formation, which does
increase $\lambda$ somewhat (though not more than the quoted error).

\begin{table}[!t]
\begin{center}
\begin{tabular}{|cc|c|c|}
\hline
&&$\lambda$ &average $\lambda$\\
\hline
$\mu=0,\sigma=1.1$ & $\Delta_{12}$ &$0.011\pm 0.005$ & 0.011\\
&$\Delta_{23}$ & $0.011\pm 0.005$ & \\
\hline
$\mu=0.5,\sigma=0.3$ & $\Delta_{12}$ &$0.021\pm 0.0007$ & 0.022\\
&$\Delta_{23}$ & $0.024\pm 0.001$ & \\
\hline
$\mu=5,\sigma=0.34$ & $\Delta_{12}$ &$0.006\pm 0.002$& 0.007\\
&$\Delta_{23}$ & $0.007\pm 0.002$& \\
\hline
$\mu=20,\sigma=0.19$ & $\Delta_{12}$ &$0.046\pm 0.009$& 0.032\\
&$\Delta_{23}$ & $0.019\pm 0.007$ & \\
\hline
\end{tabular}
\end{center}
\caption{Best fit Lyapunov coefficients $\lambda$ for adjacent amplitude
pairs and average $\lambda$ value for each $\mu,\sigma$ system studied.
Standard errors are given following $\pm$ signs.}
\label{t:lyap}
\end{table}

\begin{figure}[!t]
\centering
\begin{subfigure}[t]{0.47\textwidth}
\includegraphics[width=\textwidth]{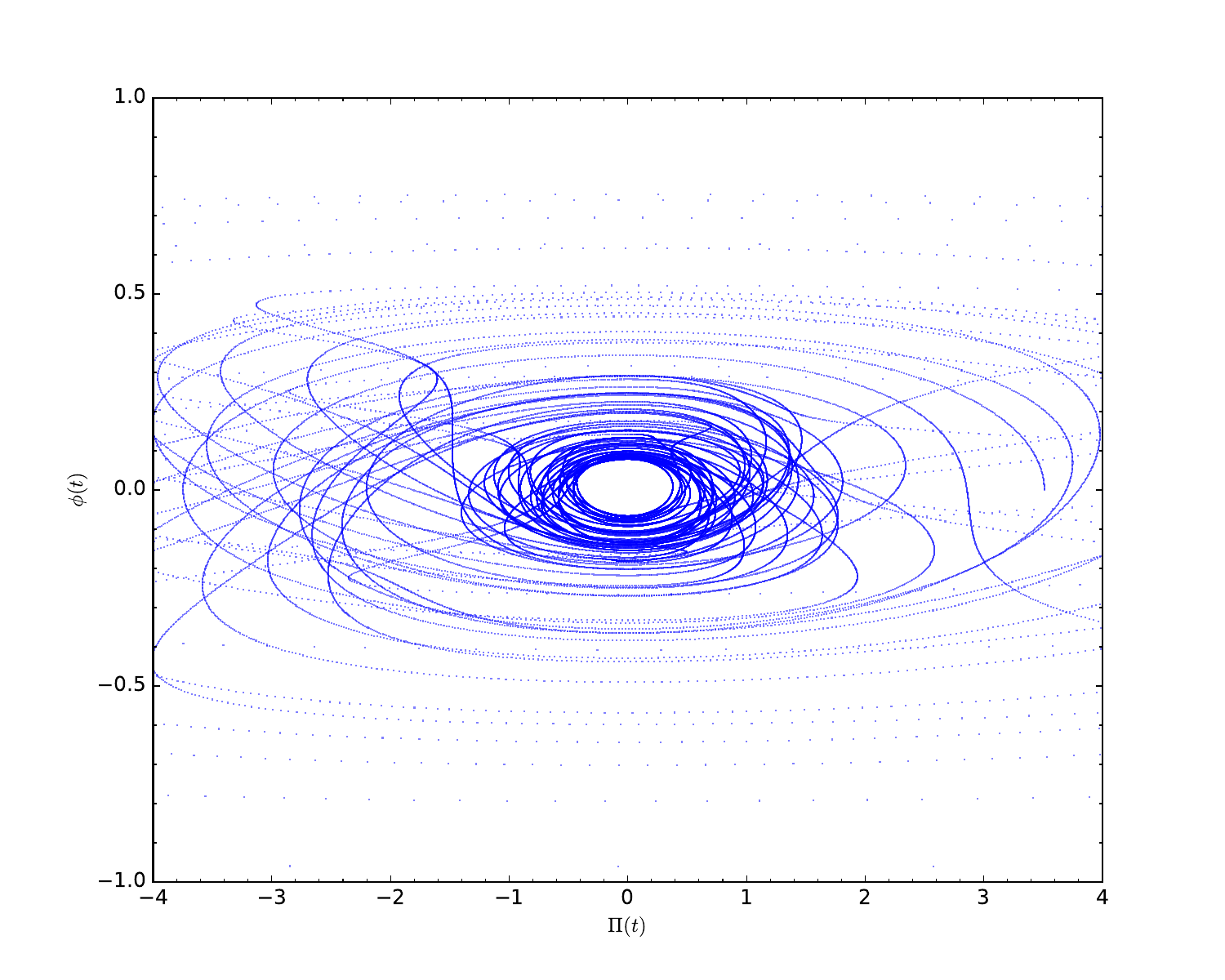}
\caption{$\mu=5,\sigma=0.34,\epsilon=3.51$}
\label{f:m5w034phase}
\end{subfigure}\hfill
\begin{subfigure}[t]{0.47\textwidth}
\includegraphics[width=\textwidth]{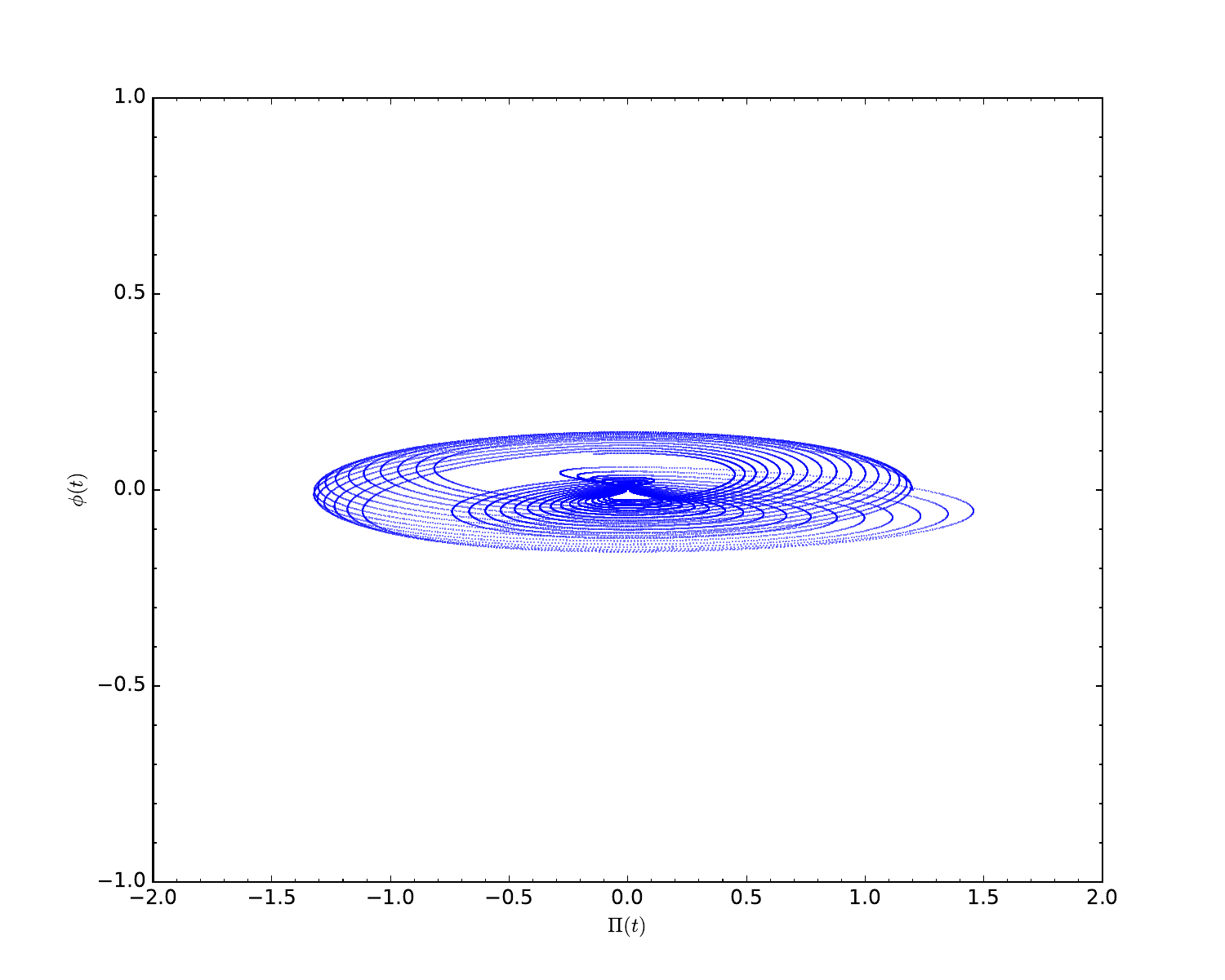}
\caption{$\mu=0.5,\sigma=0.3,\epsilon=1.20$}
\label{f:m05w03phase}
\end{subfigure}
\caption{Trajectories in $\Pi(x=0),\phi(x=0)$ phase space for one irregular
and one unstable evolution. Trajectories are shown for $t<50$.}
\label{f:chaosphasespace}
\end{figure}

Since the Lyapunov coefficients do not distinguish the irregular and unstable
cases, we also consider the phase space trajectories of the evolutions.
Following \cite{deOliveira:2012ac}, we consider the trajectory of evolutions
in $\Pi$ and $\phi$ evaluated at the origin for $t\leq 50$
in figure \ref{f:chaosphasespace}. Neither the 
$\mu=5,\sigma=0.34,\epsilon=3.51$ (figure \ref{f:m5w034phase}) or 
$\mu=0.5,\sigma=0.3,\epsilon=1.20$ (figure \ref{f:m05w03phase}) trajectories
close, though there is a clear difference. Specifically, the former trajectory
is visually disorganized (that is, strongly varying orbits) with very rapid
motion (seen in the gap between points on the trajectory between plotted
time steps). Meanwhile, the latter motion is comparatively regular, typical
of quasi-periodic motion. Figure \ref{f:m5w034phase} is typical of 
turbulence and clearly shows that these evolutions are nonperturbative, 
even though $t_H$ is large (well into the perturbative regime for unstable
initial data).

To sum up, we have identified irregular initial data that show evidence of
chaotic behavior. Specifically, several of the $t_H$ vs $\epsilon$ curves
appear qualitatively similar to analogous plots in 
\cite{Brito:2016xvw,1410.1869,1608.05402}, which were demonstrated to have
fractal-like behavior (including fractional fractal dimension in one case).
Furthermore, a number of cases of irregular initial data (and some unstable)
have positive Lyapunov exponents; phase space trajectories for irregular
initial data show very rapid motion typical of turbulence, while unstable
initial data have more regular trajectories. Taken together, this is strong
evidence for chaotic behavior for some irregular initial data, similar 
to that discussed in other studies of gravitational collapse in AdS. 
Furthermore,
this is the first evidence of chaos in the $t_H$ vs $\epsilon$ curve for
gravitational collapse of a massless scalar in AdS to our knowledge.

\begin{figure}[!t]
\centering
\begin{subfigure}[t]{0.47\textwidth}
\includegraphics[width=\textwidth]{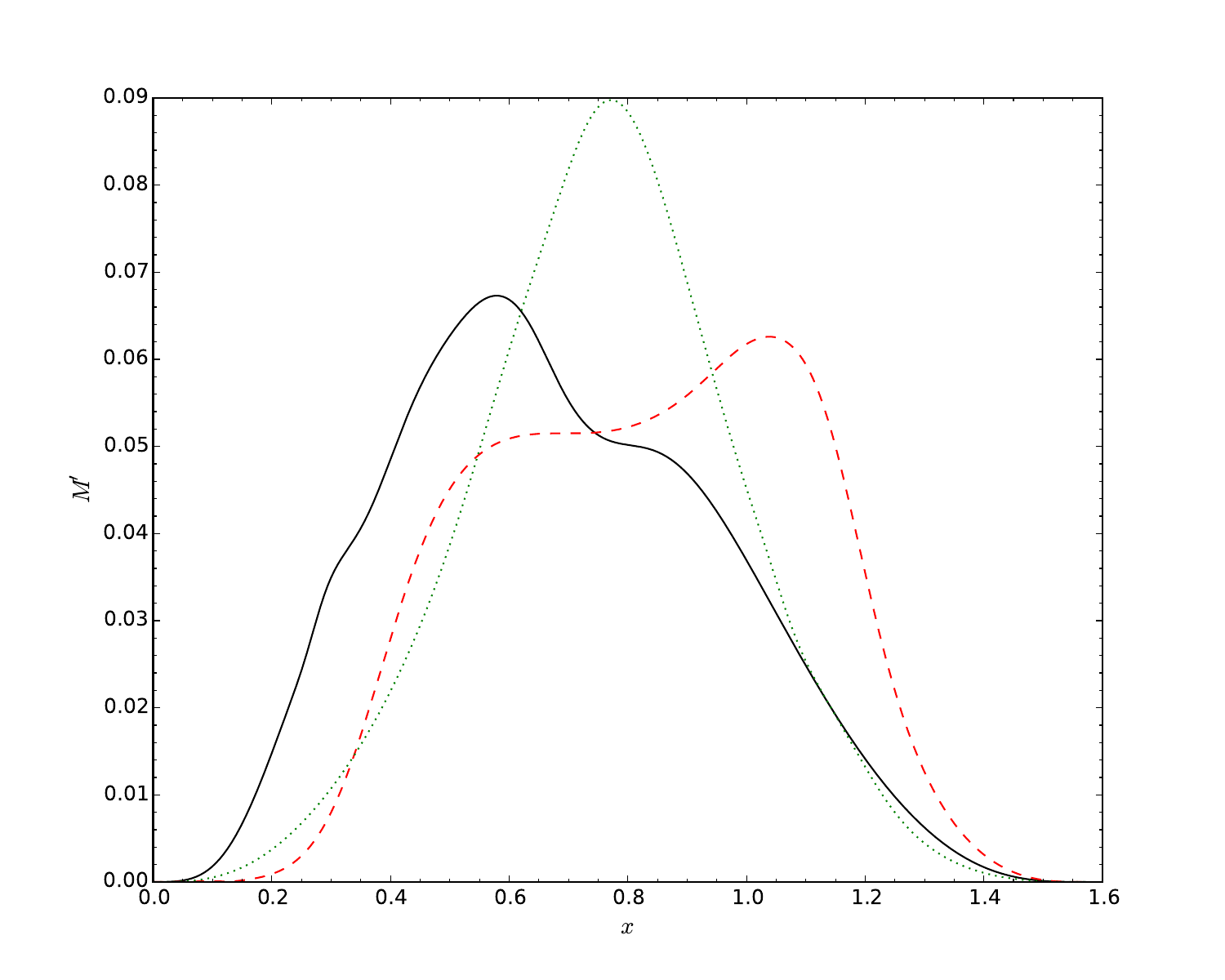}
\caption{$\mu=0,\sigma=1.1,\epsilon=1.01$, at times $t=60$ (solid black),
$t=62$ (dashed red), $t=64$ (dotted green)}
\label{f:m0hump}
\end{subfigure}
\begin{subfigure}[t]{0.47\textwidth}
\includegraphics[width=\textwidth]{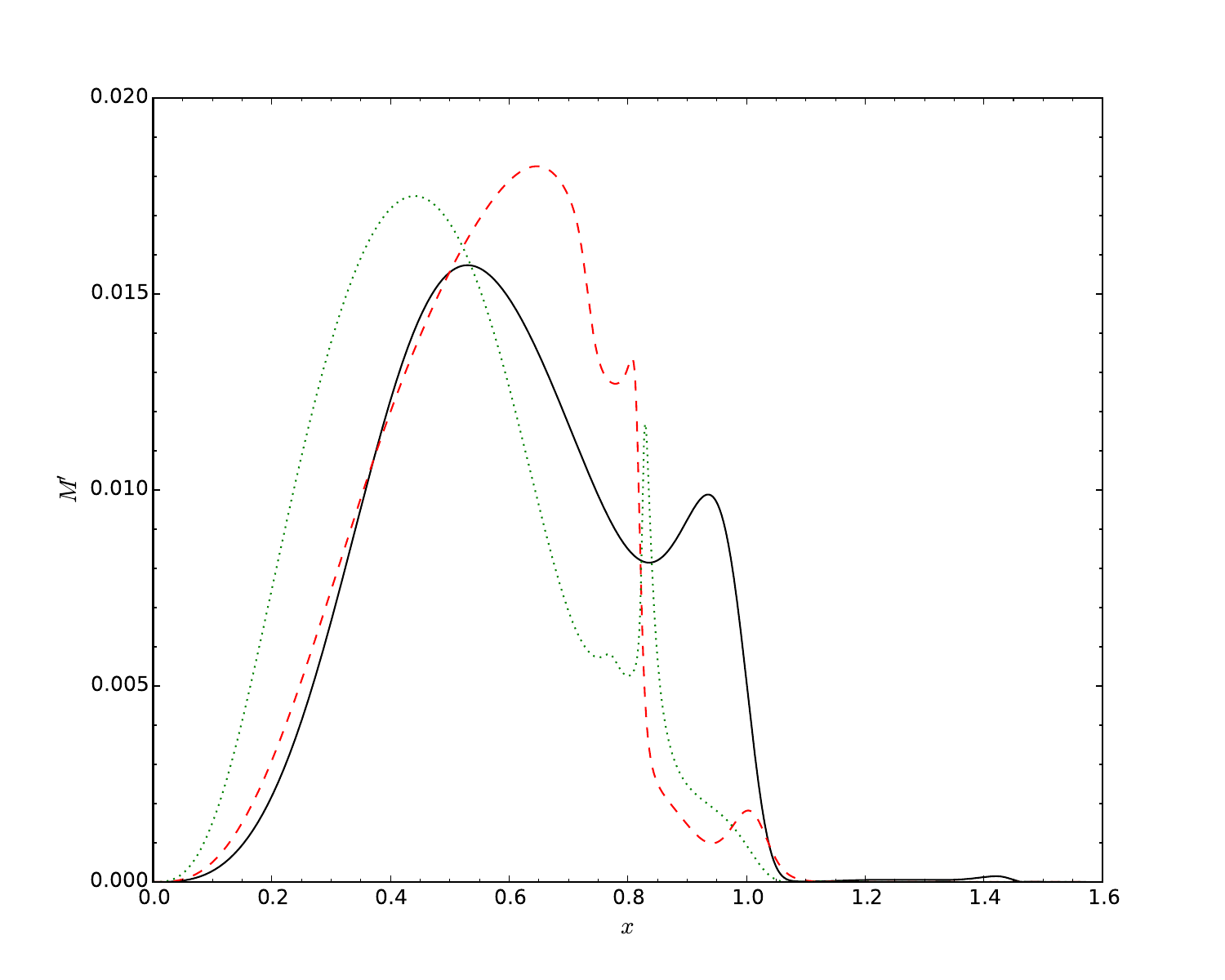}
\caption{$\mu=5,\sigma=0.34,\epsilon=3.52$, at times $t=132$ (solid black),
$t=137$ (dashed red), $t=140$ (dotted green)}
\label{f:m5ripple}
\end{subfigure}
\caption{Radial derivative of the mass function at the indicated time
for two systems that show evidence of chaos.  
Note the appearance of a secondary wave on top
of the main pulse. $(\mu,\sigma,\epsilon)$ as indicated.}
\label{f:chaosmechanism}
\end{figure}

The mechanism underlying the possibly chaotic behavior seems somewhat different
or at least weaker than the two-shell or Einstein-Gauss-Bonnet systems.
When examining the time evolution of the mass distributions of
these data, we see a single large pulse of mass energy that oscillates
between the origin and boundary without developing a pronounced peak.
However, there is also apparently a smaller wave that travels across the
large pulse.  We can see this by comparing snapshots of the mass distribution
at different times, as in figure \ref{f:chaosmechanism}.
In the massless case examined, this wave deforms the pulse,
leading to a double-shoulder appearance seen at two times in
figure \ref{f:m0hump}.
In the $\mu=5,\sigma=0.34$ case, the secondary wave is more like a ripple,
usually smaller in amplitude but more sharply localized, as toward the right
side of the main pulse in figure \ref{f:m5ripple}.  So the chaotic behavior
may be caused by the relative motion of the two waves, rather than energy
transfer between two shells.  In this hypothesis, a horizon would form when
both waves reach the neighborhood of the origin at the same time.

As a note, we have run convergence tests on several sets of irregular
initial data and find that our calculations are convergent overall, as expected
(even at lower resolution than we used).  In particular, the massless
scalar evolutions studied in table \ref{t:lyap} are convergent already at
resolution given by $n=12$ (note that we typically start at $n=14$); we 
also observe convergent behavior for the $\mu=5$ evolutions discussed
in table \ref{t:lyap}.  We have therefore validated that nonmonotonic 
behavior and even evidence of chaos occurs.
The only caveat may be for some of the apparently initial data with 
scalar mass $\mu=20$, which nonetheless appears well-behaved according to other 
indicators.  The reader may or may not wish to take them at
face value but should recall that we have presented other chaotic initial
data with rigorously convergent evolutions.  See the appendix for a more 
detailed discussion.

\section{Spectral analysis}\label{s:analysis}

As we discussed in the introduction, instability toward horizon formation
proceeds through a turbulent cascade of energy to shorter wavelengths or,
more quantitatively, to 1st-order scalar eigenmodes with more nodes.
Inverse cascades are typical of stable evolutions.
Therefore, understanding the energy spectrum of our evolutions, both initially
and over time, sheds light on the behavior of the self-gravitating scalar
field in asymptotically AdS spacetime, providing a heuristic analytic 
understanding of the stability phase diagram.

The (normalizable) eigenmodes $e_j$ are given by Jacobi polynomials as
\beq{emodes}
 e_j(x)=\kappa_j\cos^{\lambda_+}(x) P^{(d/2-1,\sqrt{d^2+4\mu^2}/2)}_j(\cos(2x))
\eeq
($\kappa_j$ is a normalization constant)
with eigenfrequency $\omega_j=2j+\lambda_+$ and
$\lambda_+=(d+\sqrt{d^2+4\mu^2})/2$ in AdS$_{d+1}$ for $j=0,1,\cdots$
(see \cite{hep-th/9905111,Nastase:2015wjb} for reviews).
Including gravitational backreaction, we define the energy spectrum
\beq{spectrum}
E_j\equiv\frac{1}{2}\left(\Pi_j{}^2-\phi_j\ddot{\phi}_j\right),\eeq
where
\begin{eqnarray}
\Pi_j&=&\left(\sqrt{A}\Pi,e_j\right),\quad
\phi_j=\left(\phi,e_j\right),\nonumber\\
\ddot\phi_j&=&\left(\cot^{d-1}(x)\partial_x\left[\tan^{d-1}(x)A\Phi\right]
-\mu^2\sec^2(x)\phi,e_j\right),
\end{eqnarray}
and the inner product is $(f,g)=\int_0^{\pi/2}dx\tan^{d-1}(x) fg$.
The sum of $E_j$ over all modes is the conserved ADM mass.

\subsection{Dependence on mass}
\begin{figure}[!t]
\centering
\begin{subfigure}[t]{0.45\textwidth}
\includegraphics[width=\textwidth]{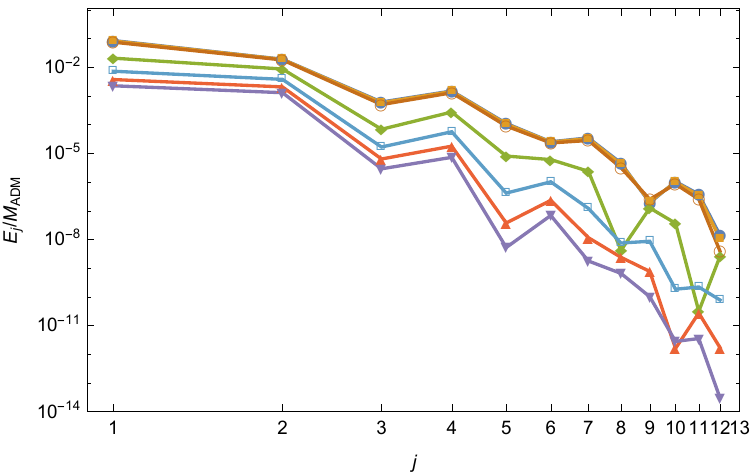}
\caption{Best fit gaussian energy spectra.} 
\label{f:gaussian}
\end{subfigure} \hfill
\begin{subfigure}[t]{0.45\textwidth}
\includegraphics[width=\textwidth]{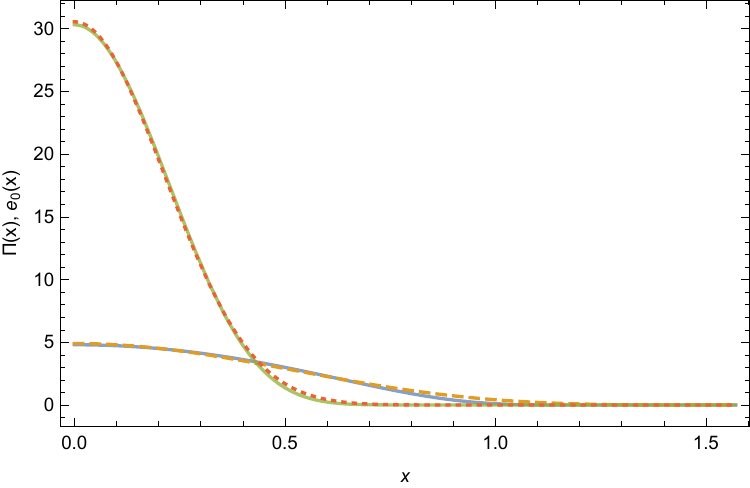}
\caption{Best fit gaussian and zeroth eigenmode.}
\label{f:gaussianfits}
\end{subfigure}
\caption{Left: Spectra of the best fit Gaussians (\ref{PiGaussianID}) to 
the $j = 0$ eigenmode for masses $\mu=0$ (blue circles), 0.5 (yellow squares),
1 (empty orange circles), 5 (green diamonds), 10 (empty cyan squares), 
15 (upward red triangles), and 20 (downward purple
triangles). Right: an overlay of the best fit Gaussian and $e_0$ eigenmode 
for $\mu = 0$ (solid blue is best fit, orange dashed is eigenmode) and $\mu = 20$ (solid green, red short dashes).}
\label{f:gaussians}
\end{figure}

The most visibly apparent feature of the stability 
phase diagram of figure \ref{f:phase}
is that the island of stability both expands and shifts to smaller widths
as the scalar mass increases.  As it turns out, the energy spectrum of the
Gaussian initial data (\ref{PiGaussianID}) provides a simple heuristic
explanation.

It is well established both in perturbation theory and numerical studies that
initial data given by a single scalar linear-order eigenmode is in fact
nonlinearly stable, and the spectra of many quasi-periodic solutions are also
dominated by a single eigenmode.  As a result, we should expect Gaussian
initial data that approximate a single eigenmode (which must be $j=0$ due to
lack of nodes) to be stable.  To explore how this depends on mass, we find the
best fit values of $\epsilon,\sigma$ for the $j=0$ eigenmode for each mass
that we consider (defined by the least-square error from the Gaussian to a
discretized eigenmode);
this is the ``best approximation'' Gaussian to the eigenmode.
Then we find the energy spectrum of that best-fit Gaussian; these are shown
in figure \ref{f:gaussian}.  From the figure, it is clear that the $j=0$
eigenmode is closer to a Gaussian at larger masses.  That is, other
eigenmodes contribute less to the Gaussian's spectrum at higher masses (by
several orders of magnitude over the range from $\mu=0$ to 20).  Simply put,
the shape of the $j=0$ eigenmode is closer to Gaussian at higher masses,
which suggests that the island of stability should be larger at larger
scalar field mass. Figure \ref{f:gaussianfits} compares the $j=0$ eigenmode
and best fit Gaussian for $\mu=0$ and 20; on inspection, there is more
deviation between the eigenmode and Gaussian for the massless scalar.

In addition, the best-fit Gaussian width decreases from $\sigma\sim 0.8$
for a massless scalar as the mass increases.  At $\mu=20$, the best-fit
width is $\sigma\sim 0.31$. This suggests that Gaussians that approximate the
$j=0$ mode well enough are narrower in width at higher masses.  An interesting
point to note is that the island of stability for $\mu=0,0.5$ is actually
centered at considerably larger widths than the best-fit Gaussian.  This may
not be surprising, since the best-fit Gaussians at low masses actually
receive non-negligible contributions from higher mode numbers; moving away from
the best-fit Gaussian can actually reduce the power in higher modes.
For example, the stable initial data shown in figure \ref{f:m0w15spect} below
have considerably less power in the $j=2$ mode.

\subsection{Spectra of different behaviors}
\begin{figure}[!t]
\centering
\begin{subfigure}[t]{0.32\textwidth}
\includegraphics[width=\textwidth]{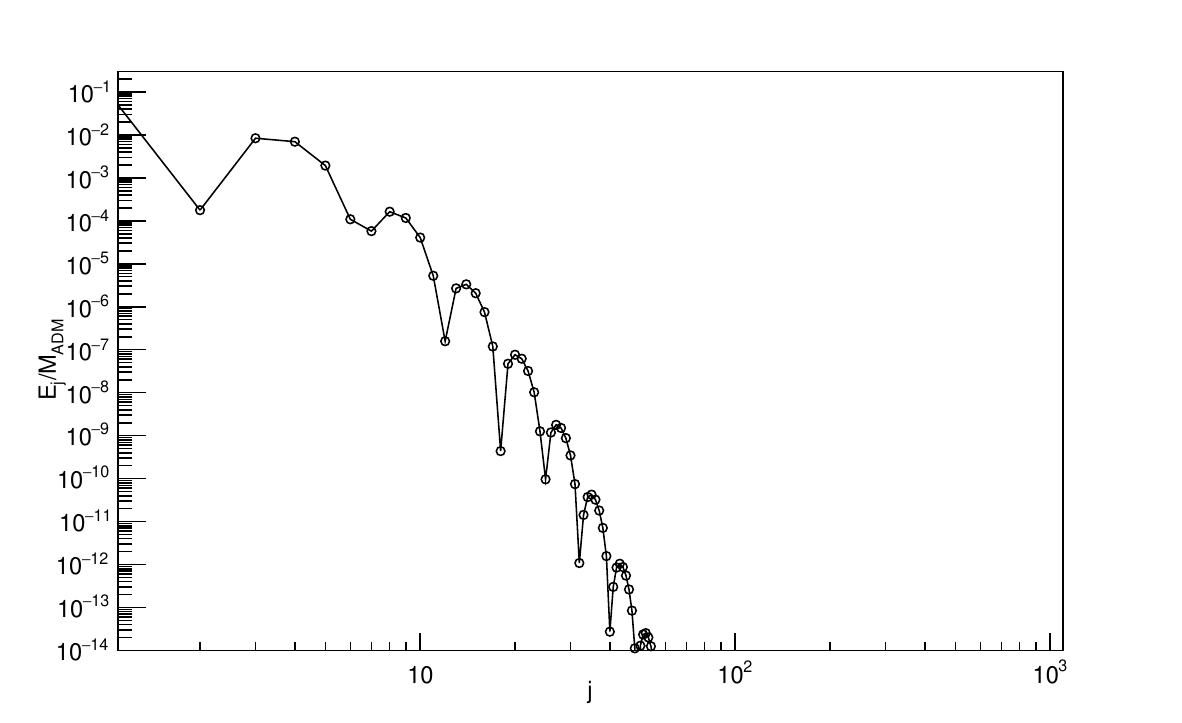}
\caption{$\mu=0,\sigma=1.5$}
\label{f:m0w15spect}
\end{subfigure}
\begin{subfigure}[t]{0.32\textwidth}
\includegraphics[width=\textwidth]{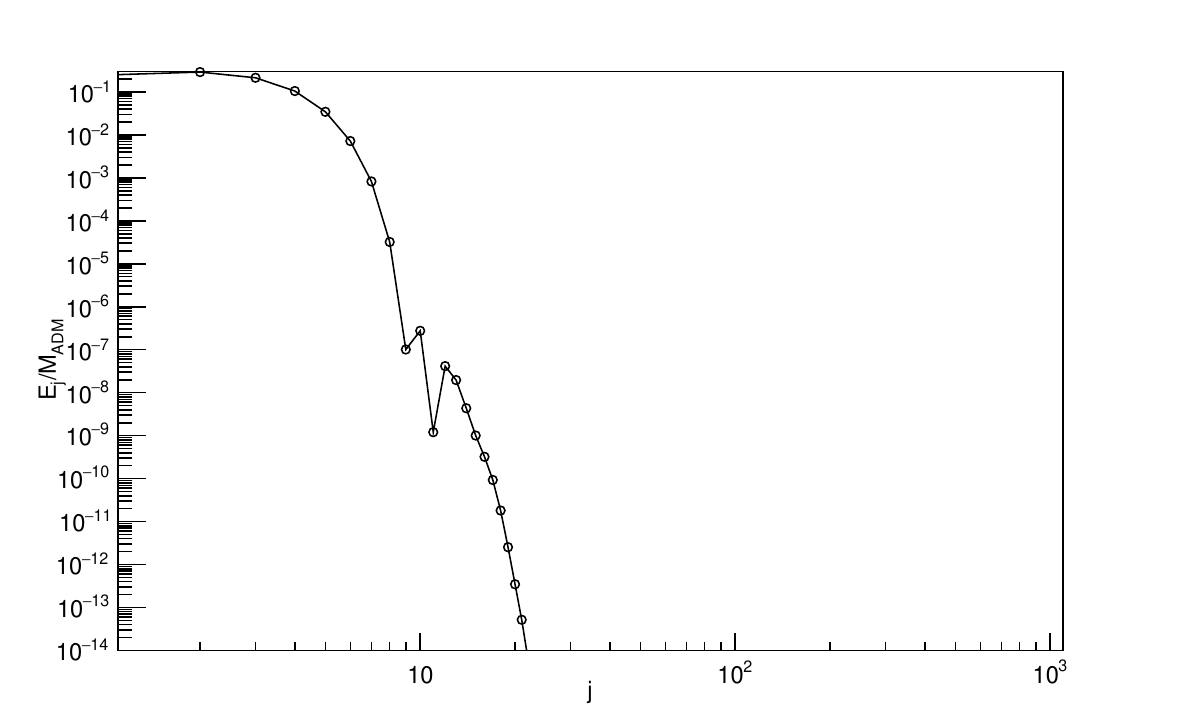}
\caption{$\mu=0,\sigma=0.25$}
\label{f:m0w025spect}
\end{subfigure}
\begin{subfigure}[t]{0.32\textwidth}
\includegraphics[width=\textwidth]{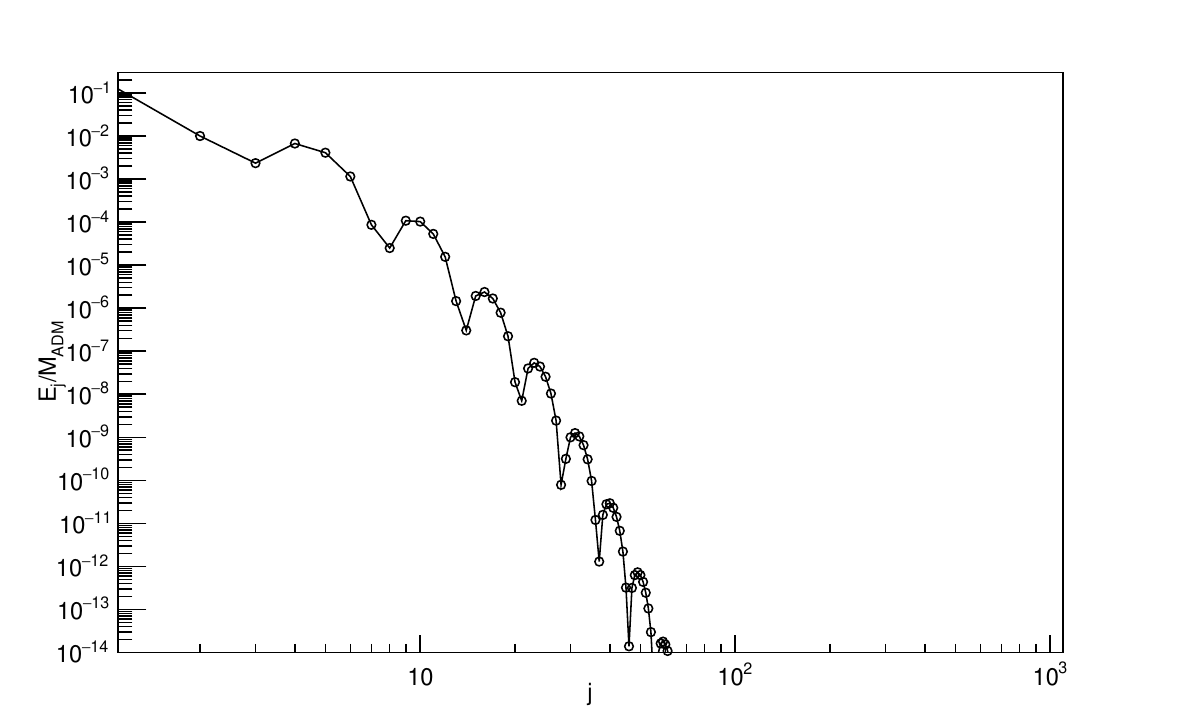}
\caption{$\mu=5,\sigma=1.7$}
\label{f:m5w170spect}
\end{subfigure}
\begin{subfigure}[t]{0.32\textwidth}
\includegraphics[width=\textwidth]{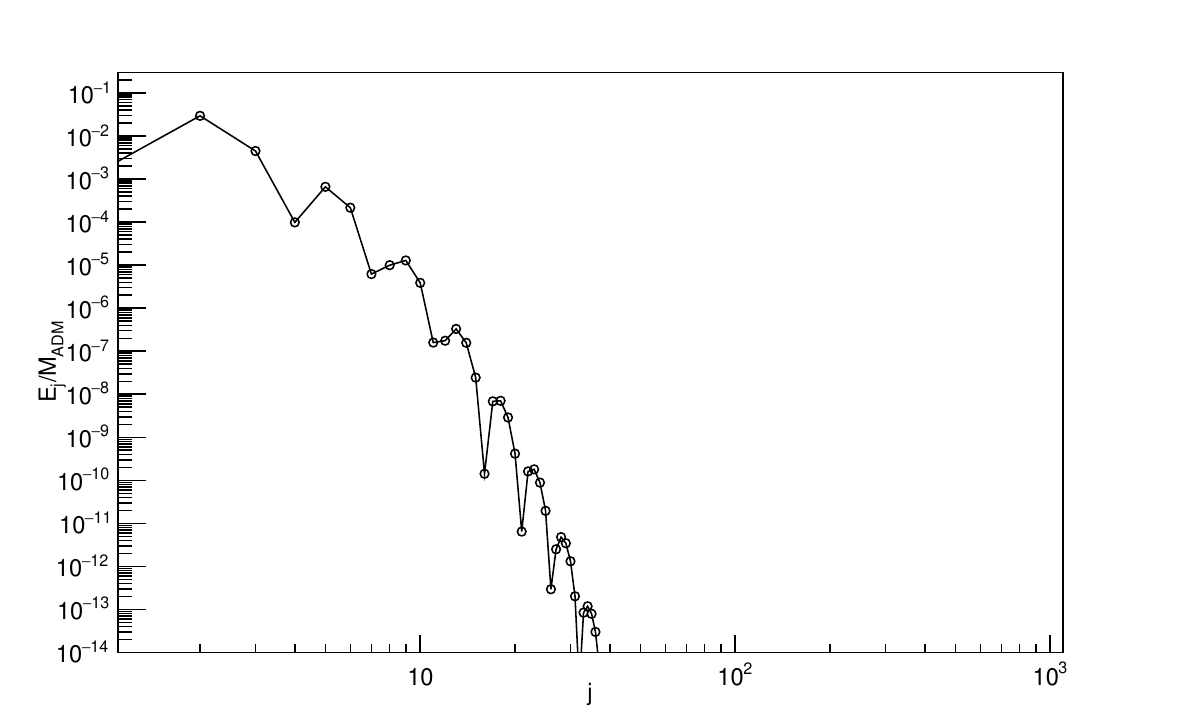}
\caption{$\mu=0.5,\sigma=1$}
\label{f:m05w1spect}
\end{subfigure}
\begin{subfigure}[t]{0.32\textwidth}
\includegraphics[width=\textwidth]{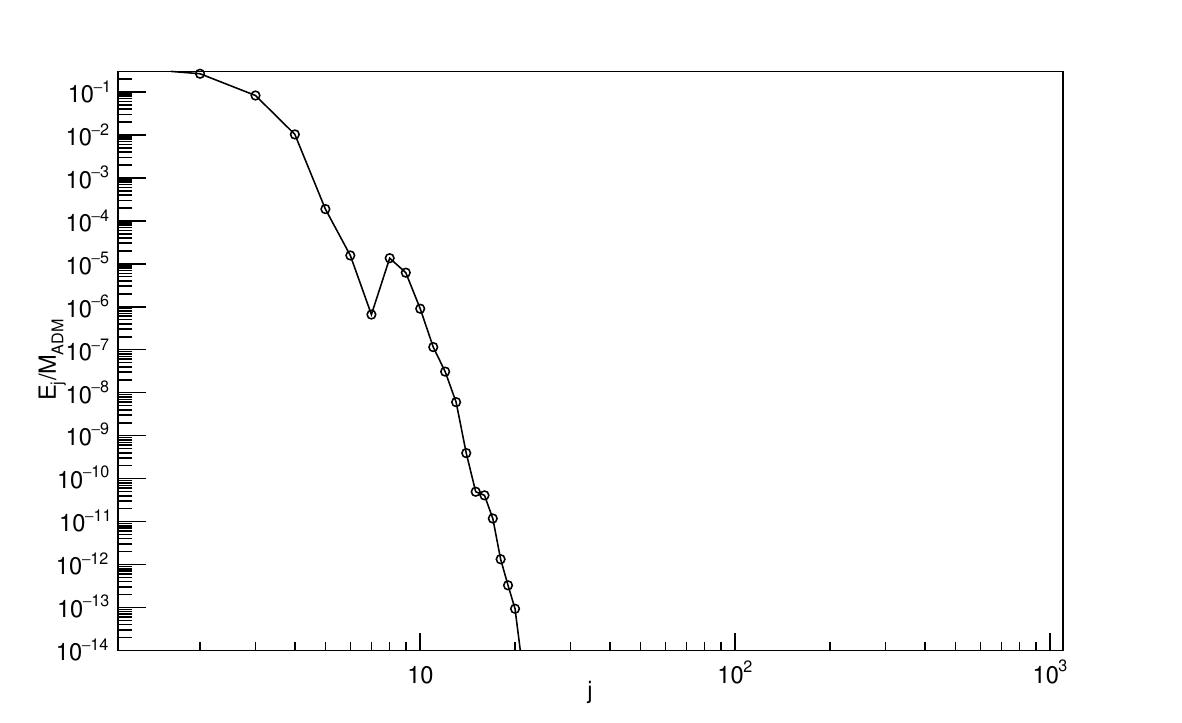}
\caption{$\mu=5,\sigma=0.34$}
\label{f:m5w034spect}
\end{subfigure}
\begin{subfigure}[t]{0.32\textwidth}
\includegraphics[width=\textwidth]{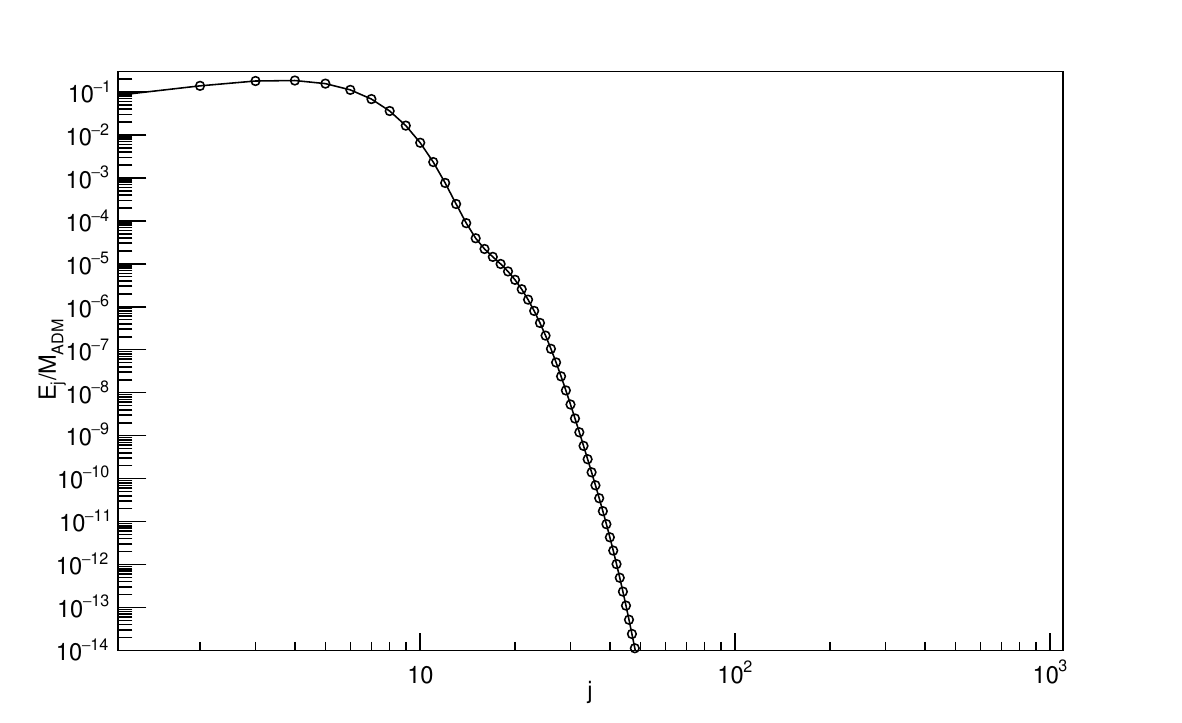}
\caption{$\mu=20,\sigma=0.16$}
\label{f:m15w150spect}
\end{subfigure}
\caption{Initial ($t=0$) energy spectra for the indicated evolutions. In order,
these represent stable, unstable, metastable, monotonic irregular,
non-monotonic irregular, and chaotic irregular initial data.}
\label{f:initialspect}
\end{figure}

A key question that one might hope to answer is whether the stability class
of a given $(\mu,\sigma)$ can be determined easily by direct inspection
of the initial data without requiring many evolutions at varying amplitudes.
The initial energy spectra for examples of each class, including monotonic,
non-monotonic, and apparently chaotic irregular behaviors, are shown in figure
\ref{f:initialspect}.  These spectra are taken from among the smallest
amplitudes we evolved in order to minimize backreaction effects.

Unfortunately, the initial energy spectra do not seem to provide such a
method for determining the stability class.  Very broadly speaking, stable
and metastable $(\mu,\sigma)$ correspond to initial spectra that drop off
fairly quickly from the $j=0$ mode as $j$ increases, while unstable and
irregular behaviors tend to have roughly constant or even slightly increasing
spectra up to $j=5$ or $10$.  However, figure \ref{f:m05w1spect} shows that
some irregular initial data have spectra that decrease rapidly after a small
increase from $j=1$ to $j=2$.  Kinks in the spectrum are more prevalent for
widths of the AdS scale or larger, while spectra for smaller widths tend
to be smoother.

\subsection{Evolution of spectra}
\begin{figure}[!t]
\centering
\begin{subfigure}[t]{0.47\textwidth}
\includegraphics[width=\textwidth]{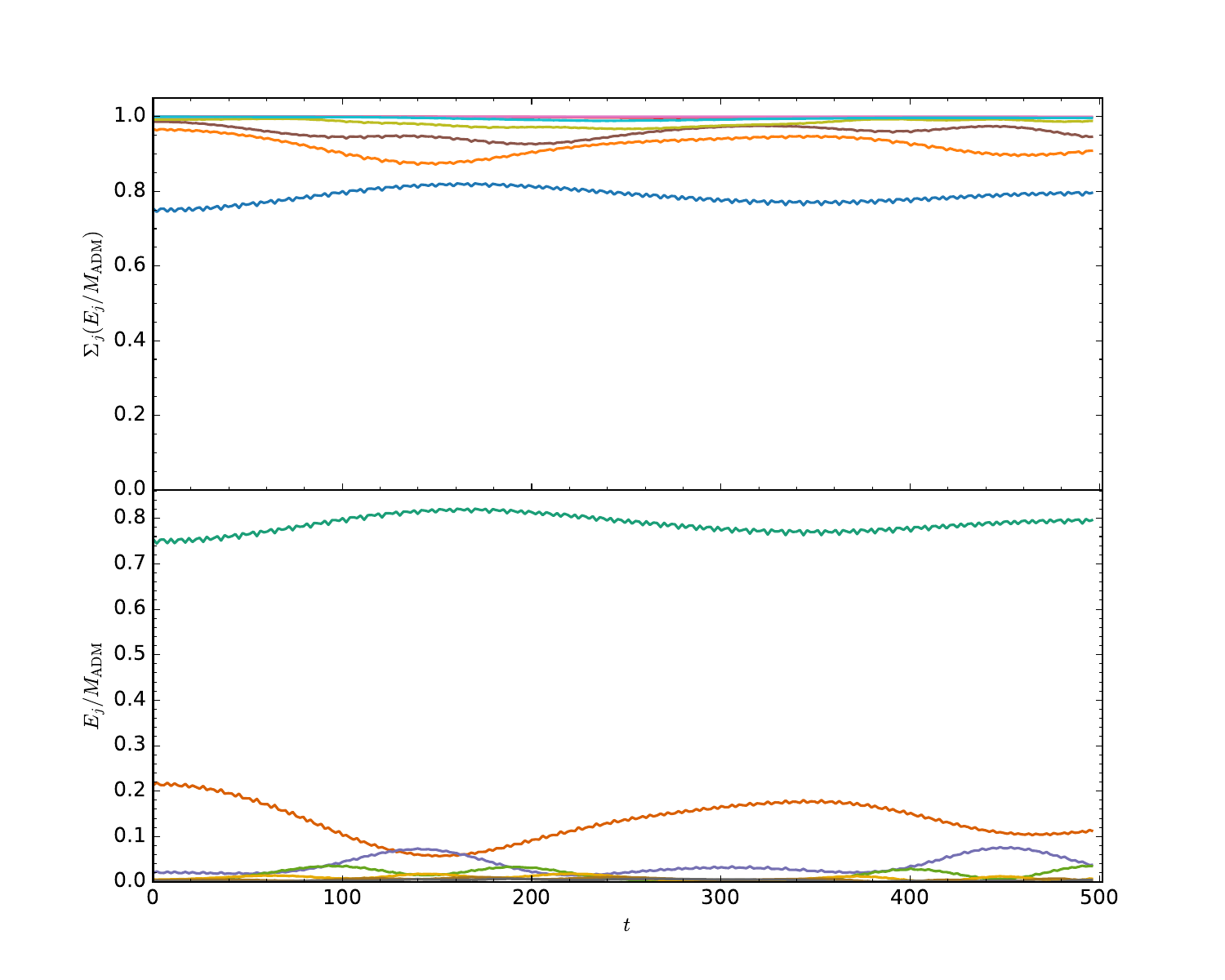}
\caption{$\mu=0,\sigma=1.8,\epsilon=0.13$}
\label{f:m0w18decomp}
\end{subfigure}
\begin{subfigure}[t]{0.47\textwidth}
\includegraphics[width=\textwidth]{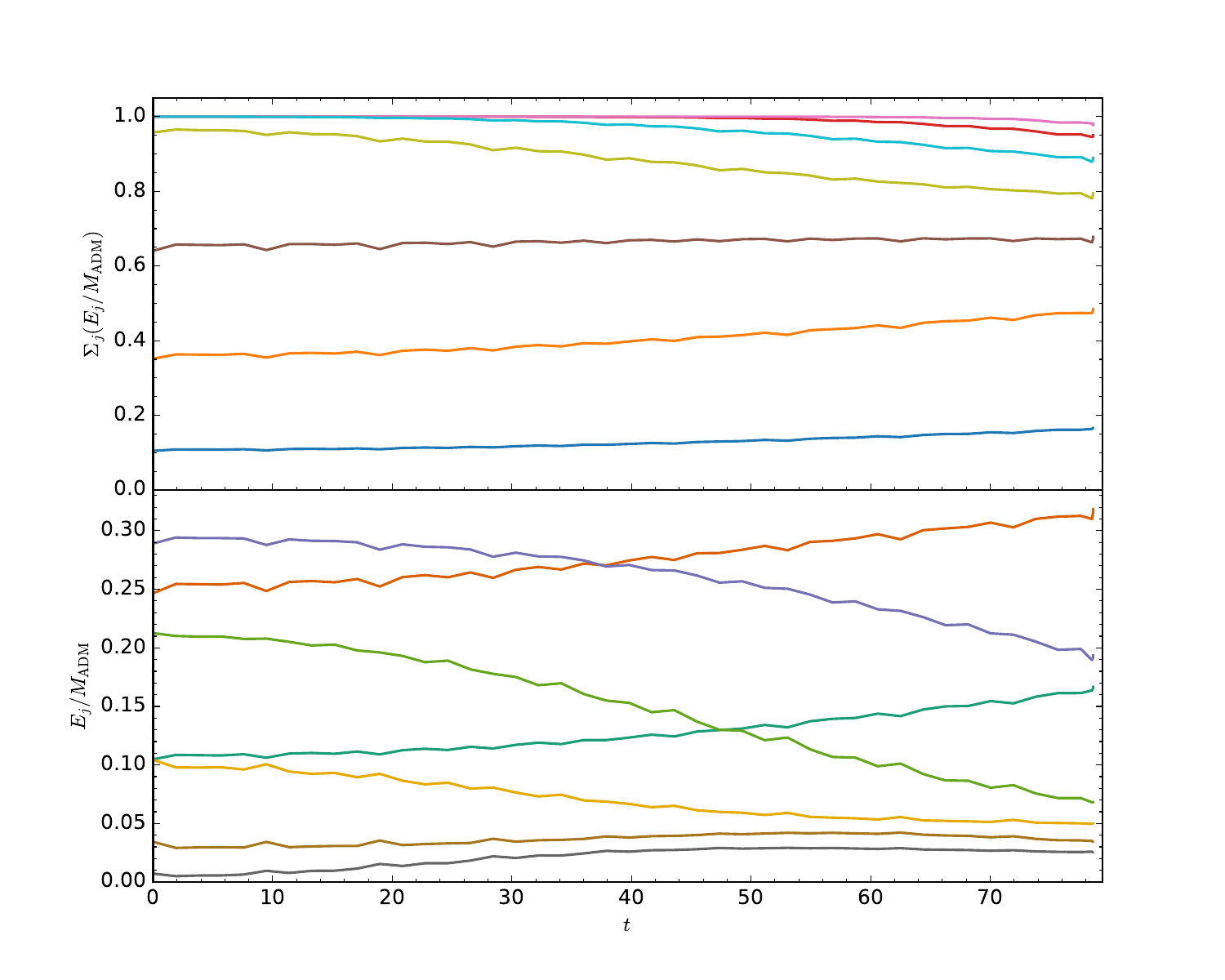}
\caption{$\mu=0,\sigma=0.25,\epsilon=2.28$}
\label{f:m0w025decomp}
\end{subfigure}
\begin{subfigure}[t]{0.47\textwidth}
\includegraphics[width=\textwidth]{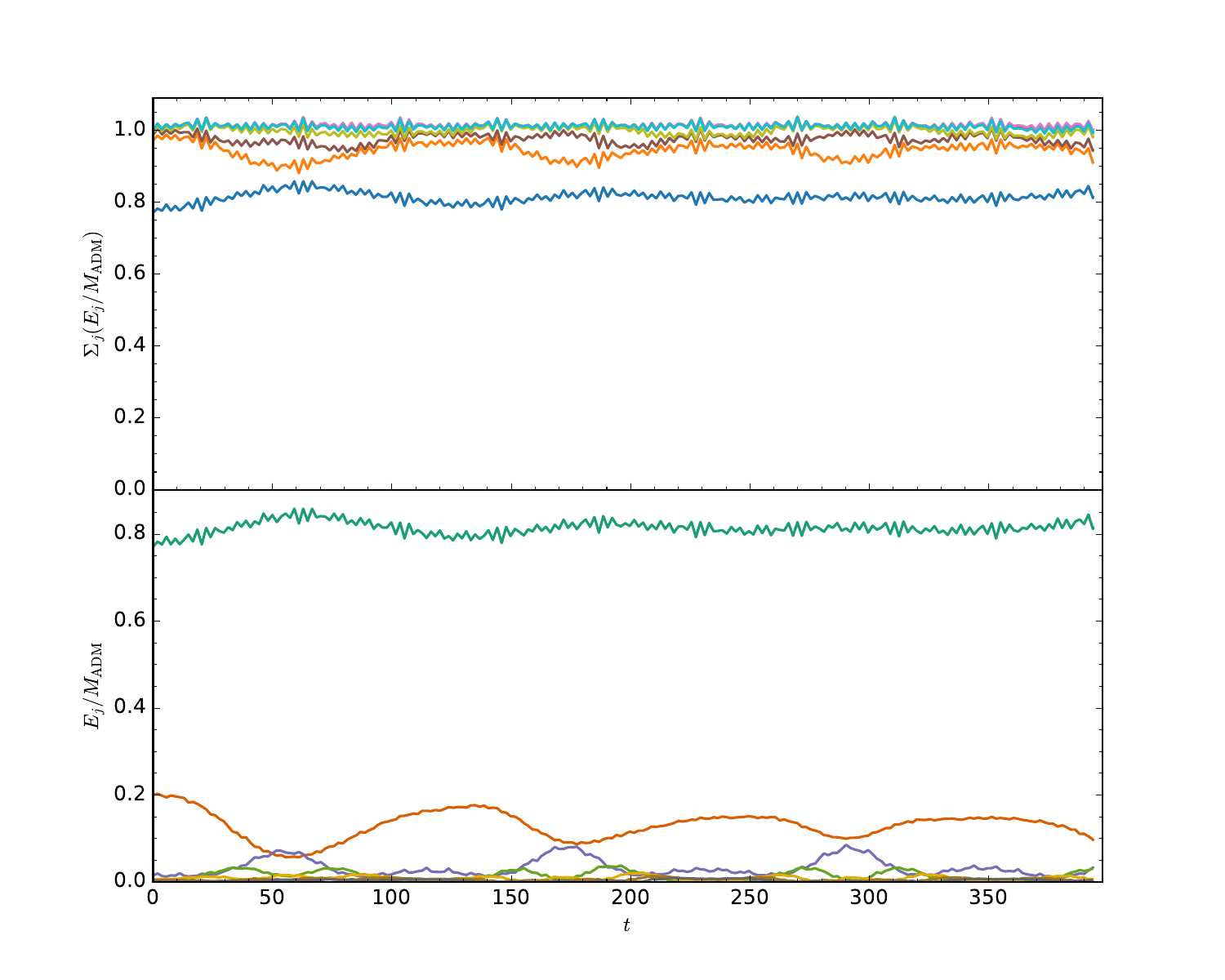}
\caption{$\mu=0.5,\sigma=1.7,\epsilon=0.216$}
\label{f:m05w17decomp}
\end{subfigure}
\begin{subfigure}[t]{0.47\textwidth}
\includegraphics[width=\textwidth]{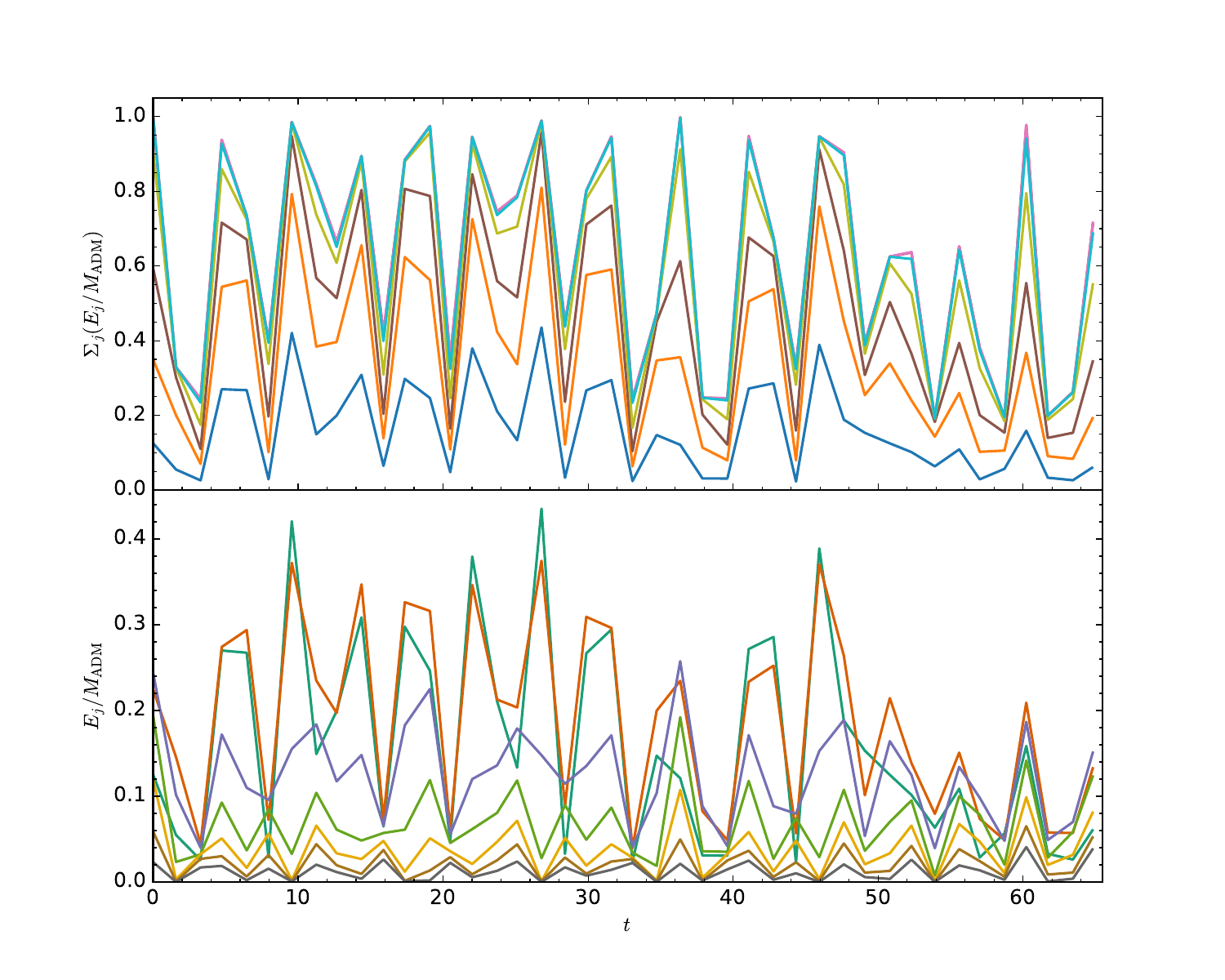}
\caption{$\mu=20,\sigma=0.19,\epsilon=6.95$}
\label{f:m20w019decomp}
\end{subfigure}
\caption{The time dependence of the energy spectra as a fraction of the
total ADM mass for the indicated
$\mu,\sigma,\epsilon$.  Lower panels show the lowest 7 modes (in colors
cyan, red, purple, green, yellow, brown, and gray
respectively).  Upper panels show cumulative energy to mode
$j=0,1,2,4,8,16,32$ (in colors blue, orange, brown, yellow, aqua, red, and
magenta). }
\label{f:evolvingspectra}
\end{figure}

While the initial spectrum for a given $(\mu,\sigma)$ pair does not have
predictive value regarding the future behavior as far as
we can tell, the time dependence of the spectrum throughout the evolution of
the system is informative.  Figure \ref{f:evolvingspectra} shows the
time-dependence of spectra for examples of the stable, unstable, metastable,
and chaotic irregular classes. In each figure, the lower panel shows the
fraction $E_j/M_{ADM}$ in each mode up to $j=6$, while the upper panel shows
the cumulative fraction $\sum_j E_j/M_{ADM}$ to the mode $2^k$ with $k=0$ to 5.

The difference between stable evolution in figure \ref{f:m0w18decomp} and
unstable evolution in figure \ref{f:m0w025decomp} is readily apparent.
As the evolution proceeds, we expect a cascade of energy into higher mode
numbers, but inverse cascades to lower modes can also occur.  The stable
evolution shows a slow pattern of cascades and inverse cascades, in fact.
On the other hand, the unstable evolution shows a nearly monotonic cascade of
energy into the highest modes along with a simultaneous cascade of energy
into the lowest modes (therefore depleting intermediate modes).  These
are common observations in the literature and are included here for
completeness.

The metastable evolution shown in figure \ref{f:m05w17decomp} is interesting
in light of the stable and unstable spectra.  The amplitude shown is
from the unstable portion of figure \ref{f:m05w17}, the part consistent
with the perturbative scaling $t_H\sim\epsilon^{-2}$.  However, the spectrum
shows a similar pattern of slow cascades and inverse cascades to the
stable initial data example, 
though on a somewhat faster time scale in this case.
While perhaps surprising, this is in keeping with the similarities noted
between the initial spectra in figures \ref{f:m0w15spect} and
\ref{f:m5w170spect}.  We have also checked that the time-dependent spectrum
at a higher amplitude with $t_H\sim 100$ follows the same pattern as
\ref{f:m05w17decomp}; in fact, it looks essentially the same but simply
ends at an earlier time.
This lends some credence to the idea that metastable initial data
are stable at lowest nontrivial order in perturbation theory, with
instability triggered by higher-order corrections.  Alternately, the
instability could be caused by an oscillatory singularity in the perturbative
theory, as discussed in \cite{1506.03519,1508.04943,1606.02712,1607.08094}
in the case of two-mode initial data.  These divergences do not appear in
the energy spectrum.

Figure \ref{f:m20w019decomp} shows the time-dependence of the spectrum
in an irregular evolution, specifically $\mu=20,\sigma=0.19$
at $\epsilon=6.95$, which is in the chaotic region of the $t_H$ vs $\epsilon$
plot in figure \ref{f:m20w019}.  There is rapid energy transfer among modes,
including cascades out of and inverse cascades into mode numbers $j\leq 32$
over approximately a light-crossing time.  It is easy to imagine that
horizon formation might occur at any of the cascades of energy into higher
modes, leading to seemingly random jumps in $t_H$ as a function of amplitude.

\begin{figure}[!t]
\centering
\begin{subfigure}[t]{0.47\textwidth}
\includegraphics[width=\textwidth]{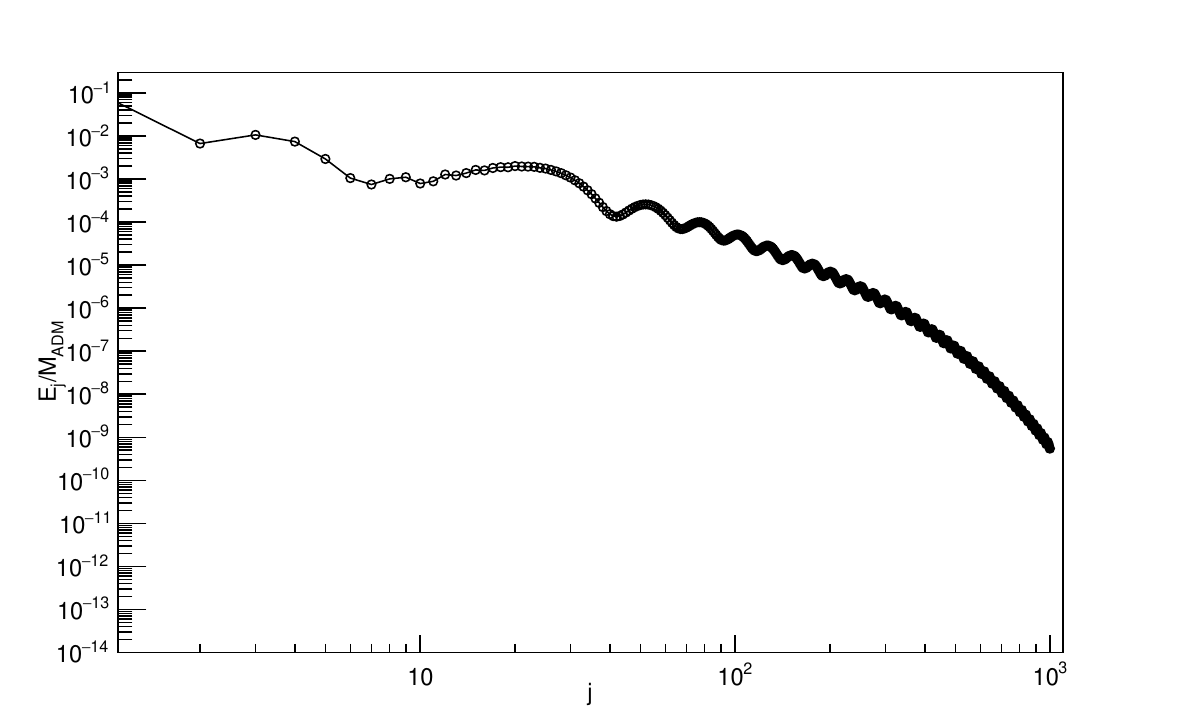}
\caption{$\epsilon=1.01$}
\label{f:m0w11e101}
\end{subfigure}
\begin{subfigure}[t]{0.47\textwidth}
\includegraphics[width=\textwidth]{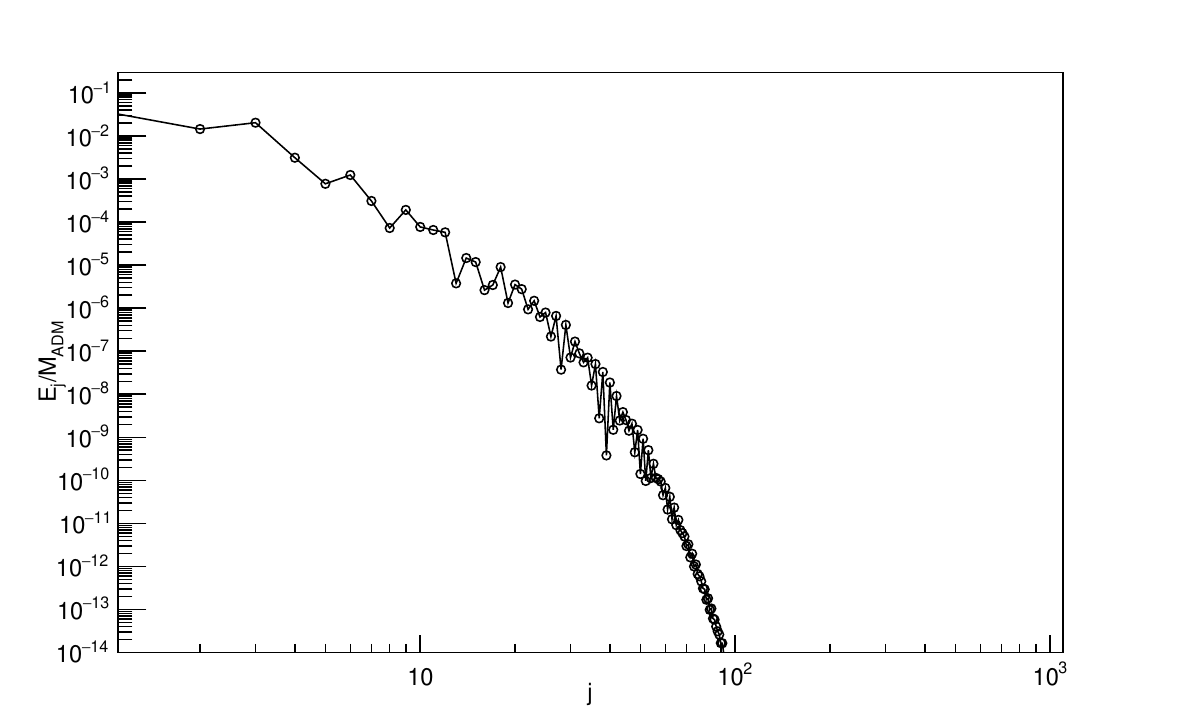}
\caption{$\epsilon=1.02$}
\label{f:m0w11e102}
\end{subfigure}
\caption{Spectra at time $t\approx 71$ for $\mu=0,\sigma=1.1$ for the two
amplitudes given.  $\epsilon=1.01$ forms a horizon at $t_H\approx 71.1$,
$\epsilon=1.02$ at $t_H\approx 248.0$.}
\label{f:chaoticspect}
\end{figure}

Finally, the time-evolved energy provides another possible measure of
approximate thermalization in the dual CFT; namely, the spectrum should
approach an (exponentially cut-off) power law at thermalization.  In most
cases, this occurs shortly before horizon formation, but there are exceptions,
such as the late time behavior of initial data below the critical mass
for black hole formation in Einstein-Gauss-Bonnet gravity \cite{1608.05402}.
When there is evidence of 
chaotic behavior, it is particularly interesting to know if
the spectra for similar amplitudes approach a power law at similar times
even if horizons form at very different times.  Figure \ref{f:chaoticspect}
shows the energy spectra for two amplitudes in the chaotic region of
the $t_H$ vs $\epsilon$ plot for $\mu=0,\sigma=1.1$.  Figure \ref{f:m0w11e101}
is the spectrum just before horizon formation for $\epsilon=1.01$,
while figure \ref{f:m0w11e102} is the spectrum at approximately the same
time for $\epsilon=1.02$, which is very long before horizon formation.  In
this example, we see that the spectrum does approach a power law for the
evolution that is forming a horizon, while the other evolution demonstrates
a more rapid decay (typically fit by a power law times an exponential in
the literature).  Therefore, this example suggests that
a power law spectrum may yield similar results to horizon formation as a
measure of thermalization in the dual CFT.

\section{Discussion}

For the first time,
we have presented the phase diagram of stability of AdS$_5$ against horizon
formation, treating the scalar field mass $\mu$ and width $\sigma$ of initial
data as free parameters.  In addition to mapping the location of the so-called
island of stability, we have gathered evidence for two non-perturbative
classes on the shorelines of the island, the metastable and irregular 
classes.
While these must either exhibit stability (no collapse below some critical
amplitude) or instability (collapse at arbitrarily small but finite amplitude)
as the amplitude $\epsilon\to 0$, they are distinguished by their behavior
at computationally accessible (finite) amplitudes.  
While perturbatively unstable
evolutions obey $t_H\propto\epsilon^{-2}$ as $\epsilon\to 0$ (and show
evidence of this behavior at finite $\epsilon$), metastable
initial data follows $t_H\propto\epsilon^{-p}$ for $p>2$ over a range of
amplitudes $\epsilon>0$. 
The irregular class is characterized by horizon formation times
$t_H$ that are not well described by a power law and sometimes exhibits
non-monotonicity or even evidence of chaos.  
Both of these classes appear across the range
of $\mu$ values that we study and at both small- and large-width boundaries
of the stable class of initial data.

At this time, it is impossible to say whether metastable initial data are
stable or unstable as $\epsilon\to 0$ (or if all metastable data behave in the
same way in that limit).  Our numerical evolutions include cases in which
the lowest amplitudes jump either to metastable scaling with smaller $p$
or to evolutions that do not collapse over the time scales we study.
We did find evidence that many metastable profiles move toward 
perturbatively unstable scaling ($t_H\propto\epsilon^{-2}$) as $\epsilon\to 0$
but more slowly than the initial data that we have classified as unstable.
It is also possible that some metastable initial data are stable in the
perturbative theory (ie, to $\epsilon^3$ order in a perturbative expansion)
but not at higher orders. We emphasize once again, however, that our interest
and therefore our classification is in small but finite $\epsilon$ behavior
(which is by definition not strictly in the perturbative regime).

The irregular class seems likely to be (mostly) stable at arbitrarily small
amplitudes based on our numerical evolutions, though we have not found a
critical amplitude for monotonic irregular initial data.  The irregular
initial data includes the quasistable initial data described in
\cite{1304.4166,1508.02709}, which has a sudden increase then decrease in
$t_H$ as $\epsilon$ decreases as well as evidence for chaotic behavior.  
In fact,
we have found evidence for weakly chaotic behavior 
for non-monotonic initial data in the form of a small but nonzero Lyapunov
coefficient and in the phase space trajectory.  Both
non-monotonicity and chaos become stronger and more common at larger scalar
masses; however, we have also found evidence of chaotic behavior for the 
massless scalar including in the $t_H$ vs $\epsilon$ curve. 
To our knowledge, this is the first evidence of chaos in this relationship
for spherically
symmetric massless scalar collapse in AdS, which is particularly interesting
because there is only one physically meaningful ratio of scales, $\sigma$
as measured in AdS units.

While we have emphasized the appearance of new behaviors outside
perturbation theory, metastable and irregular initial data are
interesting potential subjects for analysis in the multiscale
perturbation theory. A key question is if they demonstrate any unusual
behavior there or map directly onto the stable or unstable classes.

Aside from the ultimate stability or instability of metastable and
irregular initial data, several questions remain.  
For one, black holes formed in
massive scalar collapse in asymptotically flat spacetime exhibit a mass
gap for initial profiles wider than the Compton wavelength $1/\mu$
\cite{Brady1997}.  Whether this mass gap exists in AdS is not clear, and it
may disappear through repeated gravitational focusing as the field oscillates
many times across AdS; investigating this type of critical behavior will
likely require techniques similar to those of \cite{Santos-Olivan:2016djn}.
Returning to our stability 
phase diagram, the physical mechanism responsible for chaos that seems to occur
for some irregular initial data 
is not yet clear.  Is it some generalization of the
same mechanism as found in the two-shell system?  Also, would an alternate
definition of approximate thermalization in the dual CFT, such as development
of a power-law spectrum, lead to a different picture of the stability
phase diagram?
Finally, the big question is whether there is some test that could be performed
on initial data alone that would predict in advance its behavior? So far, no
test is entirely successful, so new ideas are necessary.

\begin{acknowledgments}
We would like to thank Brayden Yarish for help submitting jobs for the
$\mu=10$ evolutions.
The work of ND is supported in part by a Natural Sciences and Engineering
Research Council of Canada PGS-D grant to ND, NSF Grant PHY-1606654
at Cornell University, and by a grant from the Sherman
Fairchild Foundation.
The work of BC and AF is supported by Natural Sciences
and Engineering Research Council of Canada Discovery Grants SAPIN-2015-00046 
and SAPIN-2020-00054.
This research was enabled in part by support provided by WestGrid
(www.westgrid.ca) and Compute Canada Calcul Canada (www.computecanada.ca).
\end{acknowledgments}

\appendix
\section{Convergence Testing}

Due to the large number of evolutions we have carried out, it is not
computationally feasible to test all of them for convergence.  Therefore,
we have checked several interesting cases of irregular initial data, which 
are the most curious.  These are carried out by evolving the initial data
with a base resolution $n=14$ and again at $n=15,16$ with commensurate
time steps, as described in \cite{1508.02709}.  In the cases indicated, we
evaluated the order of convergence at lower resolutions.  We remind the
reader that the order of convergence $Q$ is the base-2 logarithm of the ratio
of $L^2$ errors (root-mean-square over all corresponding grid points) between
successive pairs of resolutions.  We also note that the initial data are 
defined analytically, so $Q$ can appear poor at $t=0$ since the errors are
controlled by round off; in some cases, $Q$ is therefore undefined and 
not plotted.

\begin{figure}[!t]
\centering
\begin{subfigure}[t]{0.47\textwidth}
\includegraphics[width=\textwidth]{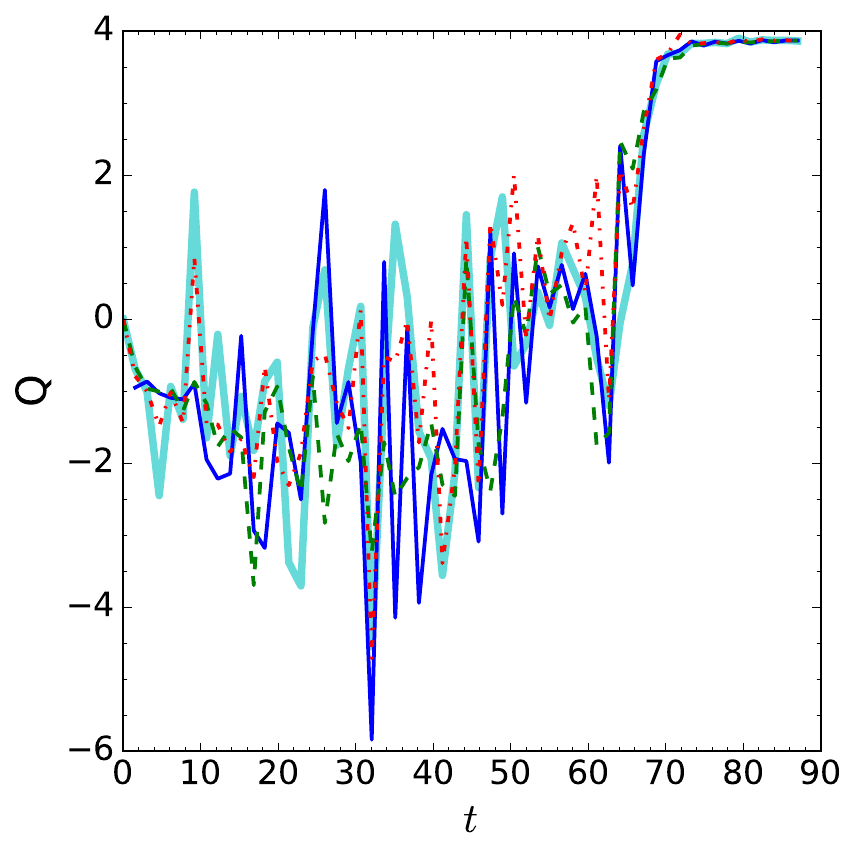}
\caption{$n=14$ base resolution}
\label{f:m05w1e112conv}
\end{subfigure}
\begin{subfigure}[t]{0.47\textwidth}
\includegraphics[width=\textwidth]{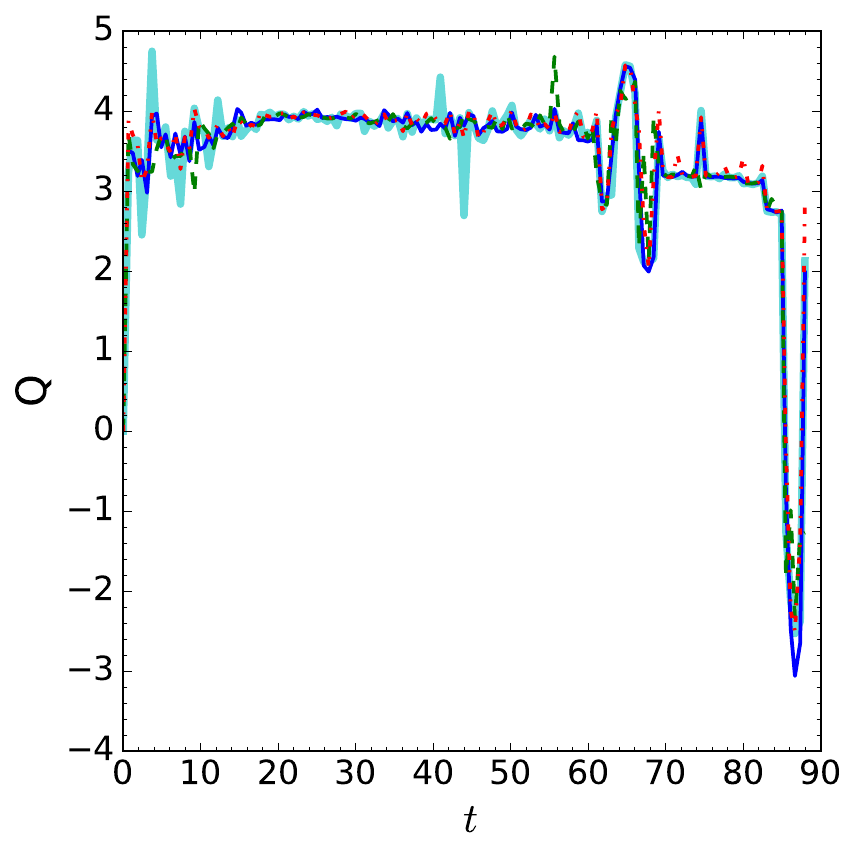}
\caption{$n=12$ base resolution}
\label{f:m05w1e112conv11}
\end{subfigure}
\caption{Convergence results for $\mu=0.5$, $\sigma=1$, $\epsilon=1.12$
showing order of convergence $Q$ vs time for $\phi,M,A,\delta$ 
(blue thin solid line, green dashed line, red dash-dotted line, cyan thick solid line, respectively). Left: Resolutions $n=14,15,16$ used.
Right: Resolutions $n=12,13,14$ used.}
\label{f:m05w1convergence}
\end{figure}

First, we carried out convergence tests for mass $\mu=0.5$, width
$\sigma=1$, and amplitude $\epsilon=1.12$, which is monotonic irregular
initial data presented in figure \ref{f:m05w1}.  
This amplitude collapses with $t_H\sim 88$. Figure \ref{f:m05w1e112conv}
shows the ($L^2$ norm) order of convergence for the field variable 
$\phi$, the mass 
function $M$, and the metric functions $A,\delta$.  While the order of
convergence is initially poor and even negative, all these variables show
approximately fourth order convergence for times $t\gtrsim 70$.  The 
reason for the initially poor convergence is that the error between 
successive resolutions is already given by (machine limited) round off.
As a demonstration, we tested the order of convergence
with base resolution $n=12$, as shown in figure \ref{f:m05w1e112conv11}.
The variables show order of convergence $Q\gtrsim 3$ already at this resolution
for most of the evolution, losing convergence only for $t>80$, where we
see approximately 4th-order convergence in the $n=14$ resolution
computations.

Two of the authors have discussed the convergence properties of evolution for
the nonmonotonic irregular initial data with 
$\mu=20,\sigma=0.1,\epsilon=11.74$, which is in an amplitude region of 
increased $t_H$ surrounded by smaller values, in detail in \cite{1508.02709}.
In short, the variables $\phi,M,A,\delta$ all exhibit fourth order 
convergence, as does $\Pi^2(t,0)$, and the conserved mass actually has 6th 
order convergence.  

\begin{figure}[!t]
\centering
\begin{subfigure}[t]{0.31\textwidth}
\includegraphics[width=\textwidth]{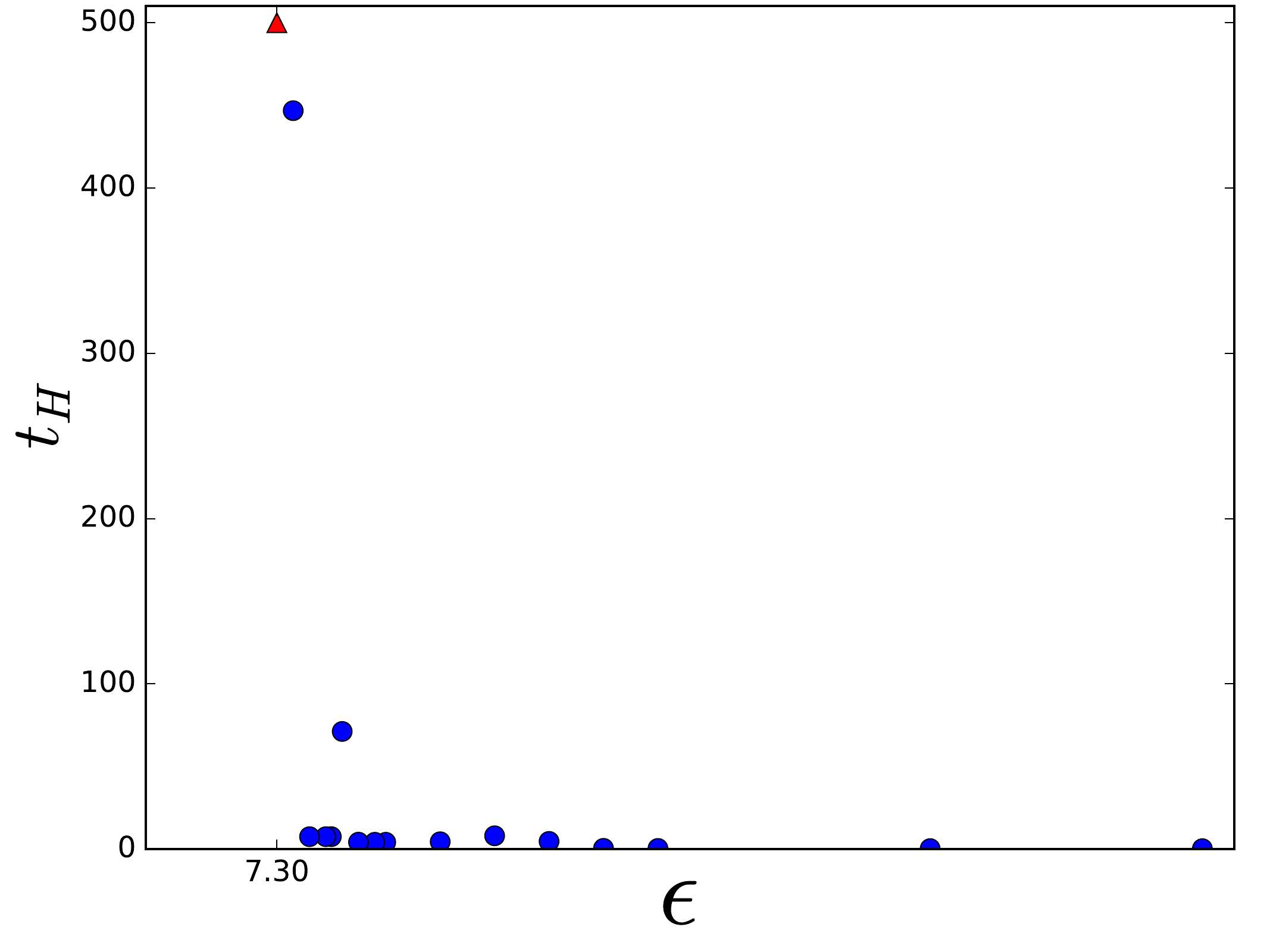}
\caption{Horizon formation}
\label{f:m15w020}
\end{subfigure}
\begin{subfigure}[t]{0.31\textwidth}
\includegraphics[width=\textwidth]{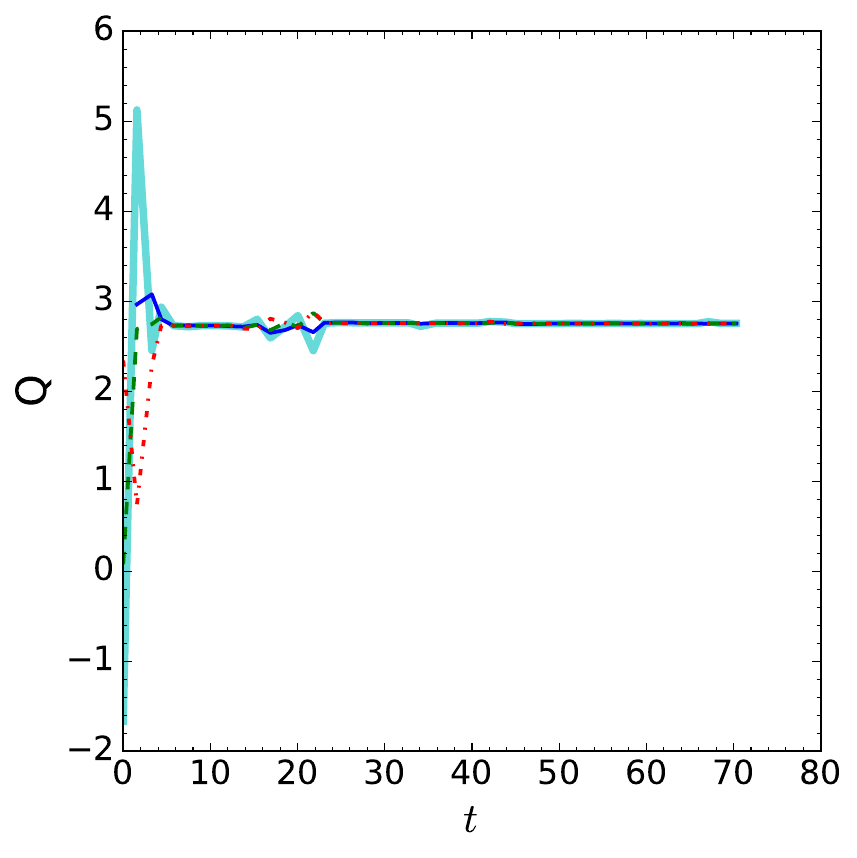}
\caption{$\epsilon=7.42$}
\label{f:m15w020A742conv}
\end{subfigure}
\begin{subfigure}[t]{0.31\textwidth}
\includegraphics[width=\textwidth]{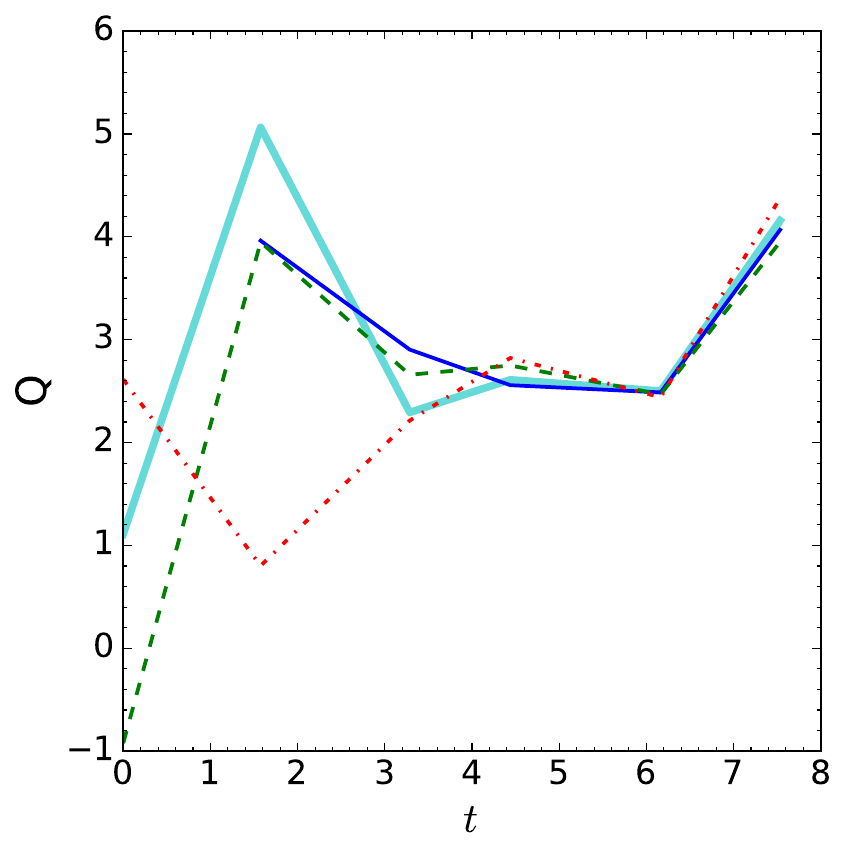}
\caption{$\epsilon=7.40$}
\label{f:m15w020A740conv}
\end{subfigure}
\caption{Convergence results for $\mu=15$, $\sigma=0.2$.
Left: $t_H$ vs $\epsilon$.  Middle \& Right: order of convergence vs time 
for $\phi,M,A,\delta$ (blue thin solid line, green dashed line, red 
dash-dotted line, cyan thick solid line, respectively)
for indicated amplitudes.
}
\label{f:m15w020convergence}
\end{figure}

Initial data for $\mu=15,\sigma=0.2$ are also nonmonotonic, as shown in 
figure \ref{f:m15w020}.  While we have not analyzed all aspects of the 
convergence, we see from the remainder of figure \ref{f:m15w020convergence}
that $\phi,M,A,\delta$ exhibit convergent behavior at better than
second order for $\epsilon=7.42$
(figure \ref{f:m15w020A742conv}, second-largest value of $t_H$ in 
figure \ref{f:m15w020}) and $\epsilon=7.40$ (figure \ref{f:m15w020A740conv},
adjacent amplitude in figure \ref{f:m15w020}).  It is important to note that
the larger amplitude also has the larger horizon formation time, contrary
to the usual monotonic behavior.  In other words, we have validated the
nonmonotonicity of this initial data through convergence testing.

\begin{figure}[!t]
\centering
\begin{subfigure}[t]{0.31\textwidth}
\includegraphics[width=\textwidth]{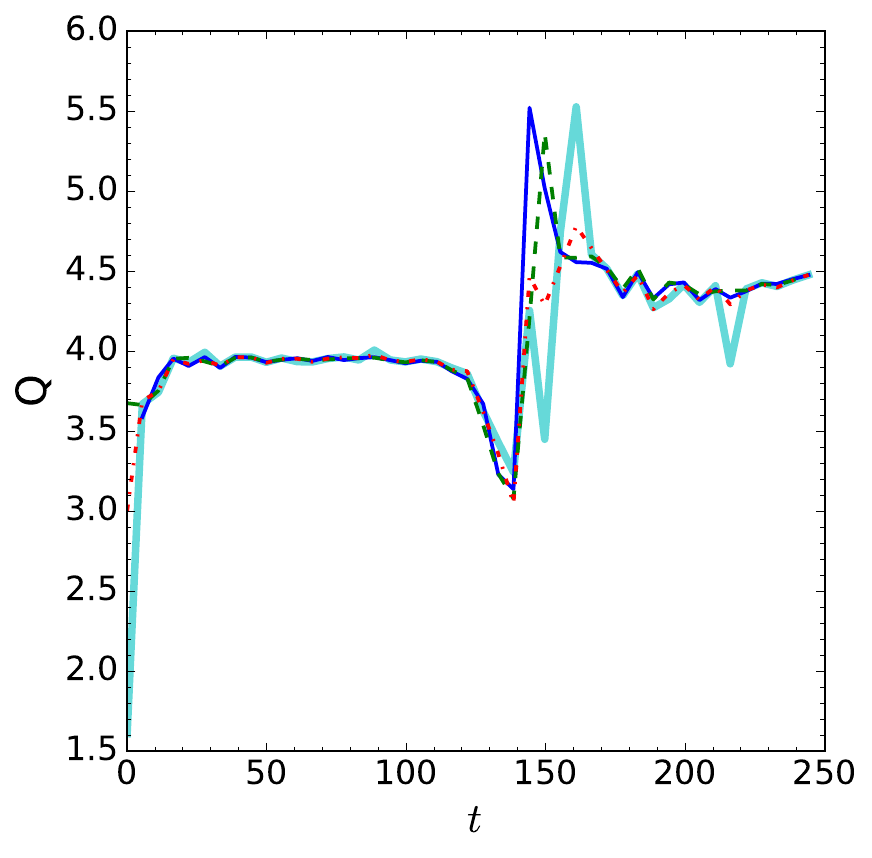}
\caption{$\epsilon=1.02$}
\label{f:m0w11A102conv}
\end{subfigure}
\begin{subfigure}[t]{0.31\textwidth}
\includegraphics[width=\textwidth]{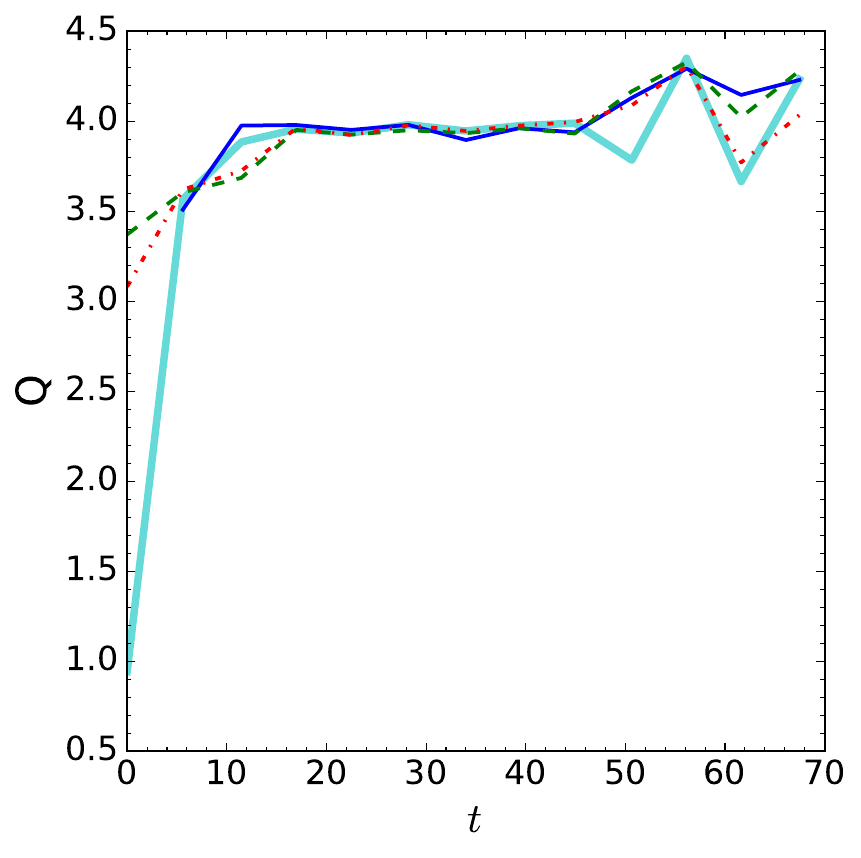}
\caption{$\epsilon=1.01$}
\label{f:m0w11A101conv}
\end{subfigure}
\begin{subfigure}[t]{0.31\textwidth}
\includegraphics[width=\textwidth]{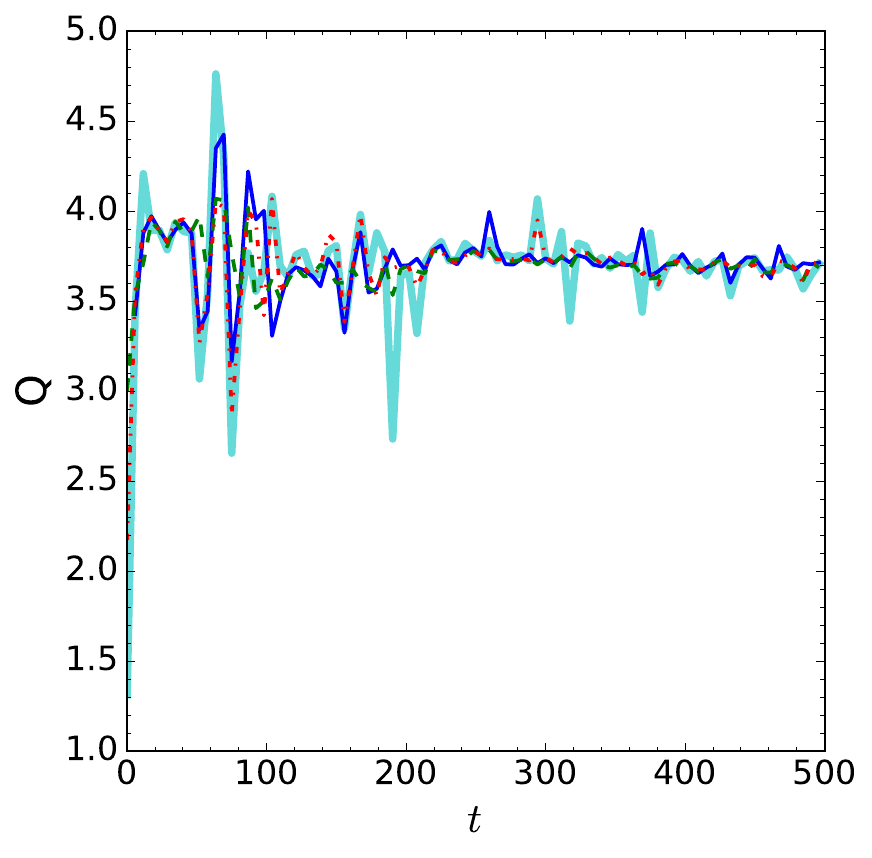}
\caption{$\epsilon=1.00$}
\label{f:m0w11A100conv}
\end{subfigure}
\caption{Convergence results for $\mu=0$, $\sigma=1.1$ for listed amplitudes
showing order of convergence $Q$ vs time for $\phi,M,A,\delta$ 
(blue thin solid line, green dashed line, red dash-dotted line, cyan thick 
solid line, respectively); resolutions $n=12,13,14$.
}
\label{f:m0w11convergence}
\end{figure}

It is most crucial to validate the convergence of chaotic evolutions.
In table \ref{t:lyap}, we noted that the Ricci scalar at the origin has
nonzero Lyapunov exponent at almost the 2 sigma level for amplitudes
$\epsilon=1.02,1.01,1.00$ for $\mu=0,\sigma=1.1$.  We show the results of
convergence tests for these amplitudes in figure \ref{f:m0w11convergence};
because these are longer evolutions, we consider the convergence at the lower
resolutions $n=12,13,14$.  After a transient start-up period, these are all
convergent with $Q>2.5$ for all variables considered at all times; for most
of the time, the order of convergence is $Q>3.5$.  It is worth noting that
one of the amplitudes does not form a horizon through $t=500$.  
These convergence tests validate both the nonmonotonic nature of the
evolution ($t_H\approx 248,71$ and $>500$ for $\epsilon=1.02,1.01,1.00$
respectively) and also the calculation of the Lyapunov coefficient.

\begin{figure}[!t]
\centering
\begin{subfigure}[t]{0.31\textwidth}
\includegraphics[width=\textwidth]{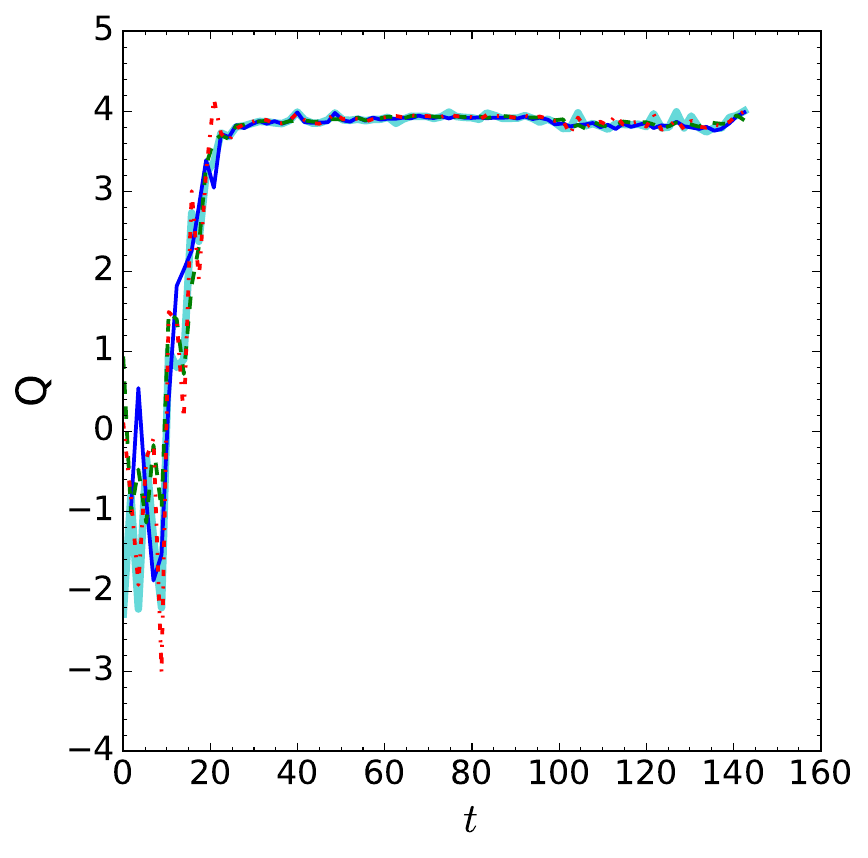}
\caption{$\epsilon=3.52$}
\label{f:m5w034A352conv}
\end{subfigure}
\begin{subfigure}[t]{0.31\textwidth}
\includegraphics[width=\textwidth]{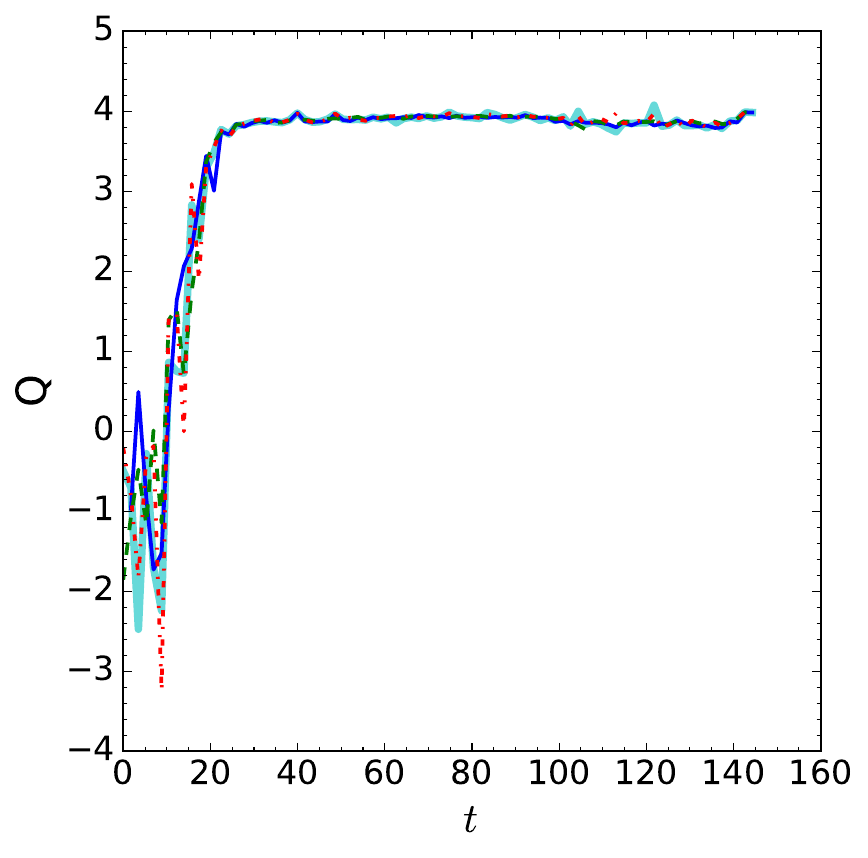}
\caption{$\epsilon=3.51$}
\label{f:m5w034A351conv}
\end{subfigure}
\begin{subfigure}[t]{0.31\textwidth}
\includegraphics[width=\textwidth]{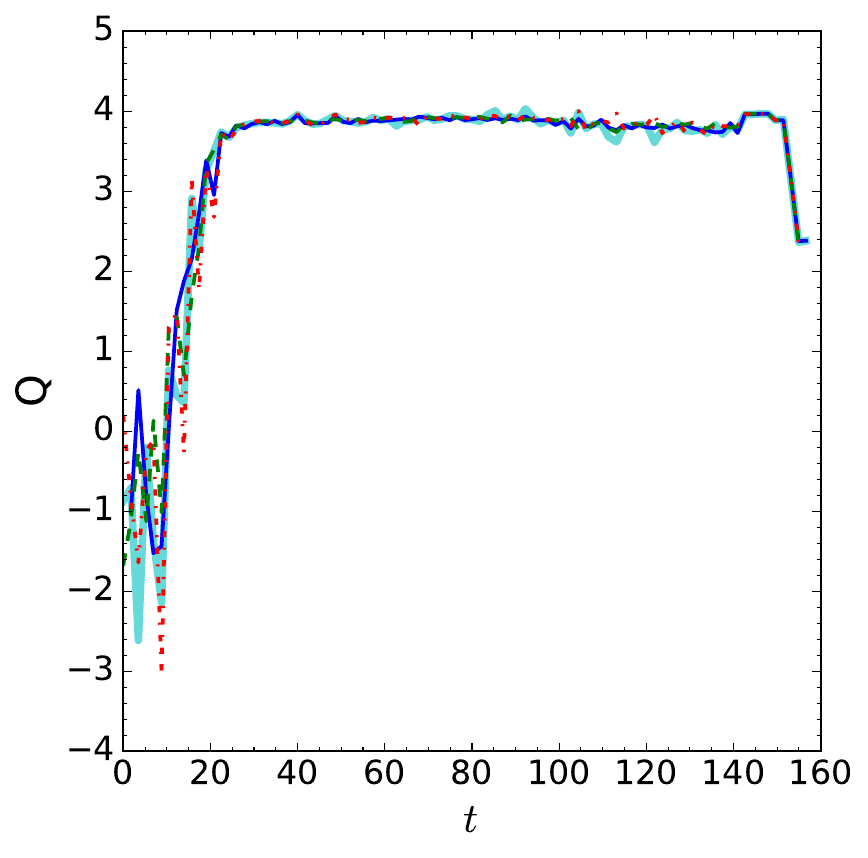}
\caption{$\epsilon=3.50$}
\label{f:m5w034A350conv}
\end{subfigure}
\caption{Convergence results for $\mu=5$, $\sigma=0.34$ for listed amplitudes
showing order of convergence $Q$ vs time for $\phi,M,A,\delta$ 
(blue thin solid line, green dashed line, red dash-dotted line, cyan thick 
solid line, respectively); resolutions $n=14,15,16$.
}
\label{f:m5w034convergence}
\end{figure}

Also in table \ref{t:lyap}, we found a nonzero Lyapunov exponent for 
$\mu=5,\sigma=0.34$ at amplitudes $\epsilon=3.52,3.51,3.50$.  The results
of convergence tests for these amplitudes appear in figure 
\ref{f:m5w034convergence}.  For $t\gtrsim 20$, these evolutions exhibit 
convergent behavior with $Q>3.5$ (and always $Q>2$).  At early times, the 
apparent poor convergence is again due to the errors being dominated by
round-off; we have carried out additional convergence tests (not shown) and
verified that these evolutions are already convergent with order of 
convergence close to $Q=4$ at base resolutions $n=12$ for $t\lesssim 20$.
Again, convergence tests validate chaotic behavior for these initial data.

\begin{figure}[!t]
\centering
\begin{subfigure}[t]{0.31\textwidth}
\includegraphics[width=\textwidth]{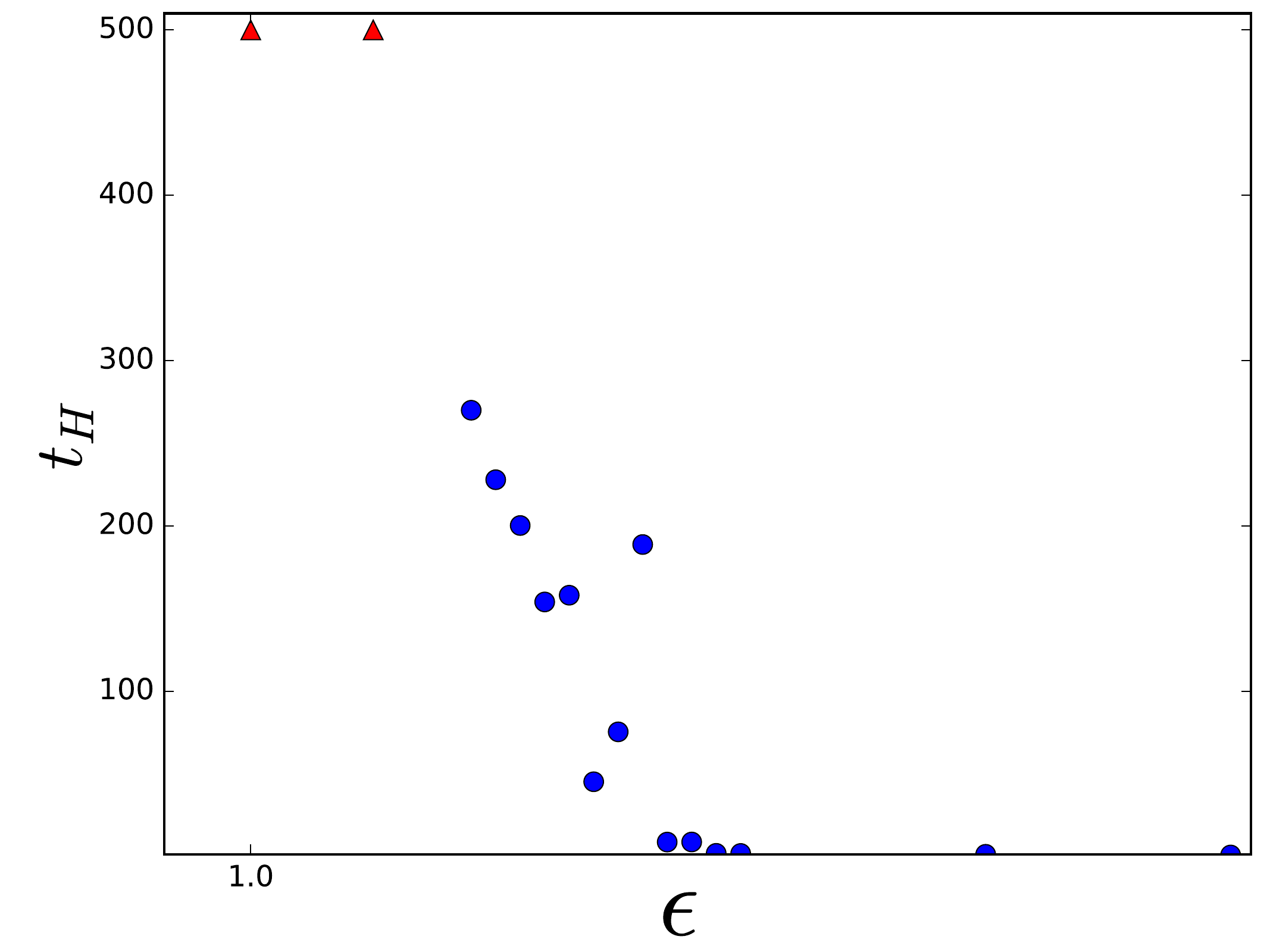}
\caption{Horizon formation}
\label{f:m1w100}
\end{subfigure}
\begin{subfigure}[t]{0.31\textwidth}
\includegraphics[width=\textwidth]{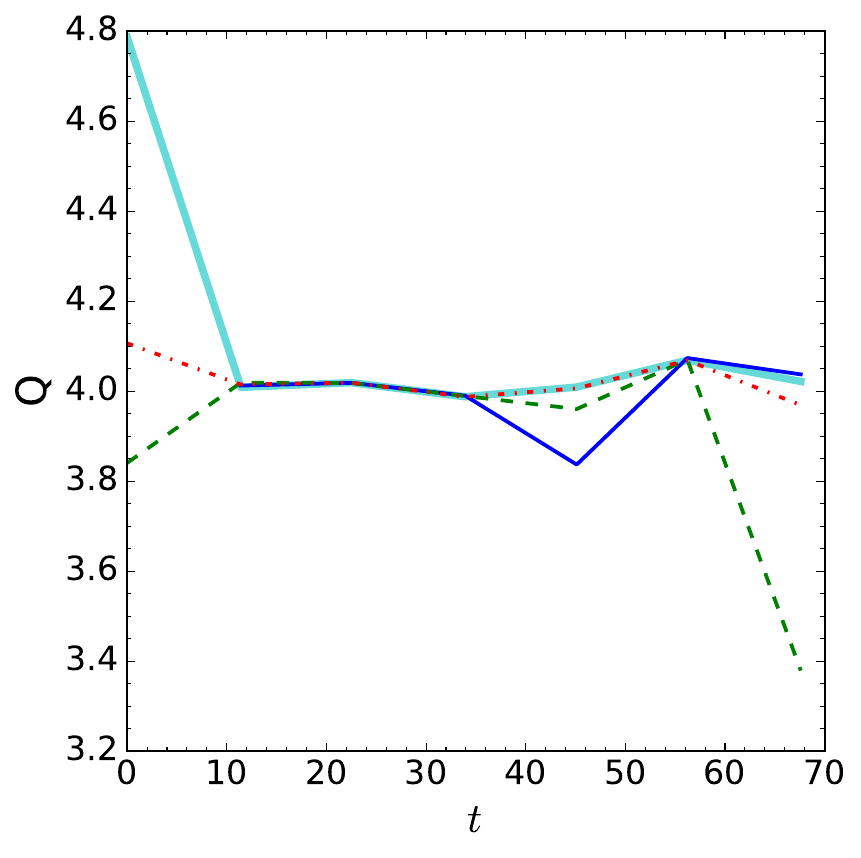}
\caption{$\epsilon=1.15$}
\label{f:m1w100A115conv}
\end{subfigure}
\begin{subfigure}[t]{0.31\textwidth}
\includegraphics[width=\textwidth]{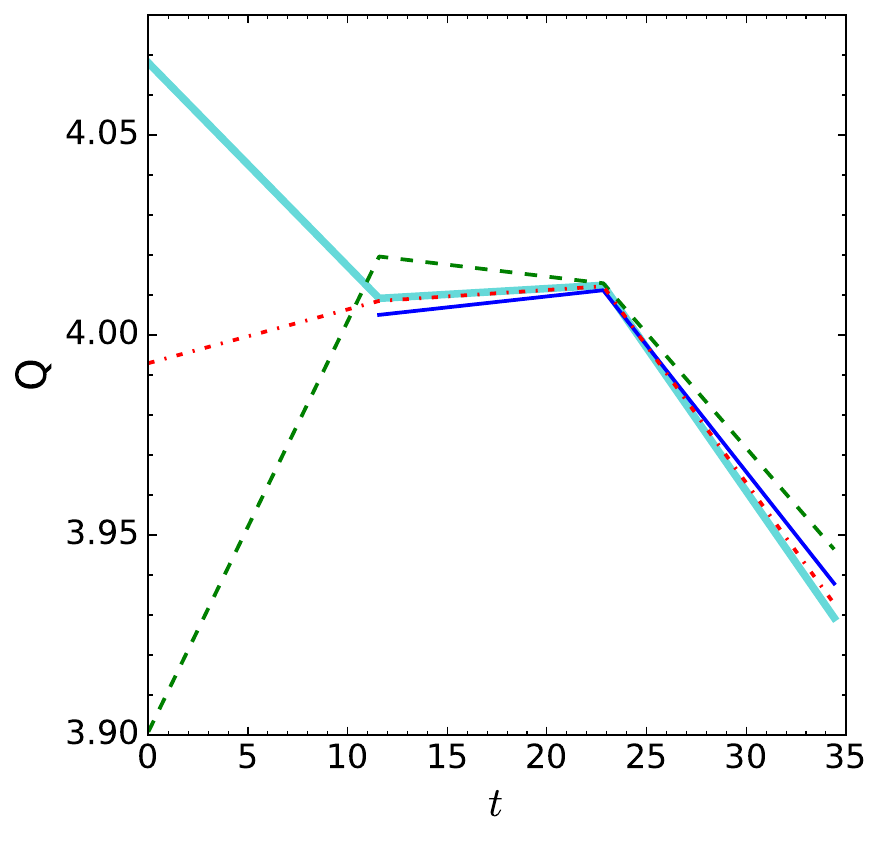}
\caption{$\epsilon=1.14$}
\label{f:m1w100A114conv}
\end{subfigure}
\caption{Convergence results for $\mu=1$, $\sigma=1$. Left: $t_H$ vs $\epsilon$.
Middle \& Right: order of convergence $Q$ vs time for $\phi,M,A,\delta$ 
(blue thin solid line, green dashed line, red dash-dotted line, cyan thick 
solid line, respectively); resolutions $n=11,12,13$.
}
\label{f:m1w100convergence}
\end{figure}

Initial data with $\mu=1,\sigma=1$ are chaotic over a narrow range of 
amplitudes.  We have carried out convergence testing for amplitudes 
$\epsilon=1.15,1.14$, which are the two amplitudes with $t_H<100$ between
amplitudes with $t_H\gtrsim 150$ in figure \ref{f:m1w100}.  The order of
convergence was poor for these amplitudes in our initial tests with base
resolution $n=14$ because the error between resolutions was dominated by 
round-off, similar to the convergence tests we discussed above for
$\mu=0.5,\sigma=1$.  In subsequent tests with lower resolutions $n=11,12,13$,
we find an order of convergence $Q\sim 4$ for most of the evolutions
(and always $Q>3$). It is important to note again
that our evolutions exhibit convergence while showing horizon formation at 
a later time for a larger amplitude in this case, again validating the
nonmonotonic behavior.

\begin{figure}[!t]
\centering
\begin{subfigure}[t]{0.47\textwidth}
\includegraphics[width=\textwidth]{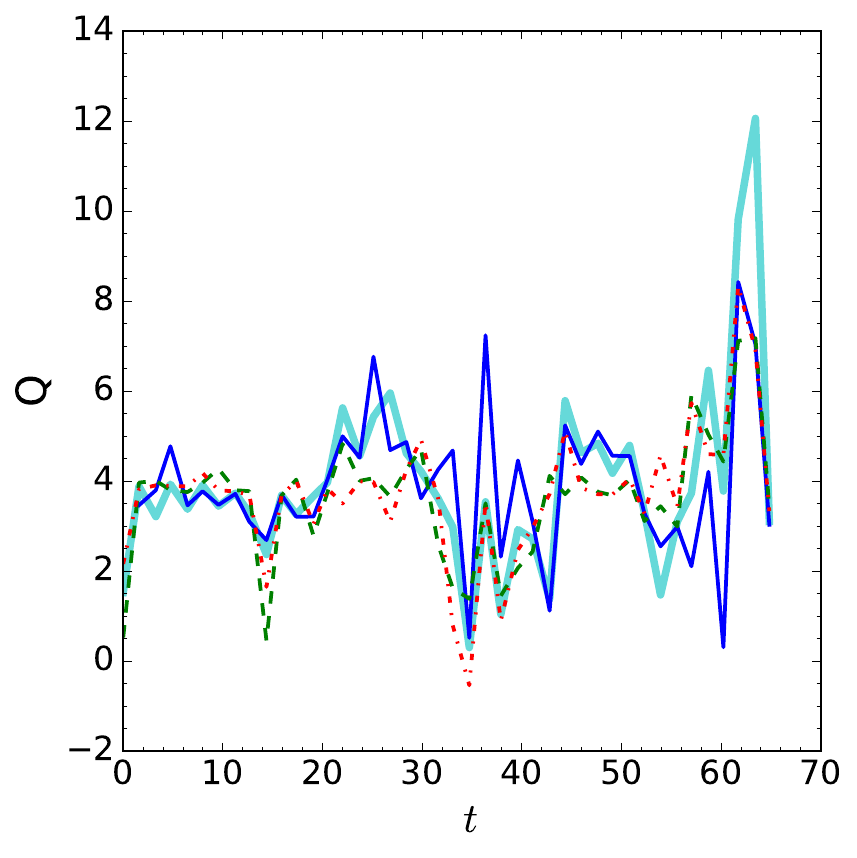}
\caption{$\epsilon=6.95$}
\label{f:m20w019A695conv}
\end{subfigure}
\begin{subfigure}[t]{0.47\textwidth}
\includegraphics[width=\textwidth]{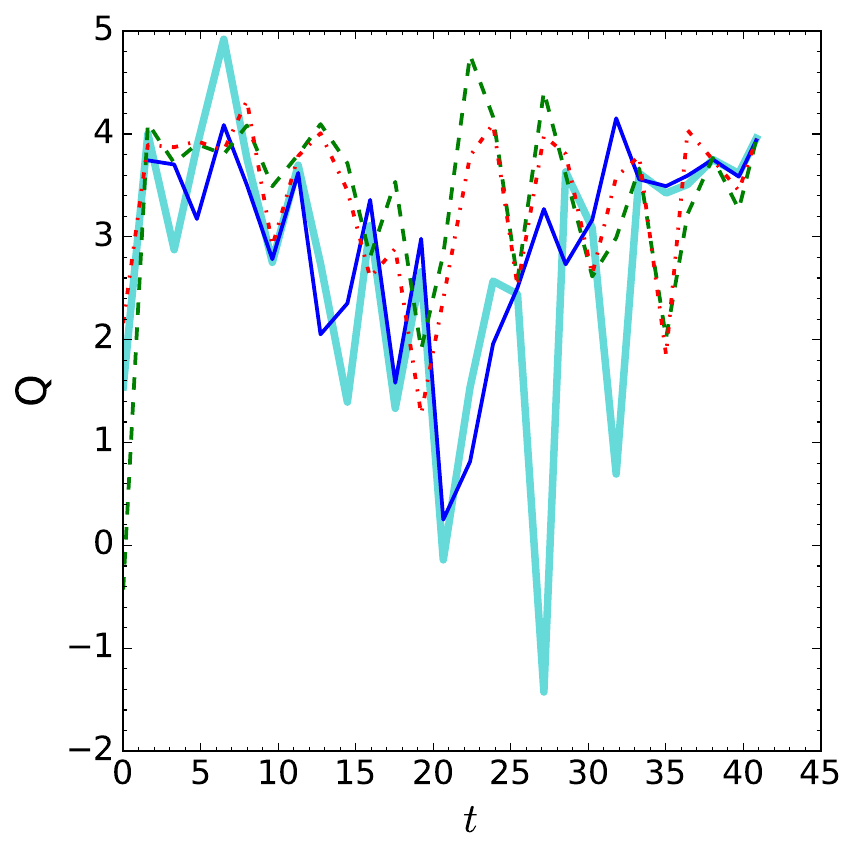}
\caption{$\epsilon=6.92$}
\label{f:m20w019A692conv}
\end{subfigure}
\caption{Order of convergence vs time for $\phi,M,A,\delta$ (blue thin solid 
line, green dashed line, red dash-dotted line, cyan thick solid line, 
respectively) for $\mu=20,\sigma=0.19$ and indicated amplitudes. 
}
\label{f:m20w019convergence}
\end{figure}

Finally, we ran convergence tests for the chaotic initial data with
$\mu=20,\sigma=0.19$ for $\epsilon=6.95,6.92$, with $t_H\approx 65.5,40.8$
respectively.  As shown in figure \ref{f:m20w019convergence}, the simulations
are close to fourth order convergence for most of the evolution, but
there are periods where the order of convergence for evolution and constraint
variables becomes negative.  This of course leads to the concern that the
evolutions should have collapsed during those periods and extend into an
``afterlife'' evolution.  We have therefore
evolved these amplitudes through these regions (approximately $t=30-40$ for
$\epsilon=6.95$ and $t=18-30$ for $\epsilon=6.92$) at high resolution ($n=18$).
If the evolutions are truly in an afterlife, this higher resolution 
calculation may include horizon formation.  We do not observe this.  Another
tell-tale of would-be horizon formation is a decrease in the time step size
by an order of magnitude or more followed by an increase.  We monitor
the time step size every 500 time steps through this evolution but do not
observe a decrease in time step size by more than a factor of 2.  As a result,
we believe the values of $t_H$ found are reliable, though the reader may 
wish to consider them with some caution.  In other words, while convergence
testing is the gold standard to validate our numerical evolutions, there are
other indicators of reliability, which these evolutions satisfy.  It is
also worth noting that the rapid energy transfer characteristic of figure
\ref{f:m20w019decomp} for $\epsilon=6.95$ begins immediately and is therefore
seen in a convergent region of the evolutions, particularly for $t\lesssim 14$.

Nonetheless, we emphasize that we have found convergent evolutions for
irregular initial data at scalar masses from $\mu=0$ to $20$.  It is
important to note that we have validated nonmonotonic behavior in plots of
$t_H$ vs $\epsilon$.  Convergence testing also specifically validates the
evolutions used to find a nonzero Lyapunov coefficient (at nearly the 2$\sigma$
level) for massless scalar collapse.

\bibliography{massive2}

\end{document}